\documentclass[11pt,a4paper,DIV=11,numbers=noenddot,parskip=half]{scrartcl}
\pdfoutput=1

\usepackage[T1]{fontenc}
\usepackage{lmodern}
\usepackage{tipa} 
\usepackage{amsmath}
\usepackage{mathtools}
\usepackage{amssymb}
\usepackage{graphicx}
\usepackage[absolute]{textpos}
\usepackage[noadjust]{cite}
\usepackage[affil-it]{authblk}
\usepackage{xcolor}
\usepackage{booktabs}
\usepackage{bold-extra}
\usepackage[labelformat=simple]{subcaption}
\usepackage{url}
\usepackage{microtype}
\expandafter\def\expandafter\UrlBreaks\expandafter{\UrlBreaks\do\_}
\usepackage{fancyvrb} 
\VerbatimFootnotes    
\usepackage[font=small,labelfont=bf,format=plain,margin=0.05\textwidth]{caption}
\usepackage[pdftitle={Automatic amplitude generation with DiaGen},
  pdfauthor={Micha\l{} Czakon, Marco Niggetiedt, and Johann Usovitsch},
pdfkeywords={Feynman diagrams, automated amplitude
generation, graph theory, integral families,
computer algebra, integration-by-parts reduction},
  bookmarks=true, linktocpage,
  colorlinks=true, allbordercolors=white, allcolors=blue]{hyperref}

\newcommand{\code}[1]{\texttt{#1}}

\newcommand*{\kira}{\code{Kira}}
\newcommand*{\diagen}{\texorpdfstring{\code{DiaGen}}{DiaGen}}
\newcommand*{\gitlab}{\code{GitLab}}
\newcommand*{\form}{\code{FORM}}

\newcommand{\pheaderline}{{\footnotesize\begin{tabular}{@{}r@{}}
HU-EP-26/40-RTG\\ P3H-26-097\\ TTK-26-44\\ ZU-TH 51/26
\end{tabular}}}

\title{Automatic amplitude generation with DiaGen}

\author[,a]{Micha\l{} Czakon%
  \thanks{E-mail: \href{mailto:mczakon@physik.rwth-aachen.de}{mczakon@physik.rwth-aachen.de}}}

\author[,b]{Marco Niggetiedt%
  \thanks{E-mail: \href{mailto:marco.niggetiedt@physik.uzh.ch}{marco.niggetiedt@physik.uzh.ch}}}

\author[,c]{Johann Usovitsch%
  \thanks{E-mail: \href{mailto:jusovitsch@googlemail.com}{jusovitsch@googlemail.com}}}

\affil[a]{Institute~for~Theoretical~Particle~Physics~and~Cosmology,
          RWTH~Aachen~University, 52056~Aachen, Germany}

\affil[b]{Physik-Institut, Universit\"at Z\"urich, Winterthurerstrasse 190, CH-8057 Z\"urich, Switzerland}

\affil[c]{
Institut f\"ur Physik, Humboldt-Universit\"at zu Berlin,
10099 Berlin, Germany
}

\date{}

\begin{document}

\maketitle
\thispagestyle{empty}
\vspace{-3\baselineskip}

\begin{abstract}
\diagen{} is a \texttt{C++} program for automated multiloop amplitude generation with integrated identification of Feynman integral topologies. Given a user-defined quantum field theory model and a scattering process, the program generates compatible diagrams at a specified loop order, assigns fields and momentum routings, determines fermionic signs and combinatorial symmetry factors, and produces \form{}-readable amplitudes for subsequent algebraic processing.

The central feature is the construction of a symmetry-aware prototype map during diagram generation. Propagator structures are classified using mass-labelled graph isomorphisms and, for vacuum and two-point integrals, canonical momentum representations, while admissible subprototypes are related to parent prototypes. The resulting mappings organize scalar integrals into a compact set of representatives and provide transformations needed for subsequent matching. \diagen{} exports these data, together with kinematic information, in a form suitable for integration-by-parts reduction with \kira{}.

We present the diagram and prototype conventions, explain the workflow from model input to integral families, and report benchmark timings for representative multiloop processes. We also describe the inverse Kirchhoff reconstruction of scalar diagrams from integral-family sectors and the subsequent determination of consistent edge orientations. The scope of the implemented equivalences and the associated computational trade-offs are stated explicitly.

Since a given input always yields byte-identical output, \diagen{} can also serve as a deterministic, verifiable component in workflows driven by AI coding agents: the agent prepares inputs and orchestrates the calculation, while the exact combinatorics of diagram generation and topology identification are delegated to a reproducible program.
\end{abstract}


\begin{textblock*}{34mm}(\dimexpr1in+\oddsidemargin+\textwidth-34mm\relax,12mm)
\raggedleft\noindent
\pheaderline
\end{textblock*}

\newpage

\section*{PROGRAM SUMMARY}

\textit{Program title:} \diagen{}

\textit{Developer's repository link:} \url{https://gitlab.com/diagen-cpp/diagen}

\textit{Licensing provisions:} GNU General Public License 3 (GPL)

\textit{Programming language:} \texttt{C++}

\textit{External routines/libraries used:} \texttt{GiNaC} [1,2] with \texttt{CLN}, \texttt{yaml-cpp} [3], \texttt{GMP}, \texttt{zlib}. Optional tools include \texttt{Graphviz} and \texttt{Ghostscript} for drawings, \form{} [4,5] for the supplied amplitude processing, and \kira{} [6--8] for the subsequent reductions.

\textit{Does this version supersede a previous version?:} Not applicable.

\textit{Reasons for this release:} First public release of \diagen{}.

\textit{Nature of problem:}
Multi-loop amplitude calculations require a consistent transition from large sets of
field-labelled Feynman diagrams to scalar integrals organized in families suitable for
integration-by-parts (IBP) reduction. Repeated topology identification, momentum rerouting, symmetry matching, and subtopology classification can create substantial computational and process-specific overhead, particularly at high loop orders and with several mass scales.

\textit{Solution method:}
\diagen{} generates diagrams at a specified loop order for a user-defined model and process. It assigns fields, masses, and momenta, determines fermionic signs and combinatorial symmetry factors, and writes amplitudes in a \texttt{FORM}-readable format~[4,5]. In parallel, the program constructs a prototype map based on physical denominators, admissible subtopologies, and graph-based equivalences, including the symmetry and momentum-mapping information needed to express different contributions in common integral representations. User-provided \texttt{FORM} procedures can perform further algebraic transformations, including partial fractioning where required. The program also exports integral-family and kinematic information for subsequent reductions with \kira{}~[6--8]. A complementary mode reconstructs scalar diagrams from specified propagator momenta and mass labels.

\textit{References:}

[1] C.~W.~Bauer, A.~Frink and R.~Kreckel,
  \emph{{Introduction to the GiNaC framework for symbolic computation within the C++ programming language}},
  \href{https://doi.org/10.1006/jsco.2000.0362}{\emph{J.\ Symb.\ Comput.}~{\bfseries 33} (2002) 1}
  [\href{https://arxiv.org/abs/cs/0004015}{{\ttfamily cs/0004015}}].

[2] J.~Vollinga, \emph{{GiNaC: Symbolic computation with C++}},
  \href{https://doi.org/10.1016/j.nima.2005.11.155}{\emph{Nucl.\ Instrum.\ Meth.\ A}~{\bfseries 559} (2006) 282--284}
  [\href{https://arxiv.org/abs/hep-ph/0510057}{{\ttfamily hep-ph/0510057}}].

[3] J.~Beder et al.,
  \emph{{yaml-cpp: A YAML parser and emitter in C++}},
  \href{https://github.com/jbeder/yaml-cpp}
  {\nolinkurl{github.com/jbeder/yaml-cpp}}.

[4] J.~A.~M.~Vermaseren, \emph{{New features of FORM}},
  \href{https://doi.org/10.48550/arXiv.math-ph/0010025}{doi:10.48550/arXiv.math-ph/0010025}
  [\href{https://arxiv.org/abs/math-ph/0010025}{{\ttfamily math-ph/0010025}}]
  (2000).

[5] B.~Ruijl, T.~Ueda and J.~Vermaseren, \emph{{FORM version 4.2}},
  \href{https://doi.org/10.48550/arXiv.1707.06453}{doi:10.48550/arXiv.1707.06453}
  [\href{https://arxiv.org/abs/1707.06453}{{\ttfamily 1707.06453}}]
  (2017).

[6] P.~Maierh\"ofer, J.~Usovitsch and P.~Uwer,
  \emph{{Kira---A Feynman integral reduction program}},
  \href{https://doi.org/10.1016/j.cpc.2018.04.012}{\emph{Comput.\ Phys.\ Commun.}~{\bfseries 230} (2018) 99}
  [\href{https://arxiv.org/abs/1705.05610}{{\ttfamily 1705.05610}}].

[7] J.~Klappert, F.~Lange, P.~Maierh\"ofer and J.~Usovitsch,
  \emph{{Integral reduction with Kira 2.0 and finite field methods}},
  \href{https://doi.org/10.1016/j.cpc.2021.108024}{\emph{Comput.\ Phys.\ Commun.}~{\bfseries 266} (2021) 108024}
  [\href{https://arxiv.org/abs/2008.06494}{{\ttfamily 2008.06494}}].

[8] F.~Lange, J.~Usovitsch and Z.~Wu,
  \emph{{Kira 3: integral reduction with efficient seeding and optimized equation selection}},
  \href{https://doi.org/10.1016/j.cpc.2025.109999}{\emph{Comput.\ Phys.\ Commun.}~{\bfseries 322} (2026) 109999}
  [\href{https://arxiv.org/abs/2505.20197}{{\ttfamily 2505.20197}}].

\newpage

\pagenumbering{Alph}
\clearpage

\tableofcontents
\pagenumbering{roman}

\clearpage

\pagenumbering{arabic}

\section{Introduction}
\label{sect:introduction}

Multi-loop scattering amplitudes are essential for precision predictions in high-energy physics. The demand for higher-order calculations is documented, for example, in the community ``precision wishlist''~\cite{Huss:2025nlt}. As loop order, external-particle multiplicity, and the number of mass scales increase, a central practical challenge is not only to generate the diagrams but also to organize their scalar-integral content consistently. A single integral topology may occur in many diagrams and with different loop-momentum routings. Identifying these repetitions early can reduce subsequent bookkeeping and simplify the interface to integral reduction.

Two complementary aspects of automation are important for this problem: the construction of the amplitude from model information and the organization of its integral content. In the broader software landscape, \texttt{QGRAF}~\cite{Nogueira:1991ex,Nogueira:2021wfp} and \texttt{FeynArts}~\cite{Hahn:1998yk,Hahn:2000kx} provide diagram-generation capabilities. More recent diagram generators include \texttt{feyngen}~\cite{Borinsky:2014xwa}, \texttt{FeynGraph}~\cite{Braun:2026buo}, and the generator built into \form{}~5~\cite{Davies:2026cci}. Model data can be prepared with \texttt{FeynRules}~\cite{Christensen:2009jx,Alloul:2013bka}. The resulting amplitudes can be processed in computer algebra systems such as \form{}~\cite{Vermaseren:2000nd,Ruijl:2017dtg}, including its \texttt{color} package~\cite{vanRitbergen:1998pn}, or \texttt{Mathematica}~\cite{Mathematica} with tools such as \texttt{FormCalc}~\cite{Hahn:1999mt} and \texttt{FeynCalc}~\cite{Mertig:1990an,Shtabovenko:2016sxi,Shtabovenko:2020gxv,Shtabovenko:FeynCalc10,Shtabovenko:2025lxq}. For multi-loop calculations, tools also address the intermediate tasks of topology classification, denominator mapping, or the preparation of reduction input. Examples include \texttt{DIANA}~\cite{Tentyukov:1999is}, \texttt{q2e}/\texttt{exp}~\cite{Harlander:1998cmq,Seidensticker:1999bb}, \texttt{Reduze~2}~\cite{vonManteuffel:2012np}, \texttt{TopoID}~\cite{Hoff:2016pot}, \texttt{HepLib}~\cite{Feng:2021kha}, \texttt{tapir}~\cite{Gerlach:2022qnc}, \texttt{FeAmGen.jl}~\cite{Wu:2023qbr}, the multiloop extension of \texttt{FeynCalc}, and the higher-order amplitude-generation tool \texttt{ANATAR}~\cite{Duhr:2025oil}. The organization of data across these stages requires compatible conventions for momenta, fields, and denominator indices. In particular, constructing integral families independently of diagram generation can require recovering relations that are already accessible from the diagrammatic representation.

In this work we present \diagen{} (pronounced {\fontfamily{cmr}\selectfont\textipa{["di:.a.\textscriptg En]}}).
The name derives from the German \mbox{\emph{Diagramm-Generator}}.
The program combines diagram generation with the construction of a symmetry-aware map of integral prototypes.
The program is built on a general collection of graph-theoretical algorithms, including graph generation, canonical labelling, automorphism searches, connectivity and component analysis, and subgraph construction. Some of these routines have applications beyond the workflow documented here. For instance, the library contains algorithms for selecting the heavy subgraphs required in a large-mass expansion by subgraphs~\cite{Smirnov:2002LME}. This functionality is not exposed as a supported workflow in the present release. The library also provides algorithms for constructing cuts of scalar diagrams, decomposing them into subdiagrams, and adapting their momentum routing. While cut-based diagram selection is already available, a dedicated interface for generating and exporting cut integrals is planned for a future release.

Instead of treating topology classification solely as a
separate step after algebraic reduction, \diagen{} records
propagator structures and their relationships during the
generation workflow.

It produces \form{}-readable expressions, identifies
graph-equivalent prototypes and admissible subprototypes,
and exports a compact set of integral-family descriptions
and momentum mappings for use with the IBP reduction program
\kira{}~\cite{Maierhofer:2017gsa,Klappert:2020nbg,Lange:2025fba}. Conversely, \diagen{} can reconstruct a scalar diagram from the propagator momenta and mass labels of an existing integral-family sector.
The original code was developed about 25 years ago and has since been used and further developed in a number of multi-loop projects~\cite{Czakon:2002wm,Awramik:2002wn,Awramik:2004ge,Awramik:2004qv,Dubovyk:2026nhx}. An account of the use of an earlier version of \texttt{DiaGen/IdSolver} in multiloop calculations can be found in Ref.~\cite{Niggetiedt:2023thesis}.

From a technical point of view, \diagen{} is implemented in \texttt{C++} and relies on well-established libraries for symbolic manipulation and structured input: \texttt{GiNaC}~\cite{Bauer:2000cp,Vollinga:2005pk} (with \texttt{CLN}~\cite{CLN}) and \texttt{yaml-cpp}~\cite{YAML}. The build system follows standard \texttt{Autotools} conventions.

The way research software is written and used is changing rapidly. AI coding agents based on large language models now write, run, and modify analysis code with little human intervention.
Their output is probabilistic: the same request can lead to different code or results. Precision calculations, by contrast, require every intermediate step to be exact and repeatable. A natural division of labour is therefore emerging, in which the agent orchestrates the workflow --- preparing input files, running programs, and checking results --- while well-defined computational steps are delegated to deterministic tools. Diagram generation is such a step. A program used in this way must produce identical output for identical input, so that any change in the output can be attributed to a change in the input or in the program, and can be detected by a plain file comparison. \diagen{} meets this requirement: repeated runs of the same version with the same input produce byte-identical amplitudes, prototype definitions, and \kira{} configurations (Section~\ref{sect:agentic}). Its inputs and outputs are plain-text files that can be read, generated, and compared by both humans and automated tools.

This article is organized as follows.
Section~\ref{sect:preliminaries} introduces the
graph-theoretic conventions and the integral
objects used by \diagen{}.
Section~\ref{sect:examples} demonstrates the
program workflow using representative examples,
including the specification of model and
kinematics inputs.
Section~\ref{sect:inverse-kirchhoff} describes the inverse Kirchhoff problem, its implementation, and the interface for drawing graphs from integral families.
Section~\ref{sect:output} explains the generated output, and
Section~\ref{sect:model} documents the template
macros used to construct propagators,
wavefunctions, and interaction vertices.
Section~\ref{sect:agentic} discusses
reproducibility and the use of \diagen{} in
agent-driven workflows.
Benchmarks are presented in
Section~\ref{sect:benchmarks}, followed by
recommendations for practical use in
Section~\ref{sect:BestPractice}.
Installation instructions are given in
Section~\ref{sect:installation}, and the scope and
limitations of the present release are summarized in
Section~\ref{sect:limitations}.
We conclude in Section~\ref{sect:conclusions}.

\section{Preliminaries}
\label{sect:preliminaries}

We fix notation and terminology used throughout the paper and summarize the
objects constructed by \diagen{}: graphs/topologies, diagrams with field and mass
assignments, momentum routings, and the integral objects used for matching and
IBP reduction.
For general graph-theoretic aspects of Feynman diagrams, see Ref.~\cite{Nakanish1971GraphTA}.

\subsection{Graphs, topologies, and diagrams}
\label{sect:preliminaries:graphs}

We use a multigraph $G=(V,E)$, allowing parallel edges and self-loops. A connected Feynman diagram has interaction nodes and explicit degree-one external nodes. Internal edges join interaction nodes, whereas external edges join an interaction node to an external node and carry the external fields and momenta. This convention makes edge orientations and momentum assignments uniform across internal and external lines~\cite{Nakanish1971GraphTA}.

In \diagen{} we distinguish the following objects:
\begin{itemize}
  \item A \emph{topology}, i.e.\ the graph structure at fixed loop order and
        external legs, without field or mass assignments.
  \item A \emph{diagram}, obtained from a topology by assigning fields to edges
        such that each vertex matches an allowed interaction of the model
        (including statistics such as fermionic signs).
  \item A \emph{scalar diagram}, i.e.\ a topology with mass labels assigned
        to its edges. A scalar diagram can be obtained from a diagram by
        forgetting the field labels while retaining the masses and the
        momentum routing, if assigned. It can also be constructed directly
        from propagator data without a model-specific diagram.
\end{itemize}
For bookkeeping, \diagen{} equips topology edges with an orientation. Physical
expressions are invariant under consistent changes of these conventions.
An \emph{incidence} specifies one endpoint of an edge at a vertex. The degree
of a vertex counts its incidences, so that a self-loop (an edge with both
endpoints at the same vertex) contributes two. In reconstructing a topology
from propagators, we refer to an endpoint not yet assigned to a vertex as an
\emph{unassigned incidence} or \emph{incidence slot}.

\subsection{Momentum routing via spanning trees}
\label{sect:preliminaries:momenta}

External momenta $\{p_i\}$ are fixed by the process definition and satisfy one
overall momentum-conservation relation, which we use to eliminate one momentum.
A momentum routing assigns to each internal edge $e$ a momentum $q_e$ such that
momentum conservation holds at each internal node.

\diagen{} constructs routings using spanning trees~\cite{Nakanish1971GraphTA}.
A spanning tree $T$ is a connected acyclic subgraph containing every vertex of $G$. For a connected graph it contains $|V|-1$ edges, so the number of edges outside the tree is $l=|E|-|V|+1=E_{\mathrm{int}}-V_{\mathrm{int}}+1$. Each such edge (a chord) defines a fundamental cycle together with the unique tree path between its endpoints. We assign an independent loop momentum $k_j$ to each chord.
Loop momenta are then propagated along tree paths with signs fixed by the chosen
edge orientations. External momenta are distributed along the unique tree paths
connecting external legs to the eliminated leg.
Different spanning-tree choices lead to equivalent routings related by linear
redefinitions and loop-momentum shifts.

\subsection{Integral families, sectors, and topology minimization}
\label{sect:preliminaries:families}

After applying projectors and performing Lorentz and color algebra (in our setup
with \form{} and its package
\texttt{color}), amplitudes are expressed as linear
combinations of scalar integrals.
A standard integral family is characterized by propagators $\{D_i\}_{i=1}^{N}$
and integer indices $a_i\in\mathbb{Z}$,
\begin{equation}
F\!\left(S \mid a_1,\ldots,a_N\right)
=
\int \prod_{j=1}^{l}\frac{\mathrm{d}^Dk_j}{i\pi^{D/2}}
\;\prod_{i=1}^{N} D_i^{-a_i}\,,
\qquad
D_i=(K_i+P_i)^2-M_i^2+i\epsilon\,,
\end{equation}
where $l$ is the loop order, $D$ is the space-time dimension,
$K_i$ and $P_i$ are linear combinations of loop and external momenta,
respectively, and $M_i$ is the mass associated with denominator $D_i$.
The term $i\epsilon$ denotes the Feynman prescription, with
$\epsilon\to0^+$. The sector label $S$ is defined below. For $m$ independent external momenta, the loop-dependent
bilinears are $k_i\cdot k_j$ ($1\leq i\leq j\leq l$) and $k_i\cdot p_r$
($1\leq i\leq l$, $1\leq r\leq m$). In generic kinematics their number is
\begin{equation}
 N=\frac{l(l+1)}{2}+lm.
 \label{eq:preliminaries:bilinears}
\end{equation}
A complete integral family spans this space, with purely external scalar
products treated as coefficients. If the physical denominators are independent
but fewer than $N$, one can complete them with independent numerator bilinears or with suitable auxiliary inverse propagators. The former convention is used in the prototype representation of \diagen{} introduced below.
When physical denominators are linearly dependent, a partial-fraction step
may instead be necessary.
The sector identifier commonly used in IBP systems is
\begin{equation}
S=\sum_{i=1}^N 2^{i-1}\,\Theta\!\left(a_i-\tfrac12\right).
\end{equation}
Here $\Theta(x)$ is the Heaviside step function, equal to one for $x>0$ and zero for $x<0$. For a fixed integral family, a sector is specified by the subset of propagators with positive indices. A \emph{top-level sector} denotes a designated maximal sector of that family. It need not have the same number of physical lines as the top-level sector of another family. Subsectors follow by removing one or more denominators from the positive-index set.
In graph-based prototype generation, the corresponding operation
is the contraction of eligible propagator edges, subject to
connectivity, loop-order, and the exclusion of massless vacuum
components.

Minimizing the number of distinct topologies is essential for efficient
reductions. A classical approach is based on graph invariants such as the Nickel
index~\cite{Nickel} and efficient algorithms for its computation~\cite{Batkovich:2014bla}. Software supporting such topology minimization includes \texttt{Alibrary}~\cite{Magerya:2021:Alibrary},
\texttt{Feynson}~\cite{Magerya:2020:Feynson}, and \texttt{tapir}.

\subsection{Prototypes in \diagen{}}
\label{sect:preliminaries:prototypes}

For fast matching and symmetry handling, \diagen{} uses a slightly more
specialized integral object than a generic family, the \emph{prototype},
\begin{equation}
\mathrm{PR}\!\left(
\mathrm{ID}\mid
a_1,\ldots,a_L,-b_{L+1},\ldots,-b_N
\right)
=
\int \prod_{j=1}^{l}
\frac{\mathrm{d}^Dk_j}{i\pi^{D/2}}
\left(\prod_{i=1}^{L} D_i^{-a_i}\right)
\prod_{i=L+1}^{N}
\left(k\!\cdot\! q\right)_{\sigma(i)}^{b_i}.
\label{eq:preliminaries:prototype}
\end{equation}
Here $L$ is the number of physical propagators ($L\leq N$), and $\{D_i\}_{i=1}^{L}$ denote the
denominators of the defining prototype, with positive
denominator powers $a_i\in\mathbb{Z}_{>0}$. The
remaining entries represent a basis of independent
numerator scalar products with non-negative powers
$b_i\in\mathbb{Z}_{\geq 0}$. In the integral-index
notation, these numerator powers are represented
by the negative indices $-b_i$. The map $\sigma(i)$
identifies the scalar product associated with each numerator entry:
$(k\!\cdot\!q)_{\sigma(i)}$ denotes one of the bilinears
$k_j\cdot k_r$ or $k_j\cdot p_r$ selected to complete the basis of
Eq.~\eqref{eq:preliminaries:bilinears}.

This notation describes integrals with all physical
propagators of the defining prototype present.
The integral families exported by \diagen{} use
the more general indexing convention introduced
in Section~\ref{sect:preliminaries:families},
where the indices of physical denominators may take arbitrary integer
values. The auxiliary numerator entries, however, are expressed directly as irreducible scalar products (ISPs), rather than as inverse propagators. They are selected by elimination of the scalar products that can be expressed in terms of the physical denominators and the remaining bilinears. A different completion of the basis, for example by auxiliary denominators of the form $(k_j+p_r)^2$, can be obtained by an algebraic change of basis, with the corresponding transformations of the numerator expressions and integral indices. A user-selectable basis completion is a natural extension of the current interface.
In particular, propagators of the defining
prototype may have zero or negative indices,
corresponding to absent denominators or numerator
factors, respectively. Thus, the exported integral
representation also accommodates subsectors and
more general numerator structures. The prototype
identifier $\mathrm{ID}$ is an internal label,
not a sector number or an IBP master-integral index.

\paragraph{From scalar diagrams to prototypes.}
The conversion from a scalar diagram to a prototype removes internal edges
that do not carry loop momentum, since their denominators depend only on
external kinematics. Edges carrying identical momenta up to a sign and the
same mass are represented by a single denominator. The powers of the original denominators can be added together.
When equal momentum squares occur with different mass labels, alternative
prototypes are constructed by retaining one representative of each momentum
class for each admissible choice of mass. This provides the representations
needed for subsequent partial fractioning. In all cases the correspondence
between the surviving prototype edges and the original scalar-diagram edges
is recorded. These operations remove redundant denominators of the indicated
kind, but do not constitute a general test of linear independence of quadratic
forms.

The program tests for massless vacuum components and rejects diagrams whose
unique candidate prototype contains such a component. When several alternative
prototypes are generated, massless vacuum candidates may be retained, for
example in connection with partial fractioning. This graph-based criterion
is not a general test of scalelessness under arbitrary external kinematics.
The prototype algorithms discussed below operate on the reduced structures. In particular, the integral-comparison routines require one-particle-
irreducible prototypes (those that cannot be disconnected by cutting a
single internal edge) without self-energy insertions in the sense of the
implemented structural test.

\subsection{Graph isomorphisms and combinatorial symmetry factors}
\label{sect:preliminaries:matching}

An isomorphism between two topologies is a bijection of their vertices
preserving the multiplicities of edges joining each pair of vertices.
An isomorphism of a topology with itself is called an automorphism.
External-leg constraints and, when appropriate, field or mass labels further
restrict the allowed permutations. To avoid repeated searches over arbitrary
vertex numberings, \diagen{} constructs a canonical labelling, i.e.\ a
representative of the graph independent of its initial numbering.

The algorithm follows the partition-refinement approach to graph isomorphism~\cite{McKay1981,McKay2008}. A partition divides the vertices into disjoint
nonempty subsets, called cells. For a simple graph, a partition is equitable
if the number of neighbours in any given cell is the same for every vertex
of another fixed cell (including the cell itself). In a topology with
multiple edges and self-loops, \diagen{} refines using a generalized degree
that also distinguishes the multiplicities of the adjacencies and the
presence of self-loops. Starting from an initial ordered partition,
successive refinement splits cells whose vertices differ according to
these invariants. The ordering of cells is maintained by a deterministic
rule. If each cell contains one vertex, the partition is called discrete
and defines an ordering of the vertices. Otherwise, one vertex in a
non-singleton cell is separated into its own cell (\emph{individualization})
and refinement is repeated. Branching over the remaining choices gives a
search tree of discrete partitions. Among the leaves of this tree the
lexicographically smallest adjacency representation defines the canonical
topology. Permutations between orderings with the same adjacency matrix
are automorphisms. For scalar diagrams, induced edge orderings are compared
with their mass labels, and ties are resolved using the external-momentum
labelling. Edge orientations are recorded separately as signs in the edge map.

\begin{figure}[!t]
  \centering
  \captionsetup[subfigure]{labelformat=parens,justification=centering,singlelinecheck=false,margin=0pt}
  \begin{subfigure}[t]{0.31\textwidth}
    \centering
    \includegraphics[width=\linewidth]{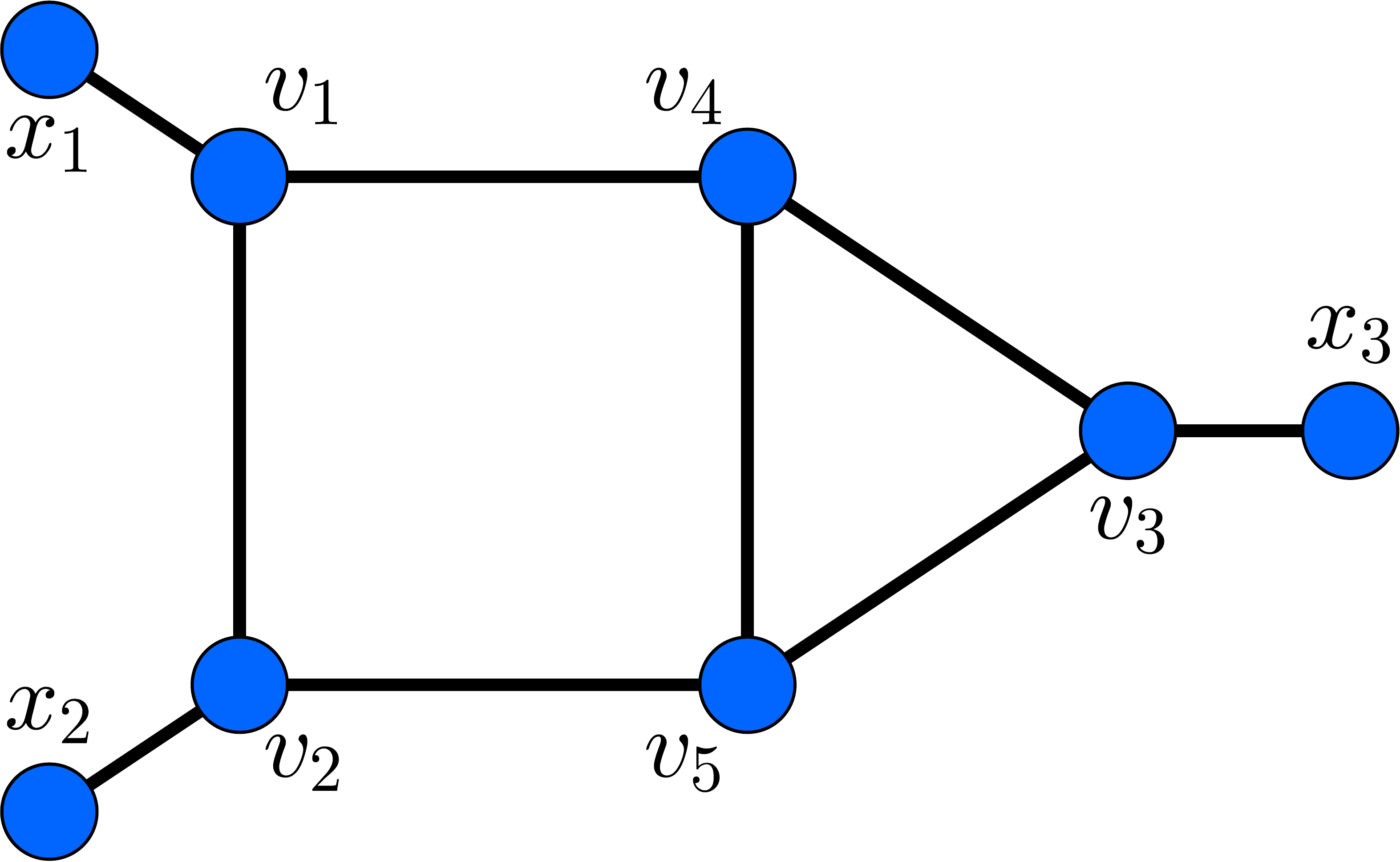}
    \caption{}
  \end{subfigure}\hfill
  \begin{subfigure}[t]{0.31\textwidth}
    \centering
    \includegraphics[width=\linewidth]{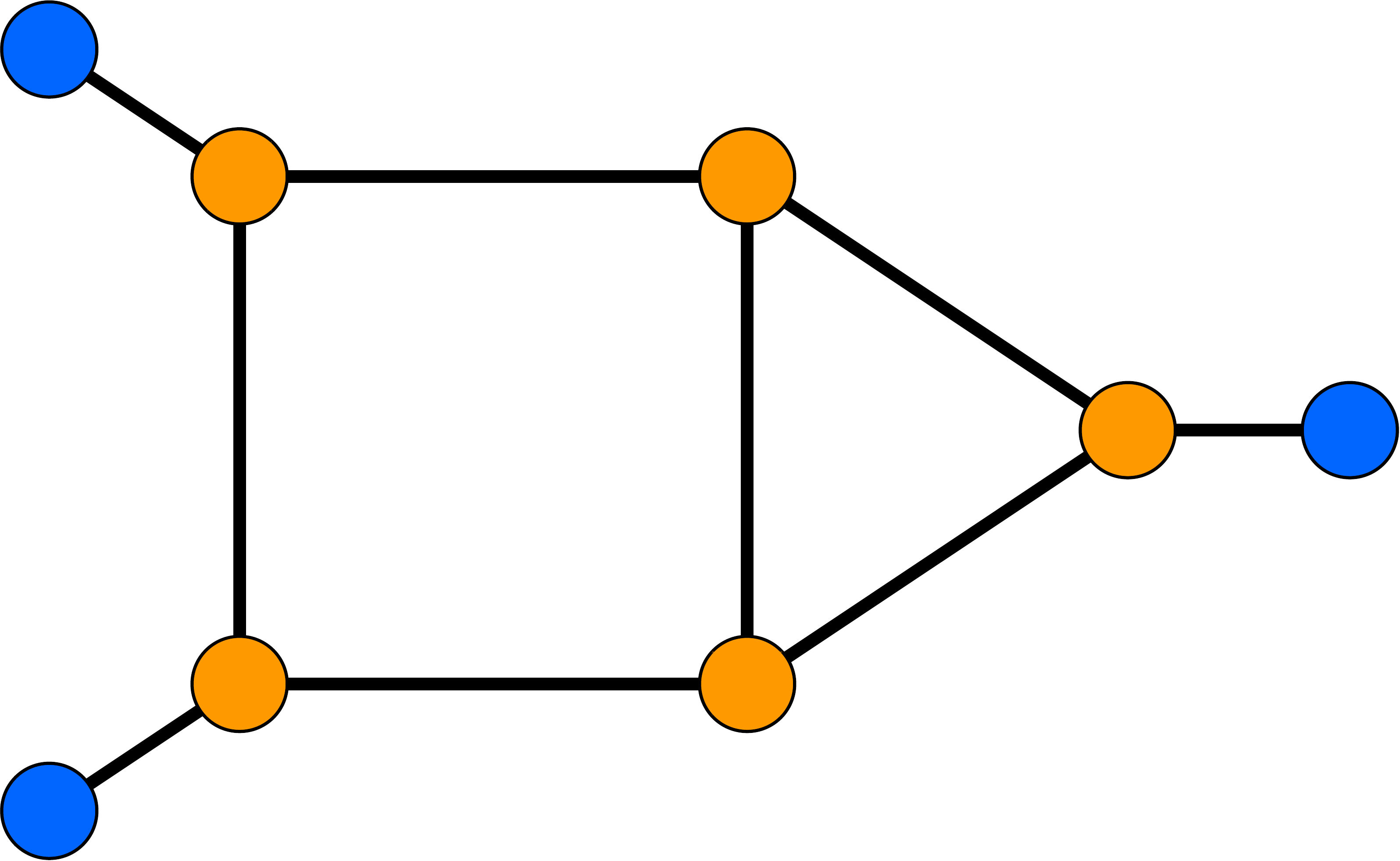}
    \caption{}
  \end{subfigure}\hfill
  \begin{subfigure}[t]{0.31\textwidth}
    \centering
    \includegraphics[width=\linewidth]{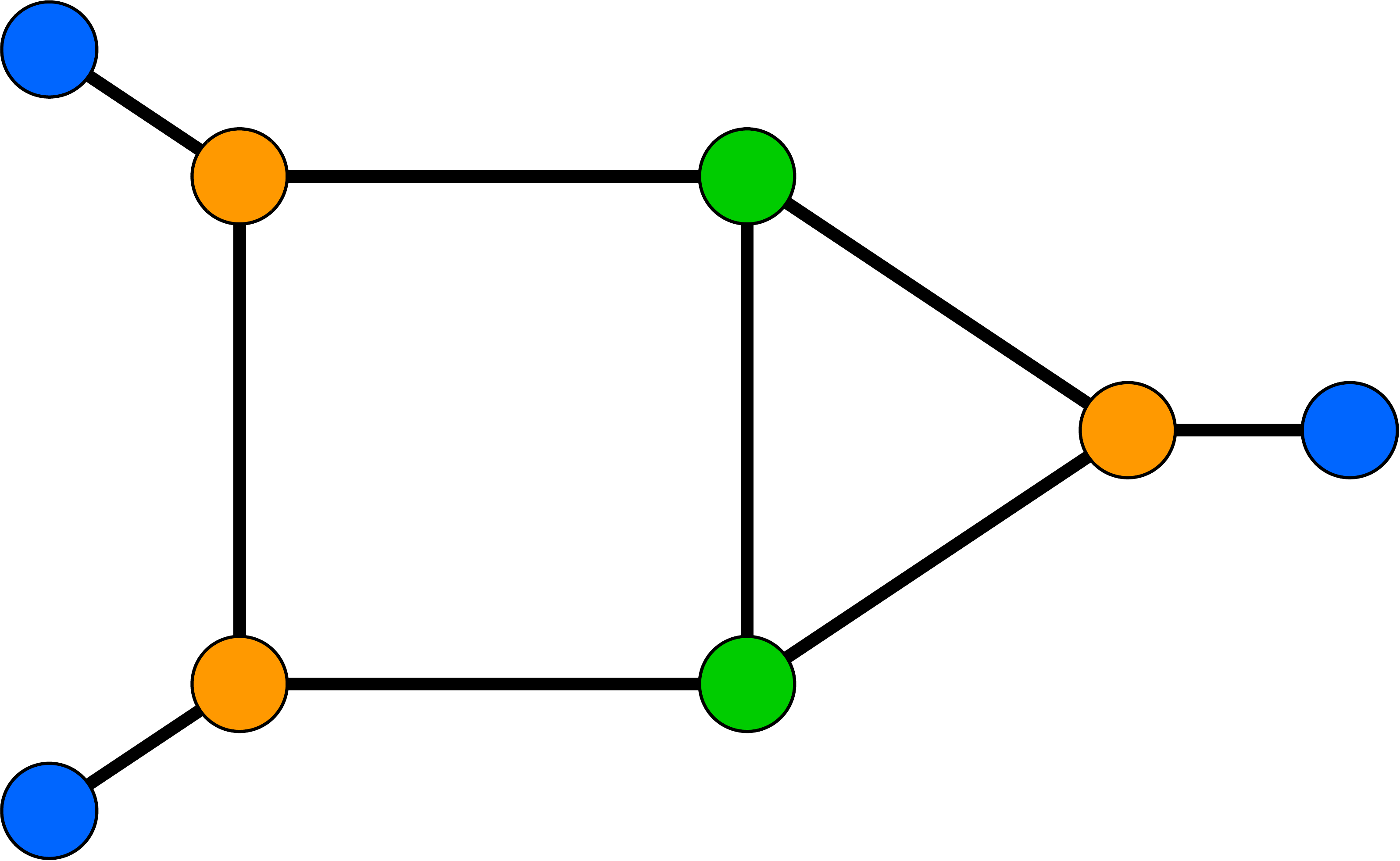}
    \caption{}
  \end{subfigure}

  \vspace{0.7em}

  \begin{subfigure}[t]{0.31\textwidth}
    \centering
    \includegraphics[width=\linewidth]{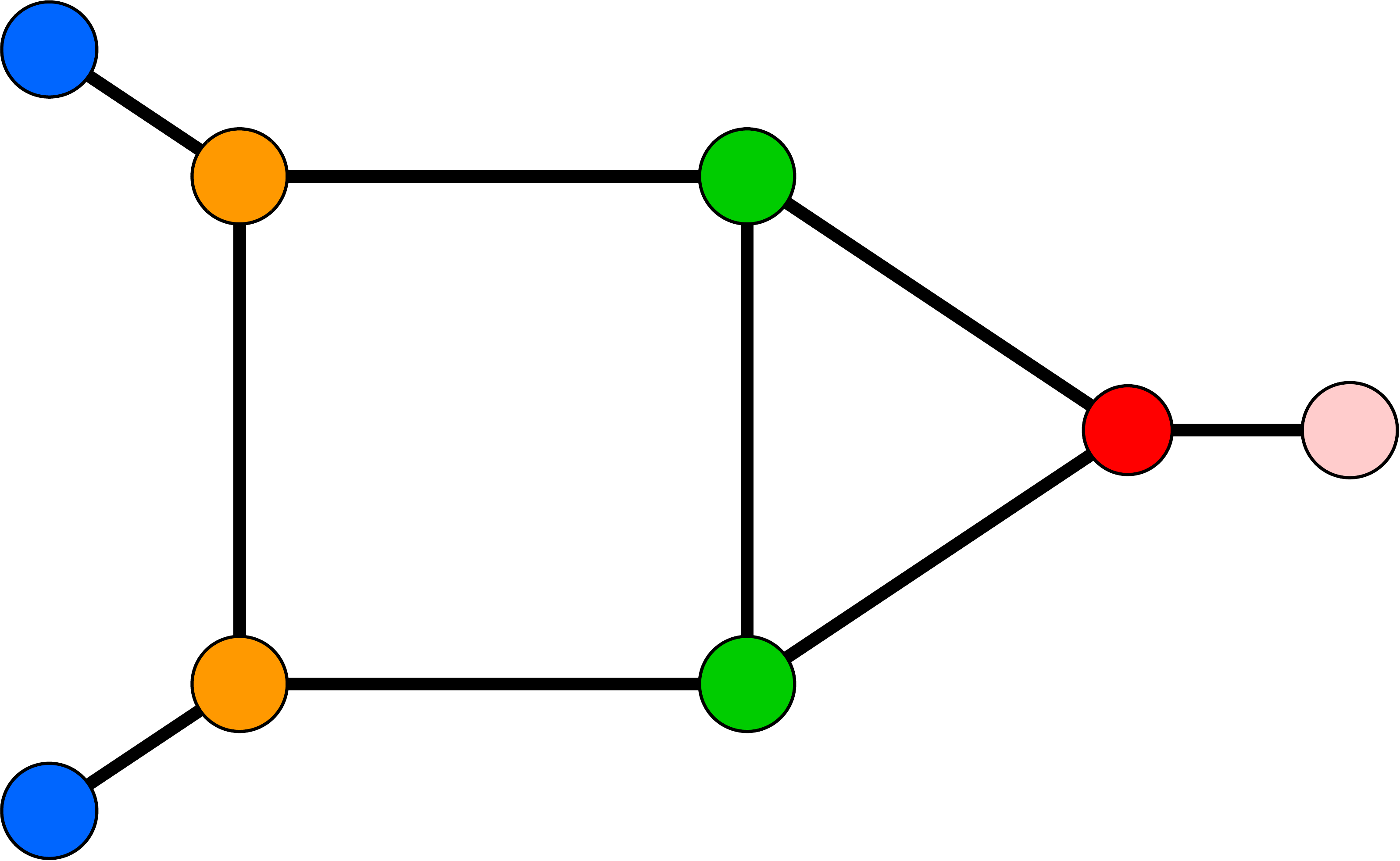}
    \caption{}
  \end{subfigure}\hfill
  \begin{subfigure}[t]{0.31\textwidth}
    \centering
    \includegraphics[width=\linewidth]{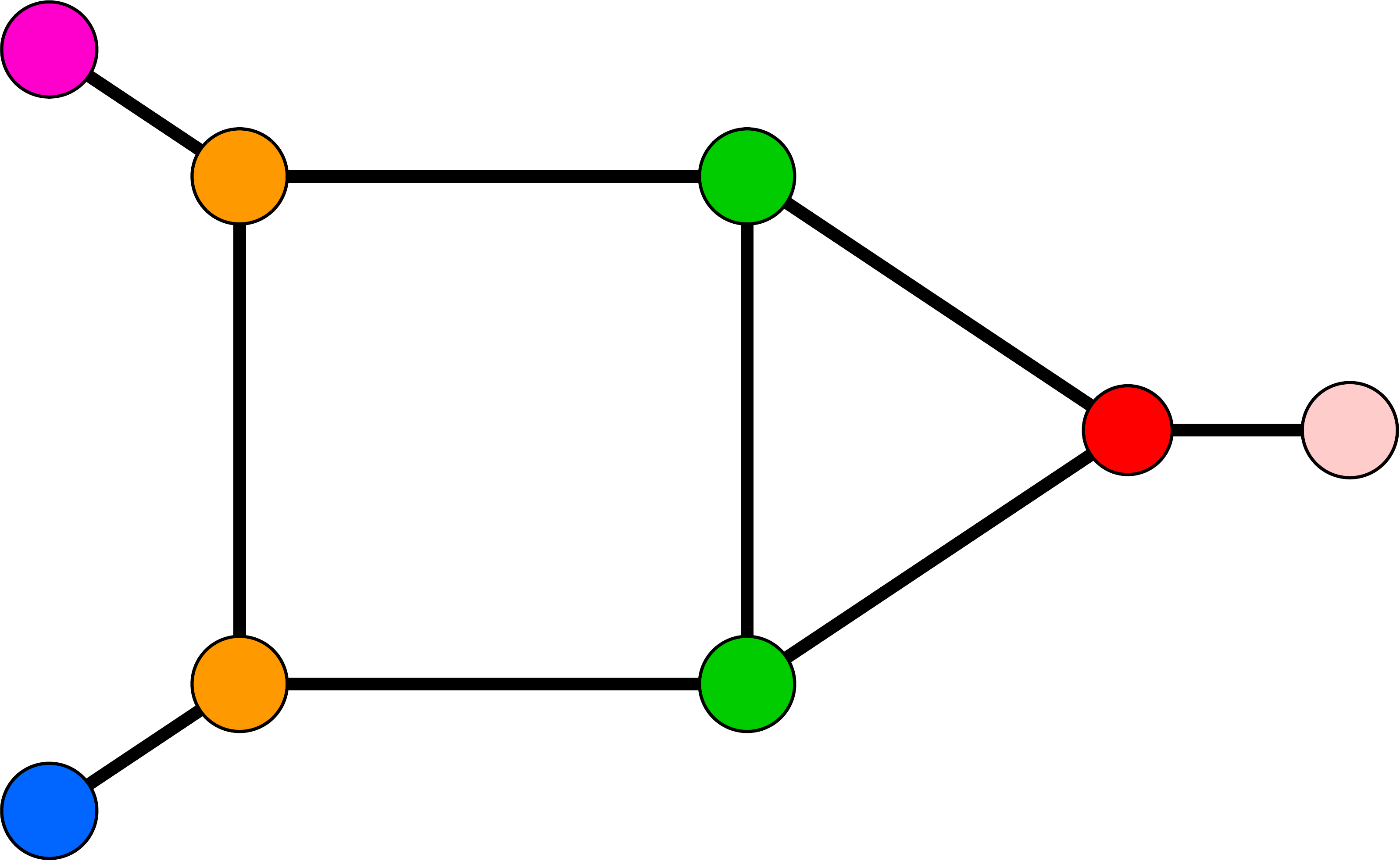}
    \caption{}
  \end{subfigure}\hfill
  \begin{subfigure}[t]{0.31\textwidth}
    \centering
    \includegraphics[width=\linewidth]{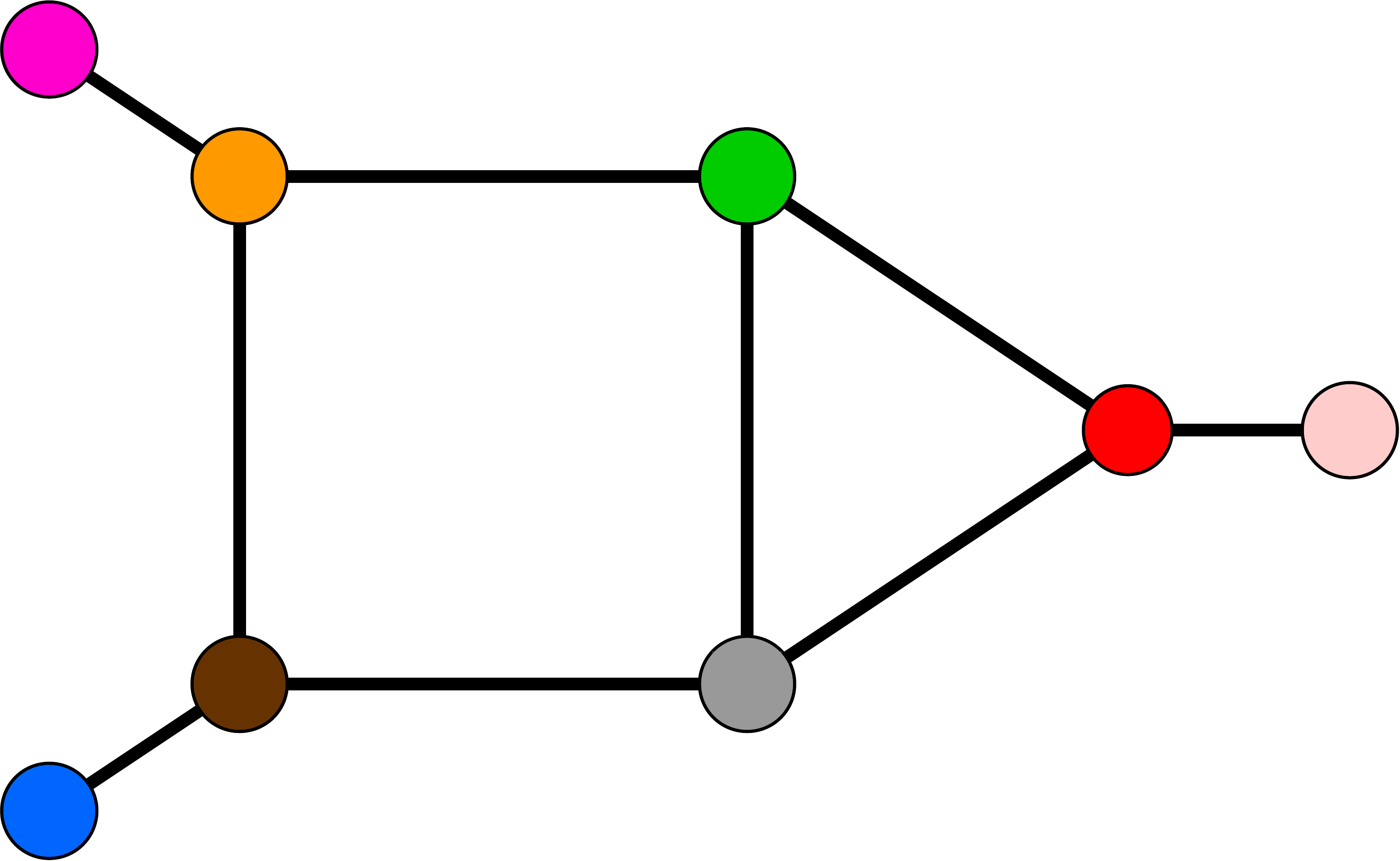}
    \caption{}
  \end{subfigure}
  \caption{Illustration of canonical topology labelling by partition refinement.
  Vertices of the same colour belong to the same partition cell. Colours
  do not denote particle species or masses. (a) Initial partition and vertex
  names. (b) Separation according to vertex degree. (c) Refinement according
  to adjacency to external vertices. (d) Equitable partition, with three
  remaining two-element cells. (e) Individualization of $x_1$ (magenta).
  (f) Discrete partition after further refinement. The alternative choice
  $x_2$ yields the same canonical adjacency representation.}
  \label{fig:canonical-labelling}
\end{figure}

\paragraph{Illustration of partition refinement.}
Figure~\ref{fig:canonical-labelling} illustrates this procedure for the
topology of the two-loop \mbox{$gg\to H$} diagram used again in
Section~\ref{sect:inverse-kirchhoff}. For the purpose of canonical topology
labelling, field assignments, mass labels and momenta are disregarded. The vertex names in panel (a) only serve to identify the vertices in this
example. Initially, all eight vertices are in the same cell. Since the
external and internal vertices have degrees one and three, respectively,
the first refinement separates these two sets (b). The internal vertices
are further distinguished by their adjacency to external vertices (c).
In the resulting partition, $v_3$ has two neighbours in the cell
$\{v_4,v_5\}$, whereas $v_1$ and $v_2$ each have only one. Splitting the cell $\{v_1,v_2,v_3\}$
and subsequently distinguishing the external vertex $x_3$ from
$x_1$ and $x_2$ gives the equitable partition in (d),
\begin{equation}
 \big\{\{x_1,x_2\},\{x_3\},\{v_1,v_2\},\{v_3\},\{v_4,v_5\}\big\},
 \label{eq:preliminaries:equitable-example}
\end{equation}
where we omit the ordering of the cells for clarity. No further
adjacency-based splitting is possible at this stage.
Individualizing $x_1$ in (e) distinguishes $v_1$ from $v_2$, which in turn
separates $v_4$ from $v_5$. The resulting partition (f) is discrete.
The alternative choice $x_2$ gives an ordering with the same adjacency
matrix: the two choices are related by the automorphism
$(x_1\,x_2)(v_1\,v_2)(v_4\,v_5)$. Thus the ambiguity can be resolved
with two leaves in this example, rather than by comparing all $8!$
possible vertex orderings. The panels are a schematic refinement sequence,
not a prescription for the order in which active cells are processed.

The external vertices are interchangeable in this example of an unlabelled
topology. For the internal vertex automorphism group relevant to diagram
generation, however, the initial partition instead contains each external
vertex as a separate singleton cell. In this example, fixing the external
vertices individually makes the internal vertex automorphism group trivial.

\paragraph{Combinatorial factors.}
The symmetries relevant to a Feynman diagram are not the same as those of
its associated scalar diagram: field assignments must be preserved, and the
specified external nodes must remain fixed. During diagram generation, the
algorithm orders the field assignments on parallel edges and self-loops so
that indistinguishable permutations are not generated repeatedly. A complete
assignment is compared lexicographically with its images under the internal
vertex automorphism group. Only the representative of each orbit is retained, and the number $g$ of vertex permutations leaving the ordered field configuration
unchanged is counted at the same time. This is the order of its stabilizer.

Let $n_{v,f}$ count occurrences of the same field $f$ in the interaction at
vertex $v$, and let $m_b$ denote the multiplicity of each class $b$ of
identical parallel propagators or identical self-loops (with the field and
conjugate-field assignments taken into account). Let $h$ count the
self-loops carrying a self-conjugate field. The combinatorial weight used
by \diagen{} is
\begin{equation}
 w_{\mathrm{comb}}=
 \frac{\displaystyle\prod_v\prod_f n_{v,f}!}
 {\displaystyle g\,2^h\prod_b m_b!}.
 \label{eq:preliminaries:symfactor}
\end{equation}
The parallel-line factors are counted once during the traversal of the
diagram. The factors of two arise because the two incidences of a
self-conjugate self-loop can be interchanged. The numerator is the
vertex-normalization factor associated with the multiplicities of identical
fields in the model interaction. It is not an additional automorphism factor.
For instance, two identical parallel lines between fixed vertices give a
factor $2!$ in the denominator, independently of possible vertex permutations.
With conventional vertex rules that already include the identical-field
factorials, the model templates must compensate the numerator in
Eq.~\eqref{eq:preliminaries:symfactor}. The familiar inverse diagram symmetry
factor is then obtained. This convention is important when comparing amplitudes
with those of another generator such as \texttt{QGRAF}. The fermionic sign is determined separately from
fermion chains and is not part of the positive combinatorial weight.

\begin{figure}[!t]
  \centering
  \captionsetup[subfigure]{labelformat=parens,justification=centering,singlelinecheck=false,margin=0pt}
  \begin{subfigure}[t]{0.24\textwidth}
    \centering
    \includegraphics[height=38mm,width=\linewidth,keepaspectratio]{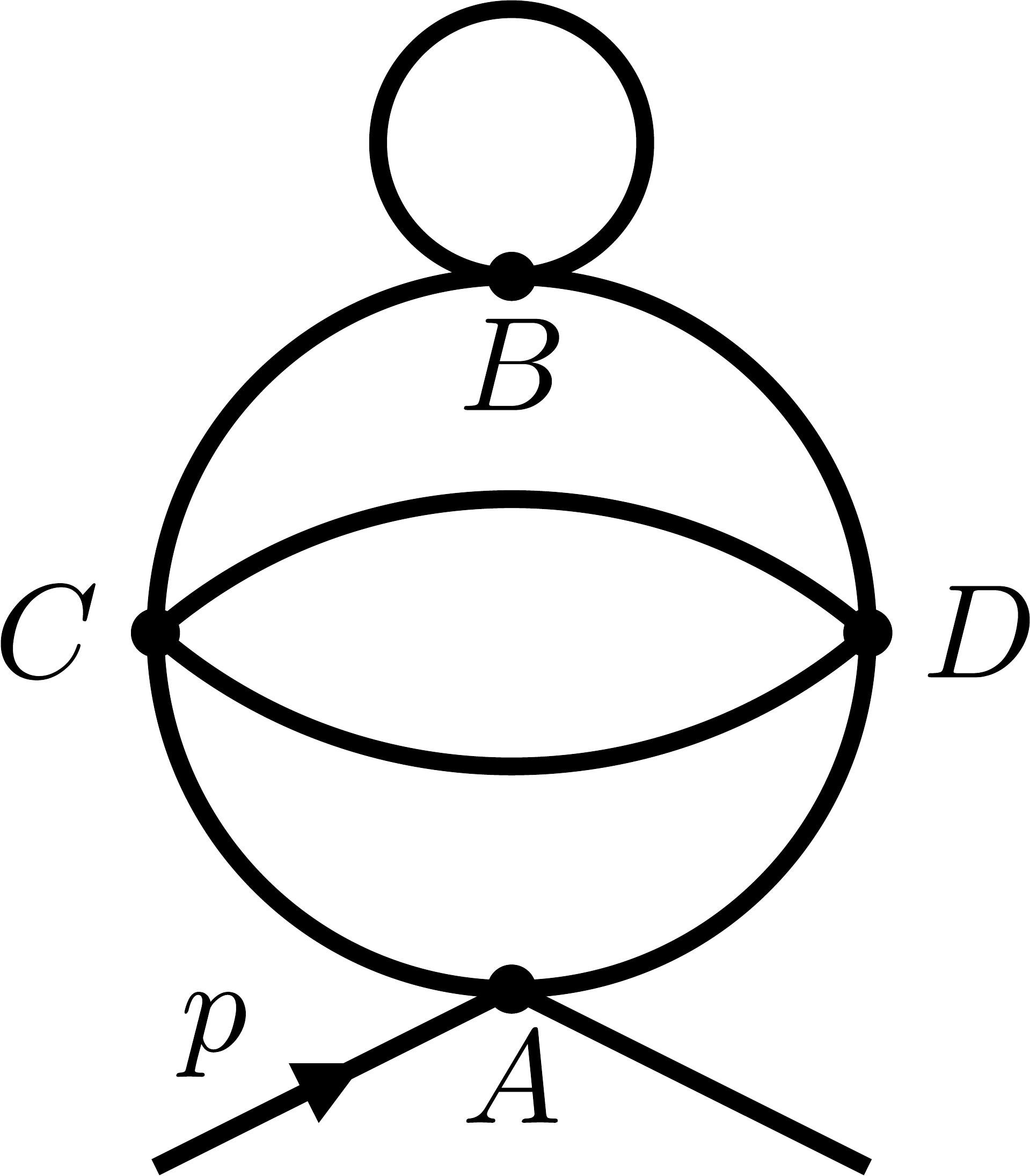}
    \caption{}
  \end{subfigure}\hfill
  \begin{subfigure}[t]{0.24\textwidth}
    \centering
    \includegraphics[height=38mm,width=\linewidth,keepaspectratio]{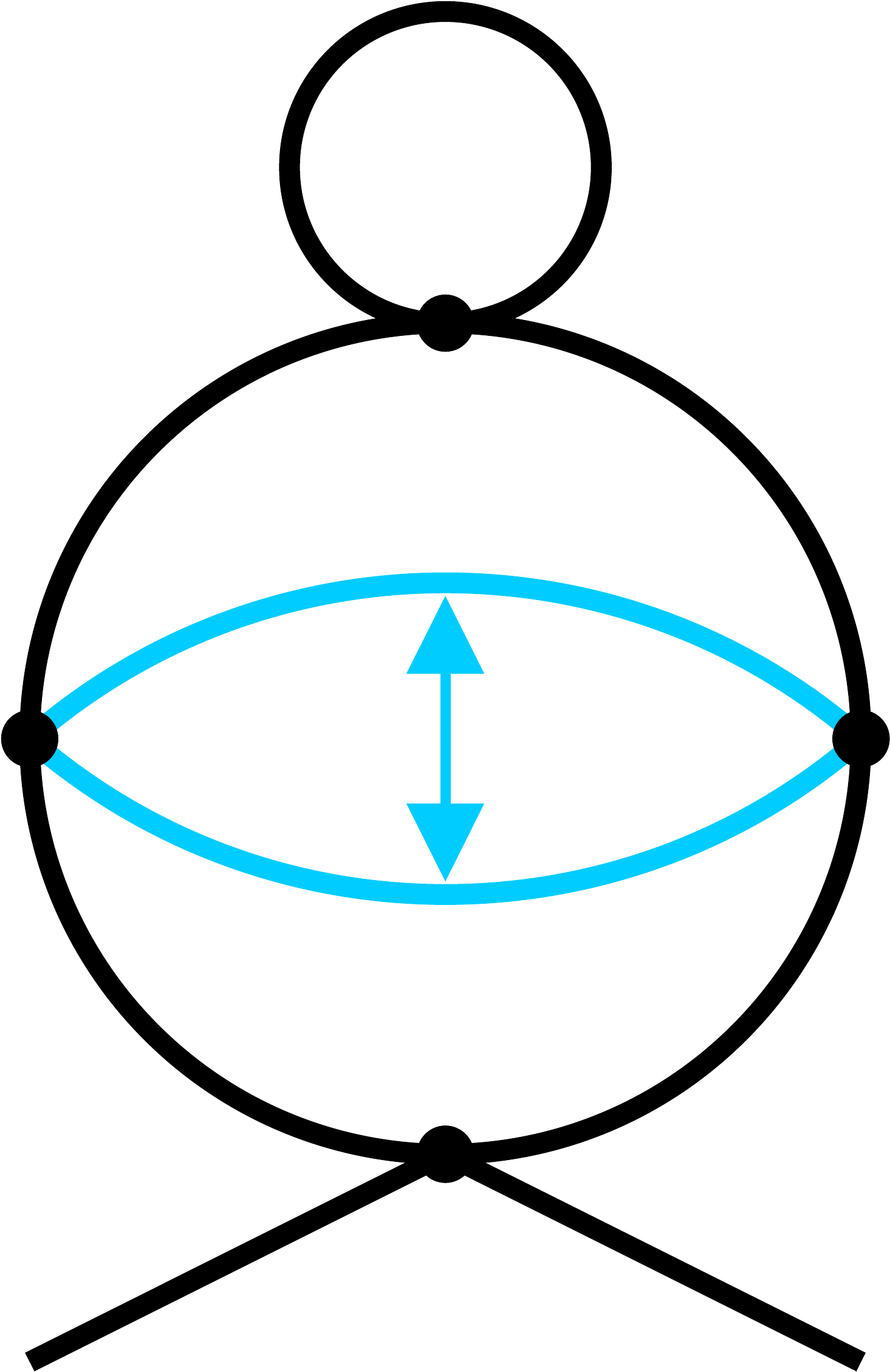}
    \caption{}
  \end{subfigure}\hfill
  \begin{subfigure}[t]{0.24\textwidth}
    \centering
    \includegraphics[height=38mm,width=\linewidth,keepaspectratio]{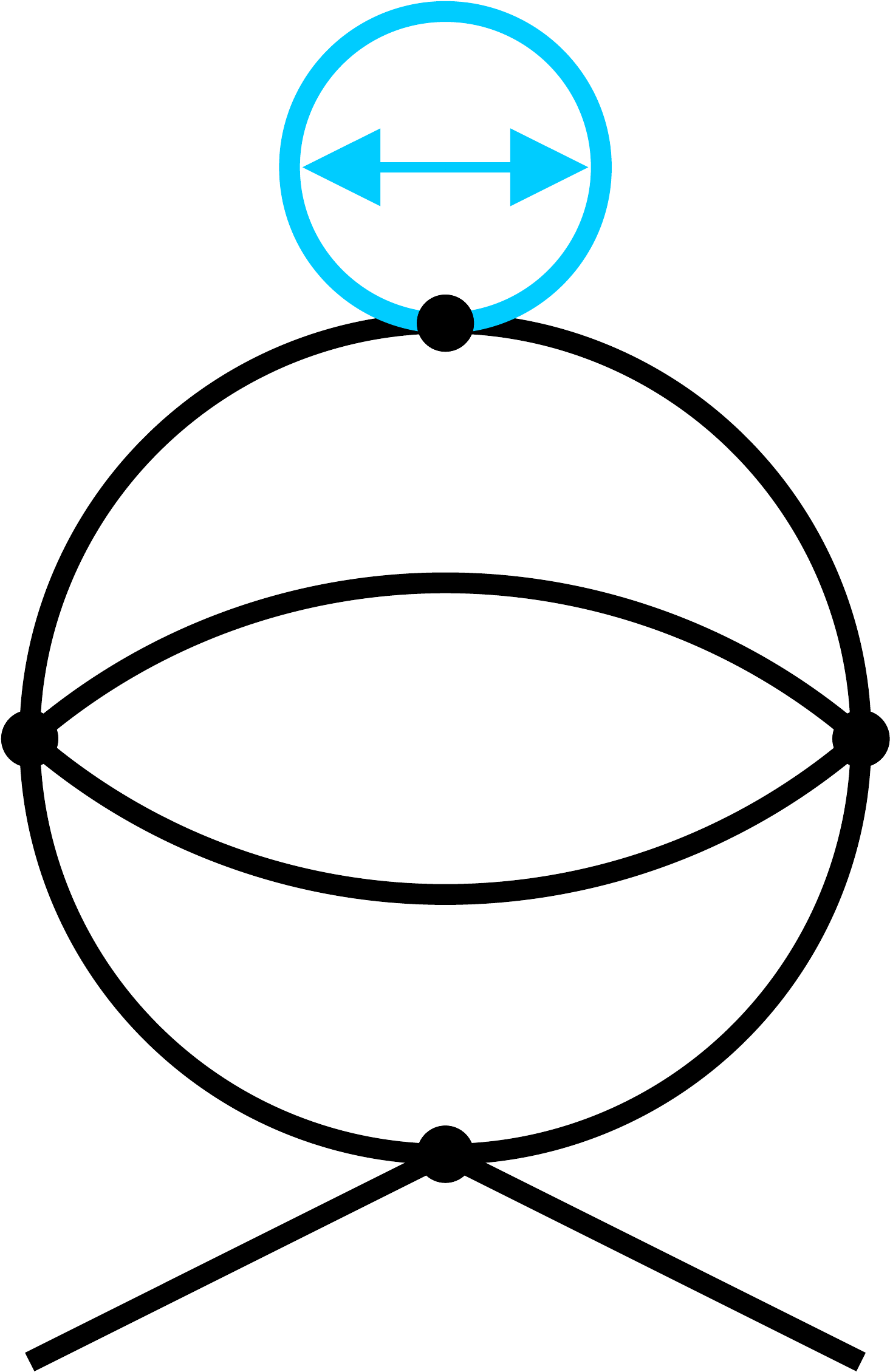}
    \caption{}
  \end{subfigure}\hfill
  \begin{subfigure}[t]{0.24\textwidth}
    \centering
    \includegraphics[height=38mm,width=\linewidth,keepaspectratio]{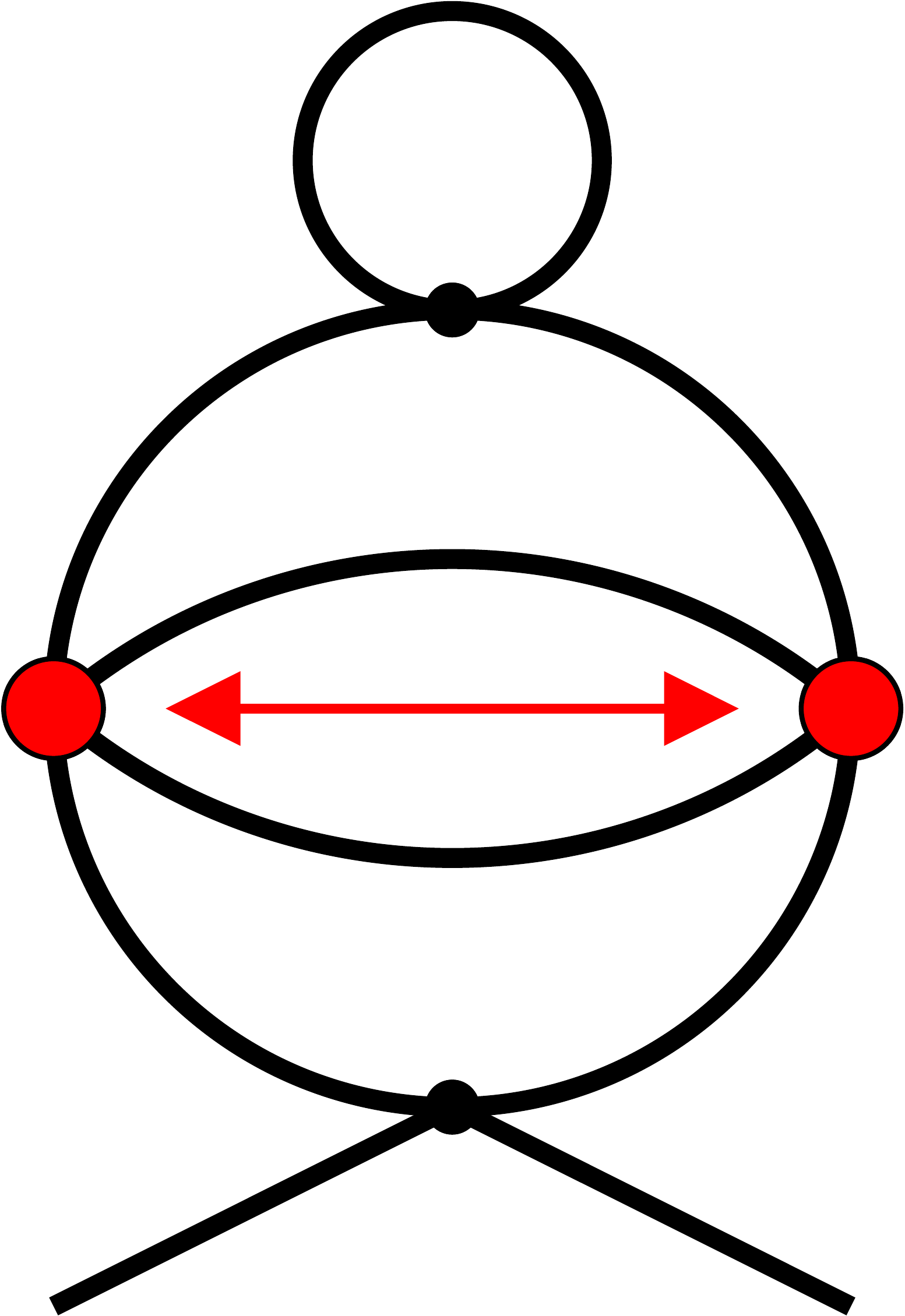}
    \caption{}
  \end{subfigure}
  \caption{Contributions to the combinatorial symmetry factor of a four-loop
  two-point diagram in massive real scalar $\phi^4$ theory. (a) The diagram
  with interaction vertices $A,B,C,D$. Momentum $p$ enters at $A$ and exits through
  the second external leg. (b) Permutation of the two identical edges
  joining $C$ and $D$ (factor $2!$). (c) Exchange of the two incidences of
  the self-conjugate self-loop at $B$ (factor $2$). (d) The internal vertex
  automorphism $(C\,D)$ (group order $2$ including the identity).
  Highlighted edges and vertices identify the respective operations. All internal propagators have the same field and mass assignment.}
  \label{fig:symmetry-factor}
\end{figure}

\paragraph{Example of a combinatorial weight.}
Consider the four-loop two-point diagram in Fig.~\ref{fig:symmetry-factor}
with four quartic vertices in a massive real scalar $\phi^4$ theory. All
seven internal propagators carry the same self-conjugate field. The two
external momenta, in the all-incoming convention, are $p$ and $-p$ and
both external edges are attached to $A$. There are three contributions to
the denominator of Eq.~\eqref{eq:preliminaries:symfactor}. The two parallel
edges between $C$ and $D$ are identical and give $\prod_b m_b!=2!$. The self-loop at $B$ gives $2^h=2$. The identity and the transposition
$(C\,D)$ constitute the internal vertex automorphism group preserving the
field assignment, so $g=2$. These contributions give the order of the
combinatorial symmetry group,
\begin{equation}
 S_{\mathrm{comb}}=g\,2^h\prod_b m_b!=2\cdot 2\cdot 2!=8.
\end{equation}
With conventional vertex rules the diagram weight is therefore $1/8$.
In the convention of Eq.~\eqref{eq:preliminaries:symfactor}, the four
identical fields at each of the four vertices give instead
$w_{\mathrm{comb}}=(4!)^4/8$. For example, if the conventional vertex rule
is $-i\lambda$ for the interaction $-\lambda\phi^4/4!$, the model template
includes the compensating factor $1/4!$ at each vertex. This recovers the
conventional contribution $(-i\lambda)^4/8$ multiplying the rest of the
integrand. The example also illustrates why counting vertex automorphisms
alone is insufficient for the symmetry factor of a diagram with multiple
edges or self-loops.

\subsection{Integral equivalences and the prototype map}
\label{sect:preliminaries:prototype-map}

\begin{figure}[!t]
  \centering
  \captionsetup[subfigure]{labelformat=parens,justification=centering,singlelinecheck=false,margin=0pt}
  \begin{subfigure}[t]{0.30\textwidth}
    \centering
    \includegraphics[width=0.70\linewidth]{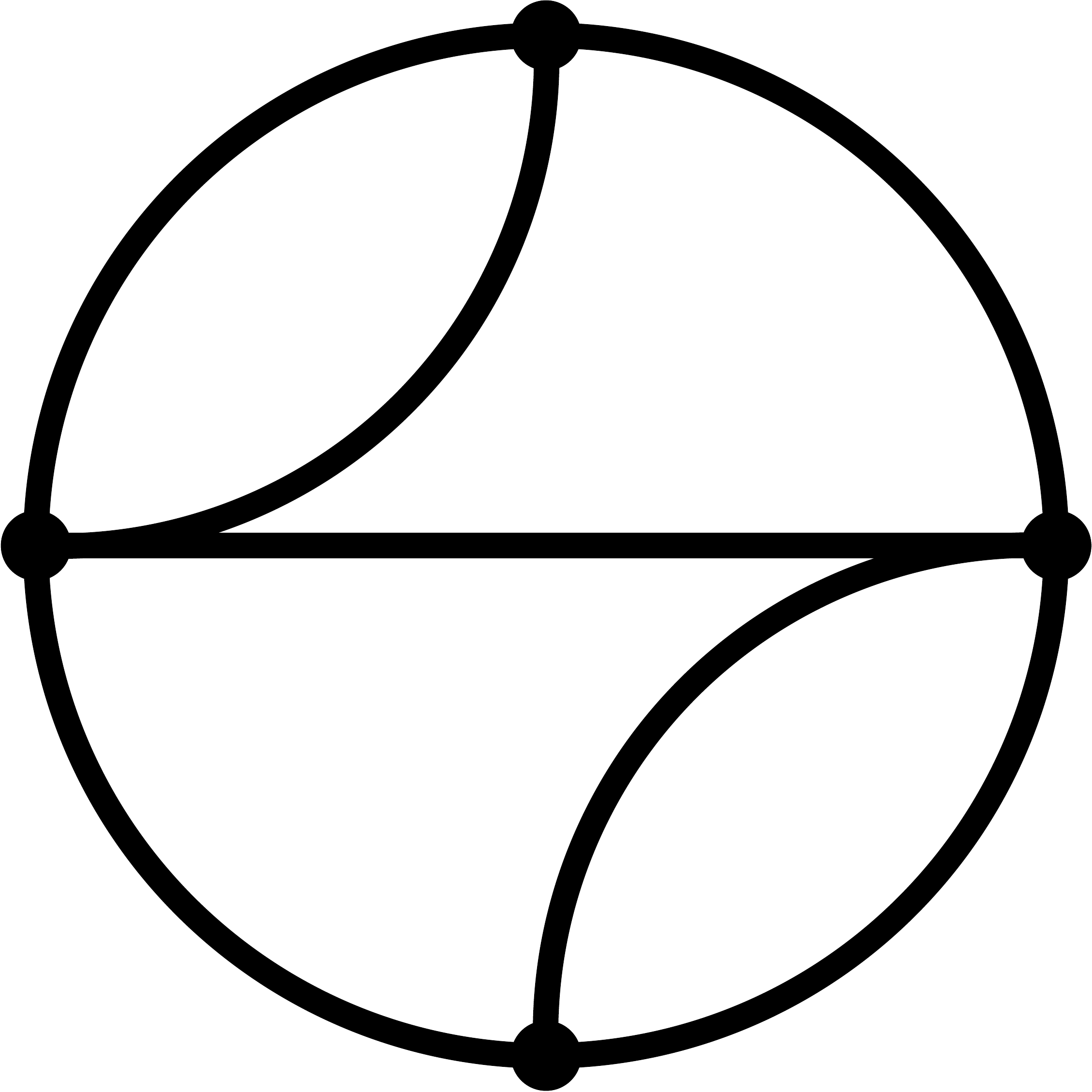}
    \caption{}
  \end{subfigure}\hspace{0.09\textwidth}
  \begin{subfigure}[t]{0.30\textwidth}
    \centering
    \includegraphics[width=0.70\linewidth]{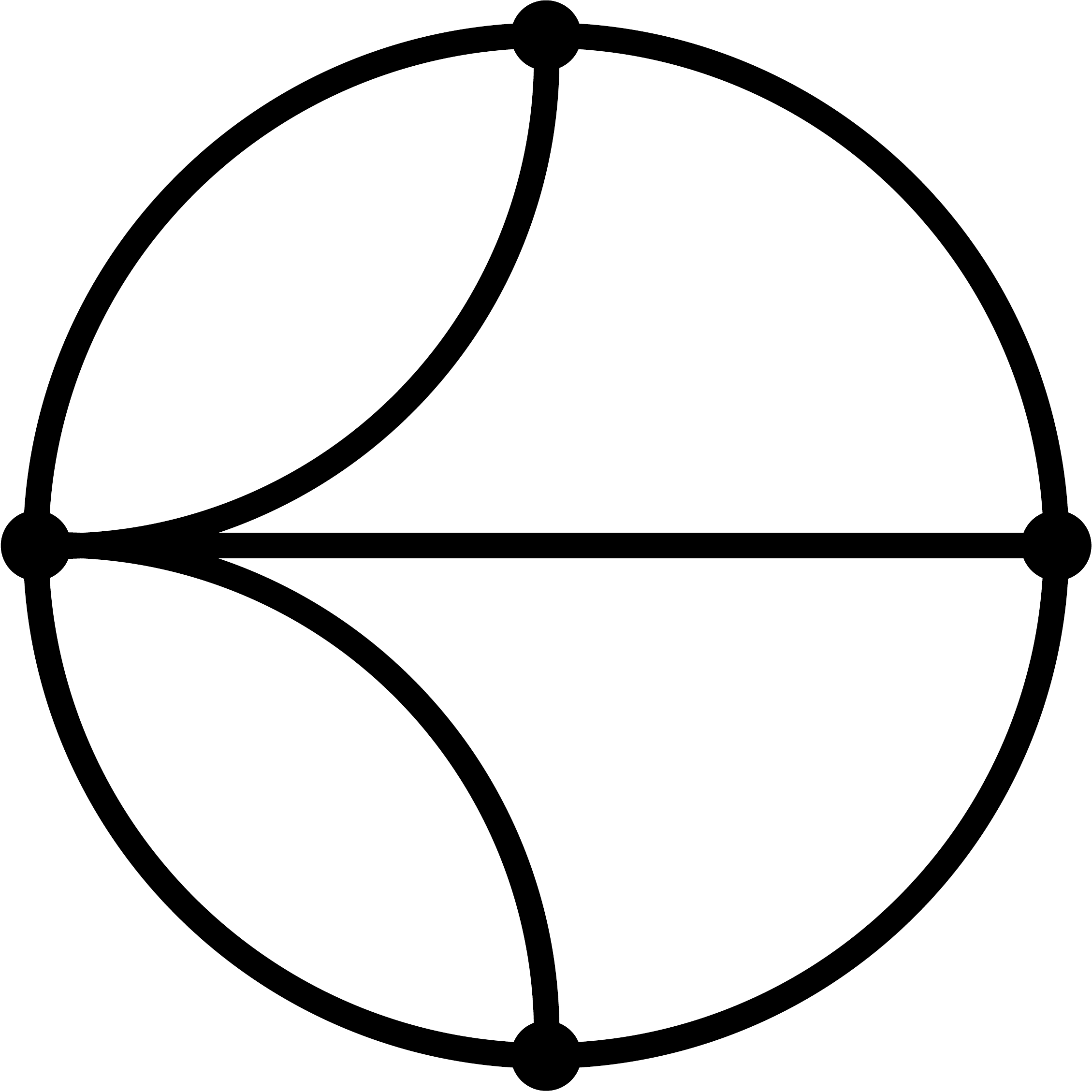}
    \caption{}
  \end{subfigure}
  \caption{Two non-isomorphic four-loop vacuum topologies related by a Whitney
  twist, following Fig.~1 of Ref.~\cite{vonManteuffel:2012np} and
  Eq.~(4.8) of Ref.~\cite{Niggetiedt:2023thesis}. Although their
  vertex-degree multisets differ, they admit the same seven denominator
  momenta, Eq.~\eqref{eq:preliminaries:equivalent-vacuum}. With matching
  masses they correspond to the same integral prototype.}
  \label{fig:equivalent-vacuum}
\end{figure}

\paragraph{Graph equivalence versus integral equivalence.}
For scalar integrals, an equivalence may exist even when the corresponding
mass-labelled graphs are not isomorphic. The relevant criterion is whether
one set of denominator momenta can be transformed into the other, up to
individual reversals and a mass-preserving permutation of the denominators.
Let $k=(k_1,\ldots,k_l)^{\mathsf T}$ and
$p=(p_1,\ldots,p_m)^{\mathsf T}$ collect the independent loop and external
momenta of one prototype, respectively, and let $k'$ and $p'$ denote the
corresponding bases of a second prototype with the same $l$ and $m$.
Write $q_e(k,p)$ for the momentum on internal edge $e$ of the first
prototype and $q'_{e'}(k',p')$ for that on edge $e'$ of the second.
Their masses are denoted by $M_e$ and $M'_{e'}$, respectively. Both
prototypes have $t$ physical denominators, and $\pi$ is a bijection
between their internal edge sets. For generic kinematics, a sufficient
condition for an equivalence of the two denominator sets is
\begin{equation}
 q_e(Ak'+Bp',Cp')=\eta_e q'_{\pi(e)}(k',p'),\qquad
 M_e=M'_{\pi(e)},\qquad \eta_e\in\{+1,-1\},
 \label{eq:preliminaries:integral-equivalence}
\end{equation}
for every internal edge $e$. The $l\times l$ integer matrix $A$ is
unimodular ($\det A=\pm1$), so that the loop integration measure is
unchanged. The $l\times m$ matrix $B$ specifies shifts of the loop momenta
by external momenta. The $m\times m$ matrix $C$ describes an allowed
invertible change of the external momentum basis that preserves the
kinematic invariants and external mass assignments. In the vacuum case,
$m=0$ and the external-momentum terms are absent. The permutation $\pi$
matches denominators of equal mass, and each sign $\eta_e$ allows for
reversal of the momentum on a matched edge. These signs are immaterial
for quadratic denominators but must be retained when constructing the
momentum map. With denominator powers permuted accordingly and the
numerator transformed by the same substitution, the integrals coincide. This criterion explains why graph isomorphism alone need
not give the smallest set of integral representatives. Consider the two
four-loop vacuum topologies in Fig.~\ref{fig:equivalent-vacuum}, an example
of a Whitney twist. Both have four vertices
and seven edges, but their vertex-degree multisets are $\{3,3,4,4\}$ and
$\{3,3,3,5\}$, respectively, so they cannot be isomorphic. Nevertheless,
both admit the same set of propagator momenta (up to individual signs),
\begin{equation}
 \big\{k_1,k_2,k_1+k_2,k_3,k_4,k_3+k_4,
             k_1+k_2-k_3-k_4\big\}.
 \label{eq:preliminaries:equivalent-vacuum}
\end{equation}
To see this, denote the left, upper, right and lower vertices of either
panel by $a,b,c,d$, respectively. In both topologies, assign $k_1,k_2$ to
the two edges from $b$ to $a$, $k_1+k_2$ to the edge from $c$ to $b$,
and $k_1+k_2-k_3-k_4$ to the horizontal edge from $a$ to $c$.
In panel (a), assign $k_3+k_4$ from $a$ to $d$ and $k_3,k_4$ to the two
edges from $d$ to $c$. In panel (b), assign $k_3,k_4$ to the two edges
from $a$ to $d$ and $k_3+k_4$ from $d$ to $c$. Momentum conservation holds
at every vertex in both routings. For equal masses, or corresponding mass
labels in these routings, the sets of quadratic denominators coincide.
The integrals are therefore identical for corresponding denominator powers
and the same numerator in this common momentum basis. The canonical
momentum comparison described below can identify this equivalence without
explicitly generating Whitney twists.

\paragraph{Vacuum and two-point prototypes.}
For these classes the program constructs a canonical integral labelling,
by comparing momentum distributions rather than merely graph connectivity.
Consider a prototype with $l$ loops and $t$ internal edges. The algorithm
tries subsets of $l$ internal edges as candidate carriers of independent
loop momenta. Momentum conservation at the internal vertices gives a linear
system for the remaining edge momenta. Choices for which the system does not
have full rank are rejected. The remaining choices enumerate the routings in
which a basis of loop momenta is carried by individual propagators. Changes
of this basis incorporate shifts by other loop and external momenta.

For multiloop vacuum prototypes, the loop-momentum names and relative signs
are permuted, and each denominator momentum is normalized up to an overall
sign. The resulting lists of coefficient vectors are sorted. The algorithm
selects a preferred distribution by first minimizing the sum of the absolute values of its momentum coefficients,
then comparing the ordered momentum patterns
and finally the mass patterns lexicographically. For two-point prototypes,
the same independent-edge search is used with the external momentum-
conservation relation imposed. In this case a recursive partition search
normalizes the propagators under momentum permutations and sign changes:
loop and external momentum names begin in separate cells. Choosing a
lexicographically maximal momentum pattern refines the cells according to
positive, zero and negative coefficients. The procedure continues on the
remaining propagators, with ties resolved using mass labels. The chosen
canonical labels contain the original edge indices and their orientation
signs, so that equal representations also provide an explicit denominator
correspondence. The vacuum case with a single independent loop uses the
simpler graph-based labelling.

For vacuum and propagator-type prototypes the implementation is designed to
provide a unique integral representation within its class of admissible
propagator structures and momentum transformations. This does not assert
uniqueness of the individual transformation (a representative can itself
have symmetries), nor does it identify integrals related solely by IBP,
partial fractions or special kinematic identities.

\paragraph{Higher-point prototypes.}
For three or more external legs the implemented comparison instead uses
canonical mass-labelled graph representations, with an additional treatment
of vacuum components. The decomposition into biconnected components, i.e.\
blocks separated by articulation vertices, permits vacuum attachments to be
compared separately from the part carrying external momentum. An articulation
vertex is a vertex whose removal disconnects a connected graph. In particular,
the position at which a vacuum component is attached need not distinguish
integral representatives. For fewer than four legs the external momentum
names do not enter the separate momentum-labelling comparison. External
mass labels are still retained. For four legs the comparison permits the
three double transpositions $(12)(34)$, $(13)(24)$ and $(14)(23)$ if they
preserve the external mass pattern. In the all-incoming convention these
preserve the usual Mandelstam invariants by momentum conservation. Beyond
four legs the external momentum labels are retained. These restrictions
reflect the implemented matching rules, rather than a general search over
all kinematically allowed transformations. Consequently, in the multileg
case equivalent integrals with different graph structures may remain in the
prototype map.

\paragraph{Construction of the prototype map.}
The map is populated concurrently with diagram generation. For each new
candidate prototype, the comparison described above is used to look for an
existing representative. If one exists, the edge correspondence gives the
momentum substitutions required to express the diagram in that representative.
Otherwise a new entry is created. For every inserted prototype, subsets of
internal edges are then considered for contraction, starting with the largest
allowed number of removed edges. Contraction merges the endpoints of an edge. If a proposed contraction would remove a self-loop, including a self-loop
created by earlier contractions in the same subset, the subset is discarded.
This preserves the loop number. The search excludes the implemented massless
vacuum and auxiliary-component cases and applies any configured selection
criteria. The \Verb^children^ option bounds the search from below in the
number of surviving denominators, with the loop number as the minimum.

Every surviving subprototype is compared with the entries already stored.
A new representative is inserted only if no match is found. An existing
subprototype may therefore be reached from several distinct parent
prototypes or from several edge contractions of the same parent. Together
with the momentum and index mappings, these relations form a directed
acyclic graph ordered by the number of physical propagators. This organization permits terms from different
diagrams to be expressed in common integral representations before IBP
reduction and allows cancellations to become visible at that stage.

\begin{figure}[!t]
    \centering
    \includegraphics[width=\textwidth]{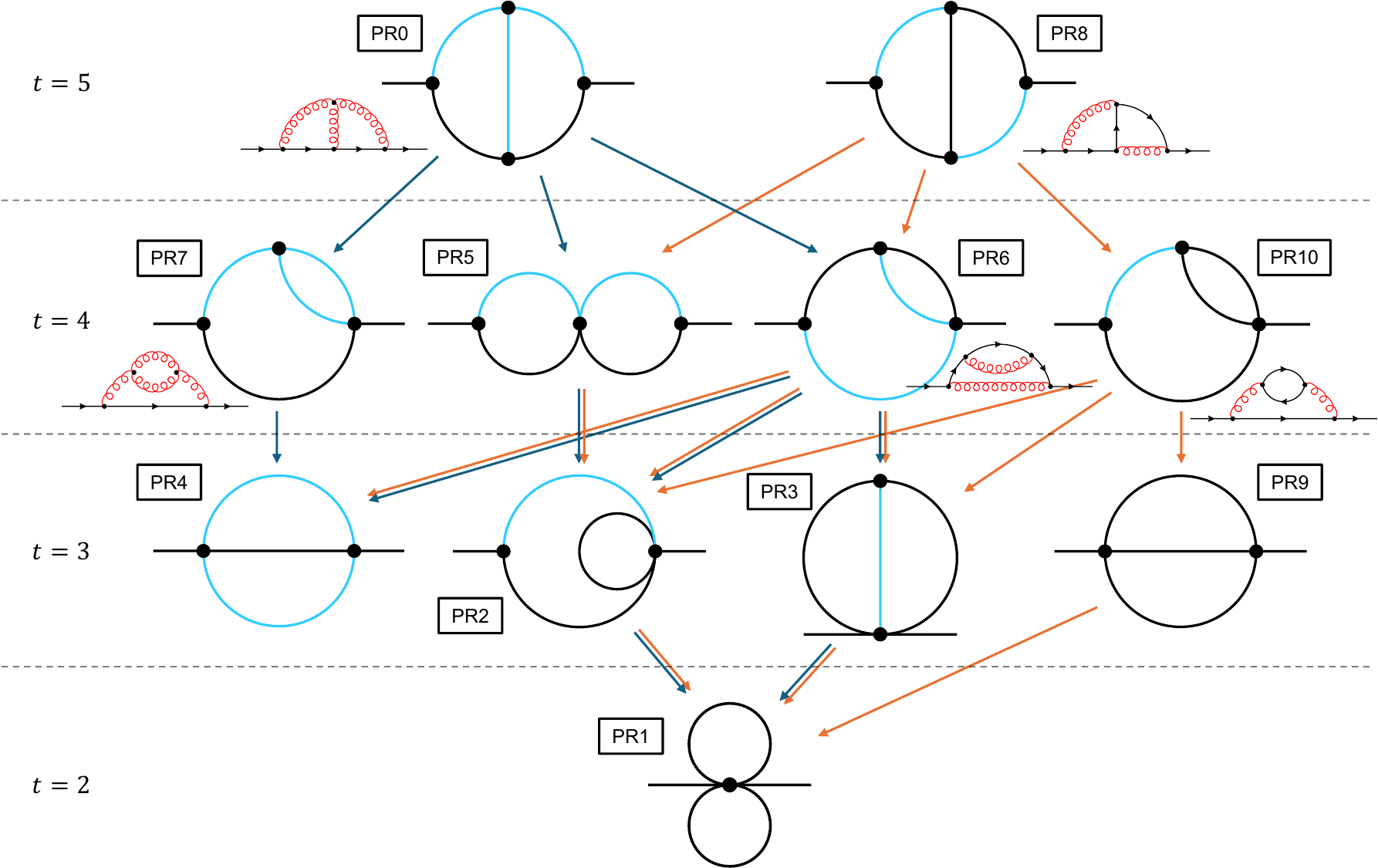}
    \caption{Successive construction of the prototype map during diagram
    generation for an illustrative set of two-loop massive-quark two-point
    diagrams. The small field-assigned diagrams are placed next to their
    corresponding prototypes. The level $t$ denotes the number of distinct
    physical propagators in a prototype. Light-blue prototype edges denote
    massless propagators and black internal edges massive ones. Blue and
    orange arrows mark contraction relations associated with the branches
    originating at PR0 and PR8, respectively. Shared descendants are stored
    only once. The two maximal prototypes PR0 and PR8 determine integral
    families into which the contributions shown can be mapped.}
    \label{fig:protomap}
\end{figure}

\paragraph{Example of the successive construction of the map.}
Figure~\ref{fig:protomap} illustrates the interplay between diagram
and prototype generation for a set of two-loop contributions to the
massive-quark two-point function. The small field-assigned Feynman diagrams
are shown next to their corresponding prototypes. Consider first the
diagram associated with PR0. After assigning momenta and converting its
scalar diagram to a prototype, \diagen{} inserts PR0 at level $t=5$
and generates its admissible subprototypes (blue arrows). In particular,
PR7, PR5 and PR6 occur at level $t=4$. Contractions of these structures
lead to lower levels. Entries with more than one incoming arrow are shared
representatives rather than separately stored copies of each contraction.
The next diagram shown, associated with PR7, has a propagator structure
that is already present as a subprototype of PR0. Consequently, its
contribution can be mapped to PR7 without introducing another entry or
repeating the construction of its descendants.

A subsequent diagram leads to PR8, a second five-propagator prototype
whose internal mass pattern is inequivalent to that of PR0. Its
subprototype relations are shown by the orange arrows. The contractions
leading to PR5 and PR6 reuse existing entries, while PR10 introduces a new
structure at level $t=4$. Further contractions also give PR9 at level
$t=3$, which was not present in the first branch. In particular, PR1
at level $t=2$ can be reached from both top-level prototypes via several
paths. The diagrams shown next to PR6 and PR10 can subsequently be matched
to these entries. For the illustrated set of contributions, PR0 and PR8
are the two maximal prototypes, and two complete integral families suffice to
accommodate these contributions, with the subprototype and momentum maps
specifying how the lower structures are embedded. The labels PR0, PR1,
\ldots{} are prototype identifiers, not sector identifiers or a loop-order
classification. The order of their assignment need not follow the order
of the levels in the figure.

The key feature is that the map is available during diagram generation:
a diagram can match a subprototype even if no earlier Feynman diagram has
had precisely that propagator structure. The two branches in the figure
also show how the construction avoids introducing a separate family for
every diagram. The number of representatives depends on the implemented
equivalence and contraction criteria and on the configured subprototype cutoff. The example does not imply that two families suffice for arbitrary
two-loop processes.

\paragraph{Momentum maps and residual symmetries.}
The canonical edge labellings provide signed edge correspondences between
matched prototypes. For parent--child relations, the original momentum
vectors on the surviving parent edges are equated to the corresponding
vectors on the child edges, including the orientation signs. After imposing
external momentum conservation, Gaussian elimination gives the substitutions
between the two momentum bases. The matching rules and scalar-product
remappings are exported in \form{} format. In addition, mass- and
external-label-preserving graph automorphisms of a prototype, including
permutations of identical parallel lines, induce symmetry transformations
within that prototype. The program retains one transformation per distinct
permutation of the internal denominators and constructs its momentum
substitution in the same way. The associated identity relates the original
integral to the transformed one, including its numerator structure. This
internal symmetry search should be distinguished from the canonical
integral comparison between different prototypes.

Algebraic relations arising from linearly dependent denominators may require
additional partial fractioning with the supplied \form{} procedures.
Further equivalences, particularly for multileg integrals, can be identified
by algorithms based on the Feynman-parameter representation, such as the
method of Pak~\cite{Pak:2011xt,Hoff2015_1000047447}. In the present workflow,
\kira{} can search for such remaining symmetries at the reduction stage.
Thus the map performs substantial topology minimization during generation,
without claiming global minimality for arbitrary integral families.
Finally, projector-based workflows mapping diagrams to form factors or
helicity amplitudes are well established (see, e.g., Refs.~\cite{Peraro:2019cjj,Peraro:2020sfm,Chen:2019wyb}).

\section{Using \diagen{}}
\label{sect:examples}

We illustrate the main interfaces of \diagen{} using two representative processes. The examples establish the input conventions and show how the generated amplitudes and prototype definitions are used in subsequent processing.

\subsection{Example A: \texorpdfstring{\mbox{$gg\to H$}}{gg -> H}}
\label{sect:examples:ggh}

\paragraph{Job file.}
The main purpose of the job file \texttt{jobs.yaml} is to define a process of incoming and outgoing particles. The following job file is provided in \texttt{examples/ggH}.
\begin{verbatim}
#jobs.yaml
jobs:
 - generate_diagrams:
    loop: 2
    children: 5
    output_form: "form"
    draw_graphs: "false"
    draw_protos: "false"
    parallel_form: 1
    incoming_fields: [A, A]
    outgoing_fields: [H]
    internal_fields_max: [[H, 0]]
    flag_loop_particles:
     - {particles: [Q], symbol: "nh"}
     - {particles: [q], symbol: "nl"}
\end{verbatim}

The options are explained in the following:
\begin{itemize}
 \item The option \Verb^loop: 2^ generates diagrams at the specified loop order, given by a non-negative integer. This option is required.
  \item The option \Verb^children: 5^ controls the lower bound on the number of physical propagators retained when generating subprototypes. A larger positive bound restricts the search and usually reduces the run time, whereas a smaller bound permits more extensive subprototype matching. In the present implementation, \Verb^children: 0^, which is also the default, selects an automatic procedure: generation starts with a large cutoff and can reset it according to the first matched diagram. This is not equivalent to requesting every admissible subprototype. The chosen setting therefore affects the completeness of the prototype map and should be reported with benchmark results. A practical procedure for choosing the cutoff is described in Section~\ref{sect:BestPractice}.
 \item The option \Verb^output_form^ selects the format of the processed amplitude output: \texttt{form} (default) or \texttt{math} (Mathematica format). The option does not change the format of the initially generated diagram expressions, which remain \form{}-compatible. The subsequent algebraic processing also continues to use \form{}.
 \item The option \Verb^draw_graphs^ draws Feynman diagrams and diagram topologies using \texttt{neato} from \texttt{Graphviz}. It takes the values \texttt{true} or \texttt{false}. The default is \texttt{false}. Drawing significantly slows down the generation.
 \item The option \Verb^draw_protos^ generates prototype drawings in the \texttt{pictures} directory, using filenames of the form \texttt{PR<N>.pdf}. The default is \texttt{false}.
 \item The option \Verb^read_protos^ loads an existing prototype map from the amplitude directory for reuse in subsequent diagram generation. Previously identified prototypes can be reused, while unmatched structures are inserted into the map according to the ordinary prototype-generation procedure. Thus, this option does not freeze the prototype map or guarantee an identical set of prototypes when changing the process. The default is \texttt{false}.
 \item The option \Verb^parallel_form^ takes a
non-negative integer. A positive value selects the
parallel \form{} processing layout and specifies
the default concurrency of the generated scripts.
Setting \Verb^parallel_form: 0^ (the default) selects serial
processing. \diagen{} generates the corresponding
scripts but does not execute \form{} during
diagram generation.
 \item The option \Verb^incoming_fields: [A, A]^ takes a list of incoming particles.
 \item The option \Verb^outgoing_fields: [H]^ takes a list of outgoing particles.
 \item The option \Verb^internal_fields_max: [[H, 0]]^ takes a list of pairs. The first entry of the pair is the name of the particle and the second entry is an integer number. The program \diagen{} discards diagrams if the number of specified particles is bigger than the integer number. Only internal particles are considered.
 \item The option \Verb^internal_fields_min: [[H, 0]]^ takes a list of pairs. The first entry of the pair is the name of the particle and the second entry is an integer number. The program \diagen{} discards diagrams if the number of specified particles is lower than the integer number. Only internal particles are considered. Both options use the canonical field name, i.e.\ the second entry of the corresponding \texttt{field} pair in \texttt{model.yaml}. For example, a field defined as \texttt{[QBar,Q]} must be selected using \Verb^internal_fields_min: [[Q,1]]^ rather than \Verb^internal_fields_min: [[QBar,1]]^.
 \item The option \Verb^min_one_field^ requires at least one internal propagator to carry one of the specified field types. For example, \Verb^min_one_field: [[Q, 1], [q, 1]]^ retains diagrams containing an internal heavy or light quark propagator. Unlike \Verb^internal_fields_min^, the conditions are combined using a logical OR, and only the field names of the pairs are used.
\item The option \Verb^flag_loop_particles^ assigns symbolic counters to closed fermion loops of specified field types. Each entry contains a list of particle names and a corresponding symbol. For example, \Verb^{particles: [Q], symbol: "nh"}^ introduces the counter \texttt{nh}, which can be used in the subsequent \form{} calculation. This option labels contributions but does not filter the generated diagrams. Use \Verb^fermion_loops_min^ and \Verb^fermion_loops_max^ to restrict their total number.
\item The optional settings \Verb^fermion_loops_min^ and \Verb^fermion_loops_max^ select diagrams according to the number of closed fermion chains. An omitted bound imposes no restriction in that direction. These options do not constrain external open fermion chains. For example, setting both bounds to \texttt{1} retains only diagrams containing exactly one closed fermion loop.
\item The options \Verb^cut_fields^ and \Verb^no_cut_fields^ select diagrams by the field content of allowed cuts. A diagram is retained if a compatible cut exists that contains all requested \Verb^cut_fields^ and contains none of the fields listed in \Verb^no_cut_fields^. For an example, the job file in \path{examples/bottom-decay} requests \Verb^cut_fields: [lepton,neutrino]^ and \Verb^no_cut_fields: [bottom]^. These options select the original diagrams. They do not export individual cut configurations or apply cutting rules to the propagators.
The options \Verb^cut_fields^ and \Verb^no_cut_fields^, as well as the particle lists in \Verb^flag_loop_particles^, use the canonical field names. For example, \Verb^cut_fields: [Q]^ selects cuts containing the field defined by \texttt{[QBar,Q]}, whereas \Verb^cut_fields: [QBar]^ does not match this canonical name.
\item The option \Verb^open_chain_photon^ selects diagrams according to whether a chosen boson is attached at an external endpoint of a selected open fermion chain. It takes the fields identifying the chain, the boson name, and \Verb^keep: with^ or \Verb^keep: without^. It does not select arbitrary internal attachments along the chain.
\begin{verbatim}
open_chain_photon:
  fields: [mu]
  boson: A
  keep: without
\end{verbatim}
Here \texttt{fields} identifies the fermion-chain
field types, \texttt{boson} specifies the boson to
check, and \texttt{keep} determines whether matching
diagrams are retained or discarded. The accepted
values are \texttt{with} and \texttt{without}.
\item The option \Verb^open_chain_mass^ replaces the mass label of internal propagators on selected open fermion chains. A chain is selected if it contains at least one of the listed fields. The internal propagators of these fields along the chain are then assigned the specified mass label instead of the mass defined in \texttt{model.yaml}. External legs and closed fermion loops are not affected.
\begin{verbatim}
open_chain_mass:
  fields: [mu]
  mass: 0
\end{verbatim}
The replacement enters both the generated propagators, through the macro \texttt{\%m}, and the scalar diagrams used for prototype matching. The value \texttt{0} sets the selected lines massless. Both \texttt{fields} and \texttt{mass} are required.
 \item The option \Verb^two_particle_vertex^ (default: \texttt{0}) specifies the exact number of two-particle vertices allowed in a diagram. All other diagrams are discarded.
 \item The option \Verb^counter_term_order^ (default: \texttt{0}) retains only those diagrams whose total counterterm order is exactly equal to this value. All others are discarded. The counterterm order may be specified in the model file.
 \item The option \Verb^output_file^ changes the name of the generated amplitude file. By default, \diagen{} uses \texttt{output.diagen}. When specifying a different filename, the supplied \form{} procedures must be adapted accordingly. In particular, \texttt{TreatAmplitude.prc} contains a reference to \texttt{output.diagen} for the momentum-mapping procedure, which is not automatically updated by this option.
\item Options that filter diagrams based on
additional information during topology generation:
 \begin{itemize}
 \item \Verb^one_particle_irreducible: false^ \\
 A topology is called \textit{one-particle irreducible} if cutting any of the internal edges does not lead to a disconnected topology. Setting this option to \texttt{true} retains only one-particle irreducible topologies. By default, this option is set to \texttt{false}.
 \item The option \Verb^no_tadpoles^ excludes topologies containing tadpole subgraphs. These are subgraphs connected to the remainder of a diagram through a single vertex and carrying no net external momentum. The default is \texttt{true}. The filter is graph-theoretic and should not be confused with the subsequent evaluation of scaleless integrals.
 \item The option \Verb^no_loops^ excludes graph self-loops, i.e.\ internal edges whose source and target are the same vertex. This restriction does not exclude diagrams containing ordinary closed cycles or diagrams of nonzero perturbative loop order. The default is \texttt{true}.
 \item \Verb^no_self_energies: false^ \\
 A topology is said to have a \textit{self-energy insertion} if there exists a proper subgraph that can be disconnected from the topology by cutting two edges such that the removed component does not contain external nodes. By default, this option is set to \texttt{false} to allow for self-energy insertions.
 \item \Verb^on_shell: true^ \\
 A topology is called \textit{on-shell} if the corresponding graph is devoid of self-energy insertions on the external edges. By default, this option is set to \texttt{true}.
\end{itemize}
\end{itemize}
The job-file parser additionally accepts the keys \Verb^allow_disconnected^, \Verb^qgraf_style^, and \Verb^qgraf_model^. They are reserved for internal and legacy use, are not part of the validated workflow described here, and should not be used in production calculations.

\paragraph{Model file.}
The user has to provide a directory \texttt{config} with a file named \texttt{model.yaml}. See for example:
\begin{verbatim}
#model.yaml
model:
 - name: "ggH"
   fields:
    - {field: [A, A],
       propagator: "DV(v%sl%e,v%tl%e,%p,%m)",
       wavefunctions: ["Eps(v%vl%l,%p,cOli%l)",
                       "EpsStar(v%vl%l,%p,cOli%l)"]}

    - {field: [CBar, C],
       anticommuting: "yes",
       propagator: "DG(%p,%m)"}

    - {field: [QBar, Q],
       anticommuting: "yes",
       fermion: "yes",
       mass: "M",
       propagator: "SF(%c,%p,%m)"}

    - {field: [qbar, q],
       anticommuting: "yes",
       fermion: "yes",
       propagator: "SF(%c,%p,%m)"}

    - {field: [H, H],
       mass: "MH",
       propagator: "DS(%p,%m,1)",
       wavefunctions: ["1","1"]}

   interactions:
    - {interaction: [QBar, Q, A],
       vertex: ["G(%c,%i3)*cOlT(cOli%l1,cOli%l2,cOli%l3)"]}

    - {interaction: [qbar,q, A],
       vertex: ["G(%c,%i3)*cOlT(cOli%l1,cOli%l2,cOli%l3)"]}

    - {interaction: [CBar, C, A],
       vertex: ["GhostV(%p1,%i3)*(i_*cOlf(cOli%l1,cOli%l2,cOli%l3))"]}

    - {interaction: [A, A, A],
       vertex: ["VVV(%p1,%i1,%p2,%i2,%p3,%i3)*
       (-1/6*i_*cOlf(cOli%l1,cOli%l2,cOli%l3))"]}

    - {interaction: [A, A, A, A],
       vertex: ["(+d_(%i1,%i2)*d_(%i3,%i4)*(-1/24*
       (cOlf(cOlj%v,cOli%l1,cOli%l3)*cOlf(cOlj%v,
       cOli%l2,cOli%l4)+cOlf(cOlj%v,cOli%l2,cOli%l3)*
       cOlf(cOlj%v,cOli%l1,cOli%l4)))+d_(%i1,%i3)*
       d_(%i2,%i4)*(-1/24*(cOlf(cOlj%v,cOli%l1,cOli%l2)*
       cOlf(cOlj%v,cOli%l3,cOli%l4)+cOlf(cOlj%v,cOli%l3,
       cOli%l2)*cOlf(cOlj%v,cOli%l1,cOli%l4)))+d_(%i1,%i4)*
       d_(%i2,%i3)*(-1/24*(cOlf(cOlj%v,cOli%l1,cOli%l2)*
       cOlf(cOlj%v,cOli%l4,cOli%l3)+cOlf(cOlj%v,cOli%l4,
       cOli%l2)*cOlf(cOlj%v,cOli%l1,cOli%l3))))"]}

    - {interaction: [QBar, Q, H],
        vertex: ["(-1)*(%m1)*delta(cOli%l1,cOli%l2)"]}
\end{verbatim}
The main options are explained in the following:
\begin{itemize}
 \item The option \texttt{fields} is used to define the fields of the particles.
 \item The option \texttt{interactions} defines the allowed interactions between particles. Vertices with up to 10 legs can be generated. However, the vertex-template macros support only single-digit leg indices (1--9), so the tenth leg cannot be accessed through these macros.
 \item The option \texttt{order: N}, where \texttt{N} is a positive integer, assigns a counterterm order to an interaction. It is used together with the job option \Verb^counter_term_order^.
 \item The option \texttt{field: [QBar,Q]} specifies the antiparticle and particle identifiers, respectively. Both identifiers may be identical for self-conjugate fields. Internally, the second entry defines the canonical field name shared by the particle and antiparticle. For example, \texttt{[QBar,Q]} has the canonical name \texttt{Q}, which is used by several diagram filters and in generated expressions.
 \item The option \texttt{anticommuting: yes} instructs \diagen{} to treat the particle as an anticommuting field. The default value is \texttt{no}.
 \item The option \texttt{fermion: yes} instructs \diagen{} to treat the particle as a fermion.
 The default value is \texttt{no}.
\item The option \texttt{mass: M} assigns a mass
label to a field. If the option is omitted or
empty, the field is treated as massless and
the internal mass label is set to \texttt{0}.
Consequently, the macro \texttt{\%m} expands
to \texttt{0} for a massless field.
 \item The option \texttt{propagator: SF(\%c,\%p,\%m)} is used to define the field propagator.
 \item The option \texttt{wavefunctions: [...]} accepts a list of two or four elements. With two entries, the first and second specify the incoming and outgoing wavefunctions, respectively. For distinct particle and antiparticle identifiers, these templates are reused for both fields.

 With four entries, separate wavefunctions can be specified: the first and fourth entries correspond to the incoming and outgoing particle, respectively, while the third and second entries correspond to the incoming and outgoing antiparticle, respectively.

 Whether a field is self-conjugate is determined by the field identifiers, not by the number of wavefunction templates. Identical particle and antiparticle identifiers define a self-conjugate field.

\item The option
\texttt{interaction: [QBar,Q,H]}
specifies the field content of an allowed
interaction vertex, using the identifiers
defined in the model.
 \item The option \texttt{vertex} takes the user-defined Feynman rule for the vertex.
\end{itemize}

The attributes \texttt{anticommuting} and \texttt{fermion} serve different purposes. The former specifies Grassmann statistics, while the latter identifies fermionic fields for the corresponding fermion-chain treatment. For example, ghost fields are anticommuting but are not assigned ordinary fermion-chain semantics.

For conventional vertex Feynman rules, the identical-field factorials in the numerator of Eq.~\eqref{eq:preliminaries:symfactor} must be compensated in the model templates. Every interaction vertex and internal line contributes one factor of the imaginary unit by default. To compensate a single such factor in a \form{} vertex template, include \Verb^(-i_)*^, where \Verb^i_^ is the predefined imaginary unit in \form{}.

\paragraph{Kinematics file.}
The file \texttt{kinematics.yaml}, located in the
\texttt{config} directory, specifies the external
kinematics and their algebraic relations.
While this file is optional for diagram generation,
it is required for the complete \form{} and \kira{}
workflow described in this paper. Without it,
\diagen{} does not generate the corresponding
kinematic declarations and replacement rules needed
for the supplied \form{} procedures, and the exported
\kira{} configuration lacks the necessary kinematic
information. Users omitting this file must provide
these definitions separately.

An example is given below:
\begin{verbatim}
#kinematics.yaml
kinematics:
  incoming_momenta: [p1, p2]
  outgoing_momenta: [p3]
  momentum_conservation: [p3, p1+p2]
  kinematic_invariants:
    - [M, 1]
    - [mQ2, 2]
    - [s, 2]
  scalarproduct_rules:
    - [[p1, p1], 0]
    - [[p2, p2], 0]
    - [[p1+p2, p1+p2], s]
  mass_replacements:
    - [M^2, mQ2]
  symbol_to_replace_by_one: s
\end{verbatim}
This file is used to generate replacement rules for diagram kinematics. In the present version, the external momenta are restricted to the label format \texttt{p1, p2, p3, \ldots}. The labels \texttt{k, k1, k2, k3, \ldots} are reserved for internal loop momenta. \diagen{} currently always eliminates the last momentum in lexicographical ordering, which has to be respected when specifying momentum conservation. For two-point functions, the incoming and outgoing momentum must be assigned the same name.
The options are briefly explained:
\begin{itemize}
 \item The option \texttt{incoming\_momenta} takes a list of incoming momenta.
 \item The option \texttt{outgoing\_momenta} takes a list of outgoing momenta.
 \item The option \texttt{momentum\_conservation} replaces the momentum defined in the first input, through the expression defined in the second input.
\item The option \texttt{kinematic\_invariants}
defines the kinematic symbols required by the
generated expressions and exported integral
families. Each entry is a pair consisting of
the symbol name and its mass dimension.
Relevant mass parameters and the invariants
appearing in \texttt{scalarproduct\_rules}
and \texttt{mass\_replacements} must be included.
Mass parameters belonging to fields not used
in the generated expressions need not be listed.
 \item The option \texttt{scalarproduct\_rules} defines all kinematic relations. Each relation is given as a pair: the first element is a pair of momenta, and the second element is the value of their scalar product.
 \item The option \texttt{mass\_replacements} is required for the interface to \kira{}: squares of masses as defined in the model must be replaced by new variables for proper integral-family definitions.
 \item The option \texttt{symbol\_to\_replace\_by\_one} can be used to set one kinematic symbol to one.
\end{itemize}

Having prepared these files, the working directory should contain the following files:
\begin{verbatim}
jobs.yaml
config
config/model.yaml
config/kinematics.yaml
\end{verbatim}
To run the diagram generation, \diagen{} is started with the name of the job file as command-line argument:
\begin{verbatim}
diagen jobs.yaml
\end{verbatim}

\subsection{Example B: \texorpdfstring{\mbox{$gg\to Q\bar{Q}$}}{gg -> QQbar}}

The following input files are provided in \texttt{examples/ggQQBar}.

\paragraph{Job file.}

\begin{verbatim}
#jobs.yaml
jobs:
 - generate_diagrams:
    loop: 1
    children: 2
    parallel_form: 1
    output_form: "form"
    draw_graphs: "true"
    draw_protos: "true"
    incoming_fields: [A, A]
    outgoing_fields: [Q, QBar]
    flag_loop_particles:
     - {particles: [Q], symbol: "nh"}
     - {particles: [q], symbol: "nl"}
\end{verbatim}

\paragraph{Model file.}
The model file for this example is obtained from that of
Section~\ref{sect:examples:ggh} by changing the model name to
\texttt{ggQQBar}, removing the Higgs field \texttt{[H,H]} and the
interaction \texttt{[QBar,Q,H]}, and adding external wavefunctions to
the heavy-quark field definition. The modified entry in the
\texttt{fields} list reads
\begin{verbatim}
    - {field: [QBar, Q],
       anticommuting: "yes",
       fermion: "yes",
       mass: "M",
       propagator: "SF(%c,%p,%m)",
       wavefunctions: ["U(%p,%c,cOli%l)","V(%p,%c,cOli%l)",
                       "VBar(%p,%c,cOli%l)","UBar(%p,%c,cOli%l)"]}
\end{verbatim}
All other field definitions and QCD interaction templates are unchanged.
The four wavefunction entries specify the incoming quark, outgoing
antiquark, incoming antiquark, and outgoing quark, respectively, following
the convention described in Section~\ref{sect:examples:ggh}.
The complete model file is supplied in \path{examples/ggQQBar/config/model.yaml}.

\paragraph{Kinematics file.}

\begin{verbatim}
#kinematics.yaml
kinematics:
  incoming_momenta: [p1, p2]
  outgoing_momenta: [p3, p4]
  momentum_conservation: [p4, p1+p2-p3]
  kinematic_invariants:
    - [M, 1]
    - [mQ2, 2]
    - [s, 2]
    - [t, 2]
    - [u, 2]
  scalarproduct_rules:
    - [[p1, p1], 0]
    - [[p2, p2], 0]
    - [[p3, p3], M^2]
    - [[p1+p2, p1+p2], s]
    - [[p1-p3, p1-p3], t]
    - [[p2-p3, p2-p3], 2*M^2-s-t]
  mass_replacements:
    - [M^2, mQ2]
  symbol_to_replace_by_one: s
\end{verbatim}

The following listings illustrate the generated output
for a representative diagram. Diagram identifiers such
as \texttt{d33} and \texttt{m33} are assigned during
generation. Repeated runs of the same \diagen{} version
with the same input produce identical output, but these
identifiers, the edge orientations, and the printed
momentum expressions may change when the model, the
job options, or the program version are changed.

\paragraph{Example output for d33.}

\begin{verbatim}
#output.diagen
*--#[ d33 :

*	topology: ./pictures/t6.pdf
*
*	internal fields: A A Q A
*
*	detailed description:
*
*	0: 0 -> 7, p1
*	1: 4 -> 7, A -> A, -p3-k
*	2: 5 -> 7, A -> A, -p1+p3+k
*	3: 1 -> 6, -p4
*	4: 4 -> 6, Q -> QBar, k
*	5: 5 -> 6, A -> A, p4-k
*	6: 2 -> 5, p2
*	7: 3 -> 4, -p3

#define nf "0"
#define nO "1"
#define nh "0"
#define nl "0"

g d33 =

	+36*(+1)
	*UBar(p3,0,cOli7)
	*G(0,v4l1)*cOlT(cOli7,cOli4,cOli1)
	*SF(0,-k,M)
	*G(0,v6l5)*cOlT(cOli4,cOli3,cOli5)
	*V(p4,0,cOli3)
	*VVV(p2,v5l6,p1-p3-k,v5l2,-p4+k,v5l5)*(-1/6*i_*cOlf(cOli6,cOli2,cOli5))
	*VVV(p1,v7l0,-p3-k,v7l1,-p1+p3+k,v7l2)*(-1/6*i_*cOlf(cOli0,cOli1,cOli2))
	*Eps(v7l0,p1,cOli0)
	*DV(v4l1,v7l1,-p3-k,0)
	*DV(v5l2,v7l2,-p1+p3+k,0)
	*DV(v5l5,v6l5,p4-k,0)
	*Eps(v5l6,p2,cOli6)

;

*--#] d33 :
\end{verbatim}

\paragraph{Example output for m33.}

\begin{verbatim}
#output.diagen
*--#[ m33 :

*	associated prototype(s):

#define NPROTO "1"
#define PROTO1 "PR19"

*	momentum shift(s):

id DS(k+p3,0,n1?)*DS(k-p1+p3,0,n2?)*DS(k,M,n3?)*DS(k-p4,0,n4?) =
  PR19*DS(k+p3,0,n1)*DS(k-p1+p3,0,n2)*DS(k,M,n3)*DS(k-p4,0,n4)
    *replace_(p1,p1,
              p4,-(-p4),
              p2,p2,
              p3,-(-p3),
              k,k);

*--#] m33 :
\end{verbatim}

\subsection{Getting-to-Know Example}
\label{sect:introductory-example}

In this example, we illustrate the complete workflow from diagram generation to the preparation of scalar integrals for IBP reduction.  We consider the one-loop process \mbox{$gg\to Q\bar{Q}$}, using the example directory \texttt{examples/ggQQBar} included in the \diagen{} source distribution. The example contains the required model, kinematics, job configuration, and \form{} procedures.

The corresponding model, kinematics, and job configuration were introduced in the preceding example.

\paragraph{Input files.}

The working directory contains the job file \texttt{jobs.yaml} and the directories \texttt{config} and \texttt{form}. The \texttt{config} directory contains \texttt{model.yaml}, specifying the field content and interactions, and \texttt{kinematics.yaml}, defining the external momenta, momentum conservation, and kinematic invariants.

The \texttt{form} directory contains the procedures used for the subsequent algebraic processing:
\begin{verbatim}
ColorFactors.prc  declarations.h    splitDS.prc
color.h           FeynmanRules.prc  TreatAmplitude.prc
\end{verbatim}

The file \texttt{FeynmanRules.prc} defines the additional algebraic rules required to process the model-dependent expressions generated by \diagen{}. The file \texttt{TreatAmplitude.prc} specifies the treatment of the amplitude, including the projectors used to extract the desired form factors or squared matrix elements.

The file \texttt{declarations.h} contains declarations of the symbolic objects introduced in the model and used during the \form{} calculation. The files \texttt{color.h} and \texttt{ColorFactors.prc} provide the color-algebra conventions and simplification procedures. For standard QCD applications, these files can typically be reused without substantial modifications.

The file \texttt{splitDS.prc} provides rules for partial fractioning scalar propagator expressions when required. In contrast to the graph-based prototype matching performed by \diagen{}, this is an algebraic preprocessing step carried out within the \form{} workflow.

For applications beyond the supplied QCD examples, the user must adapt the relevant \form{} procedures, in particular \texttt{FeynmanRules.prc} and \texttt{TreatAmplitude.prc}, to the chosen theory and observable.

\paragraph{Diagram generation.}

From the example directory, the diagram generation is started by executing
\begin{verbatim}
diagen jobs.yaml
\end{verbatim}

Upon successful completion, the working directory contains the following files and directories:
\begin{verbatim}
amplitude  config      form       kira_interface
compute    diagen.log  jobs.yaml  pictures
\end{verbatim}

The directory \texttt{amplitude} contains the generated diagrams and their associated expressions. In particular, \texttt{amplitude/output.diagen} stores the diagram contributions in a human-readable, \form{}-compatible representation.

For each diagram, \diagen{} records the field assignments, momentum routing, symmetry factor, and fermion sign. The corresponding prototype mappings specify how the scalar propagator structures are related to the selected integral representatives.

The directory \texttt{pictures} is used for visualizations of diagrams, topologies, and prototypes. Drawing is enabled in the supplied example through
\Verb^draw_graphs: true^ and \Verb^draw_protos: true^.
Since drawing introduces additional computational overhead, both options are typically disabled during large production runs.

The directory \Verb^kira_interface^ contains the generated integral-family definitions and kinematic information for subsequent IBP reduction. The subdirectory \texttt{PRall} provides a combined configuration containing the selected integral families, while directories of the form \Verb^reduce_PR<N>^ contain configurations for reductions associated with individual prototype families.

In the present example, the relevant prototype families are labelled \texttt{PR15} through \texttt{PR19}. Their numerical identifiers depend on the generated prototype map and are not universal identifiers of the corresponding integral topologies.

\paragraph{Algebraic processing with \form{}.}

The directory \texttt{compute} contains the scripts and configuration files for processing the generated amplitude:
\begin{verbatim}
auto-sum.frm  collect_integrals.sh  form.set
printPR.frm   sum_parallel.sh       tmp
\end{verbatim}

The central \form{} program is \texttt{auto-sum.frm}. It reads the generated diagram expressions, applies the user-defined algebraic procedures and projectors, and prepares the scalar-integral expressions required for the subsequent reduction.

Parallel processing is selected during diagram
generation by setting \Verb^parallel_form^ to a
positive value in \texttt{jobs.yaml}. The supplied
example uses \Verb^parallel_form: 1^, which generates
the parallel processing scripts. The \form{}
calculation must subsequently be started explicitly.

For a parallel calculation, we enter the \texttt{compute} directory and execute
\begin{verbatim}
cd compute
./sum_parallel.sh -p100 Born
\end{verbatim}

Here, \texttt{Born} denotes the projector defined in \texttt{form/TreatAmplitude.prc}. The option \texttt{-p100} requests up to 100 parallel \form{} processes and should be adjusted to the available computational resources.

After successful execution, the processed amplitude is stored in \texttt{compute/sum1l.Born.sav}.

Alternatively, serial processing requires setting
\Verb^parallel_form: 0^ in \texttt{jobs.yaml}
and regenerating the diagram output before running
the following command from the \texttt{compute}
directory:
\begin{verbatim}
form -d Projector=Born auto-sum.frm
\end{verbatim}
The serial and parallel processing layouts should
not be mixed. In particular, the serial command
must not be executed on intermediate files already
processed and removed by the parallel workflow.

\paragraph{Structure of the amplitude treatment.}
The supplied \texttt{TreatAmplitude.prc} procedures separate the process-dependent projection from the subsequent algebraic processing. The procedure takes three arguments specifying the loop order, the gauge choice passed to the Feynman-rule substitutions, and the projector name. In the \mbox{$gg\to H$} example, the projector \texttt{FF} contracts the gluon polarization vectors with a Lorentz and color tensor to extract a scalar form factor. In the \mbox{$gg\to Q\bar{Q}$} example, \texttt{Born} instead inserts the conjugate Born amplitude and implements the polarization and spin sums required for the interference with the generated amplitude. The projector name selected when starting the \form{} calculation must agree with a projector defined in the procedure.

The remaining steps are largely common to the two examples. After the projection and the loop-order-dependent phase normalization, the procedure simplifies the color structures, applies the Feynman-rule substitutions and evaluates the Dirac traces. It then normalizes scalar propagators, performs partial fractioning where necessary, applies the momentum transformations supplied by \diagen{}, and imposes momentum conservation and the kinematic relations. Factors independent of the loop momenta are extracted from the scalar propagator representation. Finally, the generated prototype-matching rules express the loop-dependent scalar products and propagators in terms of the selected integral families. When adapting an example to another process, the projector and its normalization are therefore the principal observable-dependent modifications. The common processing steps can generally be retained, provided the field conventions, kinematic substitutions and model-dependent algebra are compatible.

\paragraph{Preparation of IBP reductions.}

After the algebraic processing has finished, the scalar integrals appearing in the resulting expressions must be collected and assigned to the generated integral families.

This step is performed by executing
\begin{verbatim}
./collect_integrals.sh Born
\end{verbatim}
from the \texttt{compute} directory. The argument \texttt{Born} identifies the previously processed projector.

The script collects the relevant scalar integrals and prepares the corresponding \kira{} job files using the integral-family definitions generated by \diagen{}.

The reduction configurations are stored in the \texttt{kira\_interface} directory. For the present example, the individual configurations correspond to the prototype families \texttt{PR15} through \texttt{PR19}.

These configurations can subsequently be used to perform the IBP reductions with \kira{}, without requiring a separate manual reconstruction of the integral families from the original diagrams.

\paragraph{Choice of generation options.}

The example uses the following principal options of the \Verb^generate_diagrams^ job:
\begin{verbatim}
loop              children          parallel_form
output_form       draw_graphs       draw_protos
incoming_fields   outgoing_fields   flag_loop_particles
\end{verbatim}

The option \Verb^loop: 1^ selects the one-loop contribution to the process. The option \Verb^children: 2^ specifies a lower bound on the number of physical propagators retained during subprototype generation.

The choice of \Verb^children^ influences the extent of the prototype map and therefore the computational cost of prototype construction and matching. A larger bound generally restricts the number of subprototypes considered, whereas a smaller bound permits a more extensive search for matching relations.

For matching the physical sectors generated for a given process and loop order, a practical choice is the smallest number of physical propagators among the sectors found in a preliminary run with a sufficiently large cutoff (see Section~\ref{sect:BestPractice}). For large multiloop processes, the difference between a restricted and an extensive prototype search can be substantial, as discussed in Section~\ref{sect:benchmarks}.

The option \Verb^output_form^ selects the output format of the processed amplitude, while the initial diagram expressions remain \form{}-compatible. The settings \Verb^draw_graphs^ and \Verb^draw_protos^ control graphical output. Both are enabled in the supplied example but can be disabled to avoid unnecessary overhead.

Finally, \Verb^flag_loop_particles^ introduces symbolic counters for selected closed fermion loops. These counters can be used during the subsequent \form{} processing to separate contributions according to their fermion-loop content.

\paragraph{Adapting the examples to a new process.}
For a process in an existing theory, one can start from the closest supplied example. In \texttt{jobs.yaml}, specify the incoming and outgoing fields, the loop order, and any restrictions on the generated diagrams. In \texttt{config/model.yaml}, add any missing fields and interactions, including the external wavefunctions required by the process. The template conventions are described in Section~\ref{sect:model}. The ordering of external momenta in \texttt{config/kinematics.yaml} must agree with that of the external fields in the job file. The independent invariants, momentum-conservation relation, scalar-product rules, and mass replacements must be adapted consistently.

For a new observable, the \form{} projector in \texttt{TreatAmplitude.prc} must be specified by the user. For a new theory, the declarations, Feynman-rule substitutions and, where applicable, the color-algebra and partial-fractioning procedures must also be adapted to the model. It is useful to validate the normalization and field conventions with a tree-level or other known low-order amplitude before generating higher-loop contributions. The generated diagram expressions and integral-family definitions can then be inspected separately before starting the algebraic processing and the subsequent IBP reduction, as in the workflow above.

Together, these steps demonstrate how \diagen{} connects the generation of field-labelled Feynman diagrams with the construction of scalar integral families and the preparation of IBP reductions in a common workflow.

\section{Graph reconstruction from propagator momenta}
\label{sect:inverse-kirchhoff}

The conventional diagram-generation workflow starts with a graph and determines
momenta that obey momentum conservation at its vertices.  Conversely,
\diagen{} can reconstruct a connected scalar diagram from the momenta and mass
labels of an existing set of propagators.  We call this the \emph{inverse
Kirchhoff problem}.  In particular, it allows graph representations to be
recovered directly from integral-family sectors in the format used by
\kira{}, without supplying a field-theory model or regenerating the
amplitude.  The object reconstructed is a scalar diagram in the sense of
Section~\ref{sect:preliminaries:graphs}, not a model-specific diagram.

\subsection{Formulation of the inverse Kirchhoff problem}
\label{sect:inverse-kirchhoff:problem}

Suppose that the momenta of all proposed internal and external edges are
specified by coefficient vectors in a common independent momentum basis,
\begin{equation}
 q_e=\sum_{r=1}^{R}c_{er}\,b_r,\qquad e=1,\ldots,E.
 \label{eq:inverse-kirchhoff:momenta}
\end{equation}
Here $E$ counts all internal and external edges, $b_r$ is the $r$th
basis momentum, $R$ is the number of independent basis momenta, and
$c_{er}$ is the integer coefficient of $b_r$ in the momentum of edge $e$.
For a connected graph with external legs, the basis consists of $L$ loop
momenta and $E_{\mathrm{ext}}-1$ independent external momenta, so that
$R=L+E_{\mathrm{ext}}-1$.  The input determines the edge momenta $q_e$
and mass labels $m_e$, but not the endpoints of the edges.  The task is to
find the topology of a connected scalar diagram and its oriented incidence
matrix $B$ satisfying
\begin{equation}
 \sum_{e=1}^{E} B_{ve}\,q_e=0
 \qquad\text{for each internal vertex }v.
 \label{eq:inverse-kirchhoff:conservation}
\end{equation}
For an ordinary internal edge, the entries of $B$ at its two endpoints are
$+1$ and $-1$. All other entries vanish.  Each external edge has one free
internal endpoint and a degree-one external vertex, at which no conservation
condition is imposed.  A self-loop has cancelling incidences.  Mass labels
are carried along as edge attributes. They do not affect the momentum test.

With $E=E_{\mathrm{int}}+E_{\mathrm{ext}}$ edges, Euler's relation gives
\begin{equation}
 V_{\mathrm{int}}=E_{\mathrm{int}}-L+1=E-R,
 \label{eq:inverse-kirchhoff:vertices}
\end{equation}
where external degree-one vertices are not included in $V_{\mathrm{int}}$.
The reconstruction assumes that the supplied symbols form an independent
basis and that all reconstructed internal vertices have degree at least
three.  For vacuum graphs the corresponding relation is $V_{\mathrm{int}}=E-R+1$.

The inverse problem need not have a unique solution: different connected
multigraphs can have the same specified momentum set.  The implementation
enumerates candidate incidence assignments and selects the first connected
representative according to the ordering of scalar diagrams
for the subsequent orientation step.
This selection should not be confused with a general uniqueness theorem.

\subsection{User interface and a two-loop example}
\label{sect:draw-kira}
\label{sect:inverse-kirchhoff:usage}

The \Verb^draw_diagrams^ job reads \kira{} integral-family and kinematics
files.  By default it processes all entries in the \Verb^top_level_sectors^
lists of all families in the specified file.  The relevant example
configuration is stored in \nolinkurl{examples/kira_draw_all_top_sectors}. The paths below are relative to a working directory containing the
\texttt{config/} subdirectory.  A minimal \texttt{jobs.yaml} is
\begin{verbatim}
jobs:
  - draw_diagrams:
      integralfamilies: config/integralfamilies.yaml
      output_dir: drawn_diagrams
      output_format: both
\end{verbatim}
The \Verb^integralfamilies^ key gives the integral-family file.
The optional \Verb^kinematics^ key overrides the default path to
\texttt{kinematics.yaml} in the same directory as that file.  The
optional \Verb^sectors^ key restricts the calculation to explicitly
specified family/sector pairs. The sector must be a subsector of one
of the family's top-level sectors.  The \Verb^external_masses^ list, if
provided, assigns masses to external legs in incoming-then-outgoing order. By default these are zero.  The \Verb^output_dir^ default is
\texttt{drawn\_diagrams}.  The \Verb^output_format^ choices are
\texttt{edges}, \texttt{pdf}, and \texttt{both} (the default).

For a concrete two-loop example relevant to \mbox{$gg\to H$}, consider the
following integral family, in which the first six entries are physical
propagators and the seventh is an auxiliary denominator:
\begin{verbatim}
integralfamilies:
  - name: SIX_LINE
    top_level_sectors: [63]
    propagators:
      - ["k1", "M2"]
      - ["k2", "0"]
      - ["k1+p1", "M2"]
      - ["k1-p2", "M2"]
      - ["k1+k2+p1", "M2"]
      - ["k1+k2-p2", "M2"]
      - ["k1+k2", "0"]   # auxiliary
\end{verbatim}
The bitmask $63=2^6-1$ selects exactly the first six propagators. Therefore, the auxiliary seventh entry is \emph{not} reconstructed as an
edge.  Five selected lines have mass label \texttt{M2}, and the line
carrying $k_2$ is massless.  The accompanying kinematics file reads
\begin{verbatim}
kinematics:
  incoming_momenta: [p1, p2]
  outgoing_momenta: [p3]
  momentum_conservation: [p3, p1+p2]
\end{verbatim}
Thus the external momenta entering the graph are $p_1$, $p_2$ and
$-p_3$, with $p_3=p_1+p_2$.  The three external edges and six
selected internal propagators give $E=9$.  With independent symbols
$k_1,k_2,p_1,p_2$, Eq.~\eqref{eq:inverse-kirchhoff:vertices}
predicts five internal vertices and $L=2$ loops.

Writing the six selected propagator momenta as
\begin{equation}
\begin{aligned}
 q_1&=k_1, & q_2&=k_2, & q_3&=k_1+p_1,\\
 q_4&=k_1-p_2, & q_5&=k_1+k_2+p_1,
 &q_6&=k_1+k_2-p_2,
\end{aligned}
\label{eq:inverse-kirchhoff:six-line}
\end{equation}
the corresponding reconstruction is illustrated in
Fig.~\ref{fig:inverse-kirchhoff-steps}.  Since there are
$2E_{\mathrm{int}}+E_{\mathrm{ext}}=15$ unassigned incidences and five
internal vertices of degree at least three, every internal vertex in
this example must be trivalent.  The reconstructed graph contains two
independent cycles and no internal bridge (an internal edge whose removal
would disconnect the graph).

The supplied input also contains a \texttt{FOUR\_LINE} family with
\Verb^top_level_sectors: [60,51]^, representing additional sectors
which the default job processes in the same way.  To select individual
sectors instead, one can use
\begin{verbatim}
jobs:
  - draw_diagrams:
      integralfamilies: config/integralfamilies.yaml
      sectors:
        - {family: SIX_LINE, sector: 63}
        - {family: FOUR_LINE, sector: 60}
      output_format: edges
\end{verbatim}
Running either job with \Verb^diagen jobs.yaml^ produces, for example,
\path{drawn_diagrams/SIX_LINE_63.edges}. When PDF output is enabled, it
also writes \path{drawn_diagrams/SIX_LINE_63.pdf}.  With edge-list
output it additionally writes the combined
\path{drawn_diagrams/diagrams.edges}.  The \texttt{.edges} format records
the node count and an oriented edge list with mass labels, as well as
comments identifying the momentum and the family/sector.  External
edges are included.  The graphical output is an illustration of the
reconstructed connectivity, whereas the edge list retains the explicit
momentum annotations. PDF drawing requires \texttt{Graphviz}.
The default job reconstructs the top-level sectors
\texttt{SIX\_LINE/63}, \texttt{FOUR\_LINE/60}, and
\texttt{FOUR\_LINE/51}. A more involved example is provided in
\nolinkurl{examples/kira_draw_all_top_sectors_6L-5P}.
It reconstructs the top-level sector of a six-loop five-point family
with 18 physical propagators, six massless and twelve massive
(top-level sector 262143). The momentum assignment in this example is based on the six-loop graph used to illustrate spanning-tree routing in Fig.~4.1 of Ref.~\cite{Niggetiedt:2023thesis}. The remaining 27 entries are auxiliary
propagators for irreducible scalar products and are not drawn.

\subsection{Incidence-based reconstruction algorithm}
\label{sect:inverse-kirchhoff:algorithm}

The reconstruction begins by extracting the propagator momenta and mass
labels of the selected sector and the external momenta from the input.
After eliminating the dependent external momentum using momentum
conservation, the momentum expressions are represented as integer
coefficient vectors in the common basis of
Eq.~\eqref{eq:inverse-kirchhoff:momenta}.

\paragraph{Incidence slots.}
An internal edge contributes two endpoint incidences, while an external
edge contributes one unassigned internal incidence: the other endpoint
is fixed to its own external degree-one vertex.  Hence the initial
list of unassigned incidence slots contains each internal edge index
twice and each external edge index once, with the external indices
preceding the internal ones.  The number of internal
vertices to be constructed is fixed by
Eq.~\eqref{eq:inverse-kirchhoff:vertices}.

\paragraph{Recursive vertex search.}
The recursive search partitions the remaining slots
into subsets representing internal vertices.  Every proposed subset $A$
must contain at least three slots. Writing $e(a)$ for the edge associated
with incidence slot $a$, the local necessary condition is
\begin{equation}
 \exists\,(\sigma_a)_{a\in A}\in\{-1,+1\}^{|A|}:
 \qquad\sum_{a\in A}\sigma_a q_{e(a)}=0.
 \label{eq:inverse-kirchhoff:local-test}
\end{equation}
This condition is tested coefficient by coefficient in the chosen
momentum basis.  Fixing the sign of one chosen incidence (the overall
sign of the sum is irrelevant) leaves at most $2^{|A|-1}$ sign assignments to test.  The
subsets are required to contain the first unused incidence, avoiding
duplicate vertex-order permutations.  The search also bounds the
maximum subset size so that at least three incidences remain for each
still-unconstructed vertex. Repeated subsets caused by duplicate
internal-edge indices are skipped.  For every acceptable subset the
algorithm marks its slots as used, recurses on the remainder, and
backtracks to explore alternative subsets, including after a complete
candidate has been found.

\paragraph{Candidate scalar diagrams and representative selection.}
Once the prescribed number of vertices has been formed, a candidate
scalar diagram is constructed only if every slot has been consumed.
The incidences determine the topology, and the input mass labels
complete the scalar diagram up to the momentum routing. Disconnected
candidates are rejected. The surviving scalar diagrams are collected in
an ordered set using a graph comparison that accounts for their mass labels.
Equivalent candidates under this comparison are retained only once.
The first representative in this ordering is selected for the
subsequent orientation step. This does not imply
that the supplied momentum information determines a mathematically unique
graph.

\begin{figure}[!t]
  \centering
  \captionsetup[subfigure]{labelformat=parens,justification=centering,singlelinecheck=false,margin=0pt}
  \begin{subfigure}[t]{0.31\textwidth}
    \centering
    \includegraphics[width=\linewidth]{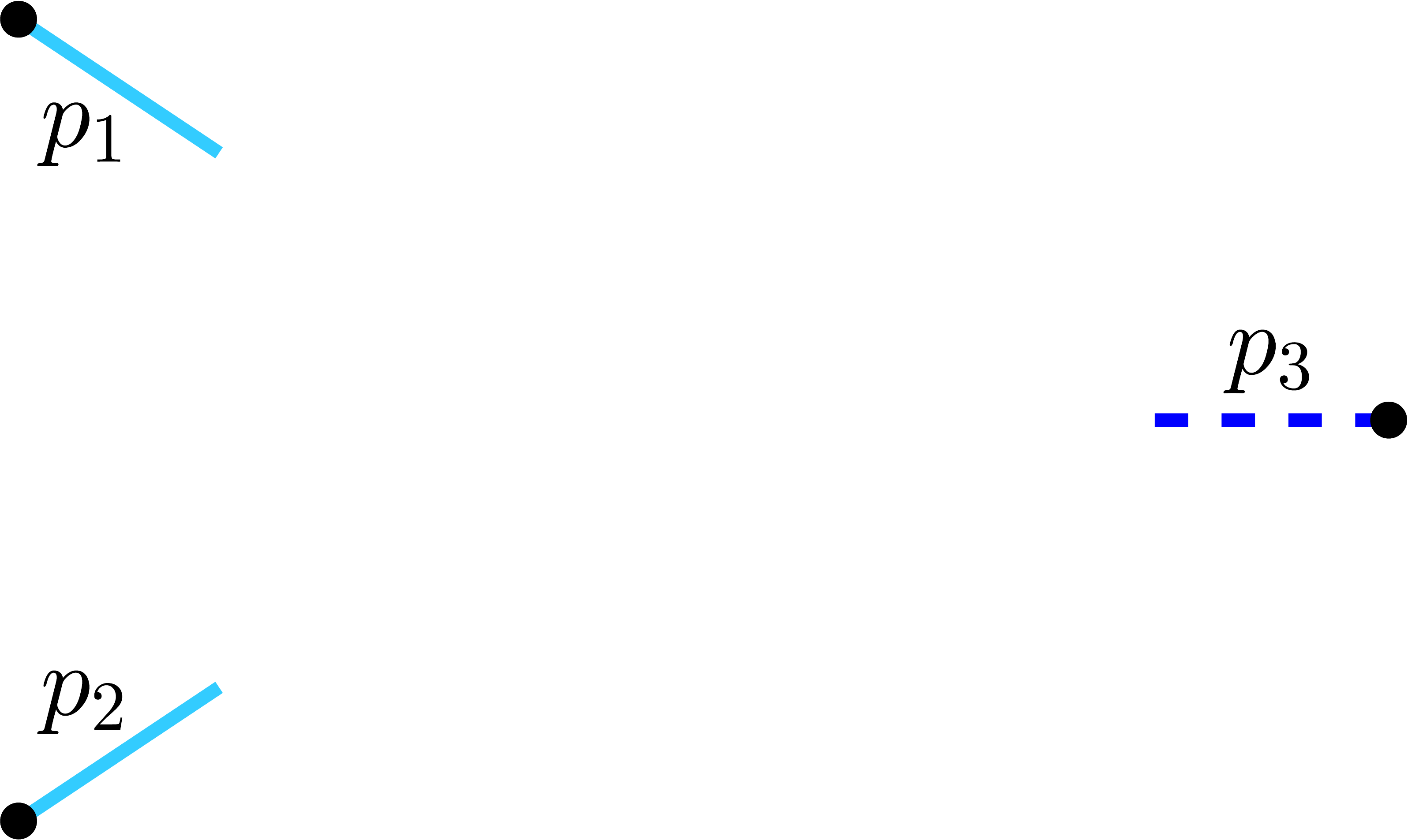}
    \caption{}
    \label{fig:inverse-kirchhoff-v0}
  \end{subfigure}\hfill
  \begin{subfigure}[t]{0.31\textwidth}
    \centering
    \includegraphics[width=\linewidth]{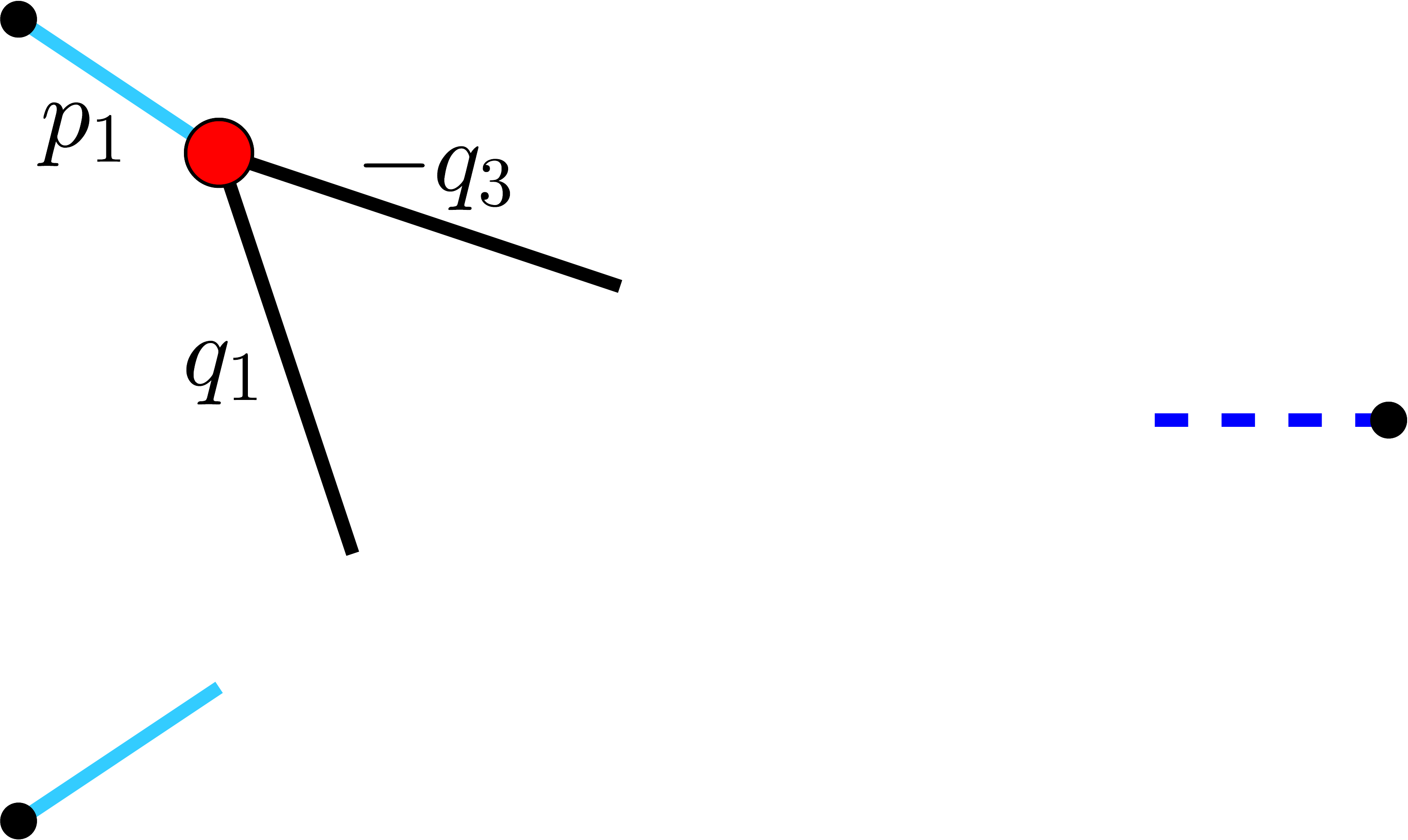}
    \caption{}
    \label{fig:inverse-kirchhoff-v1}
  \end{subfigure}\hfill
  \begin{subfigure}[t]{0.31\textwidth}
    \centering
    \includegraphics[width=\linewidth]{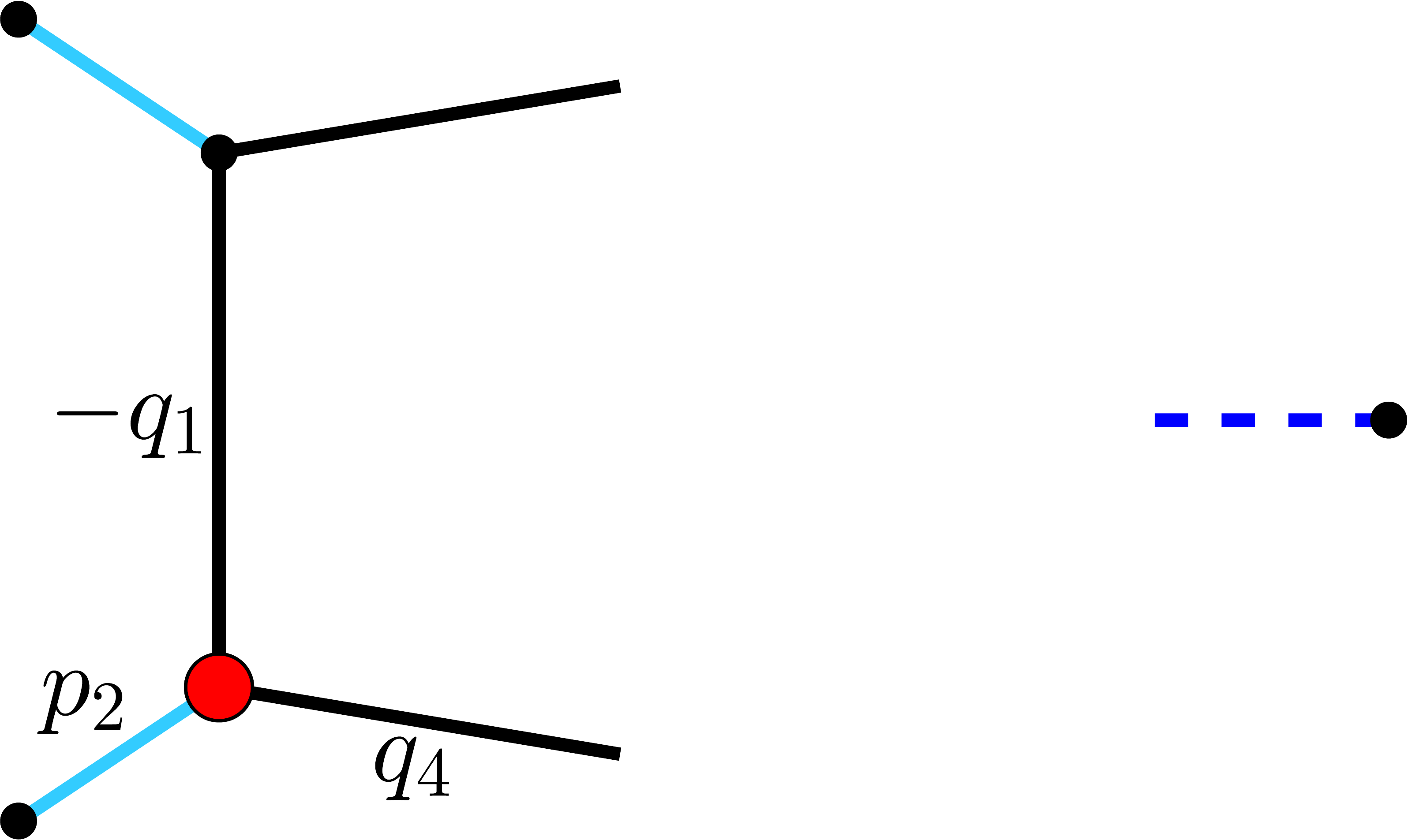}
    \caption{}
    \label{fig:inverse-kirchhoff-v2}
  \end{subfigure}

  \vspace{0.7em}

  \begin{subfigure}[t]{0.31\textwidth}
    \centering
    \includegraphics[width=\linewidth]{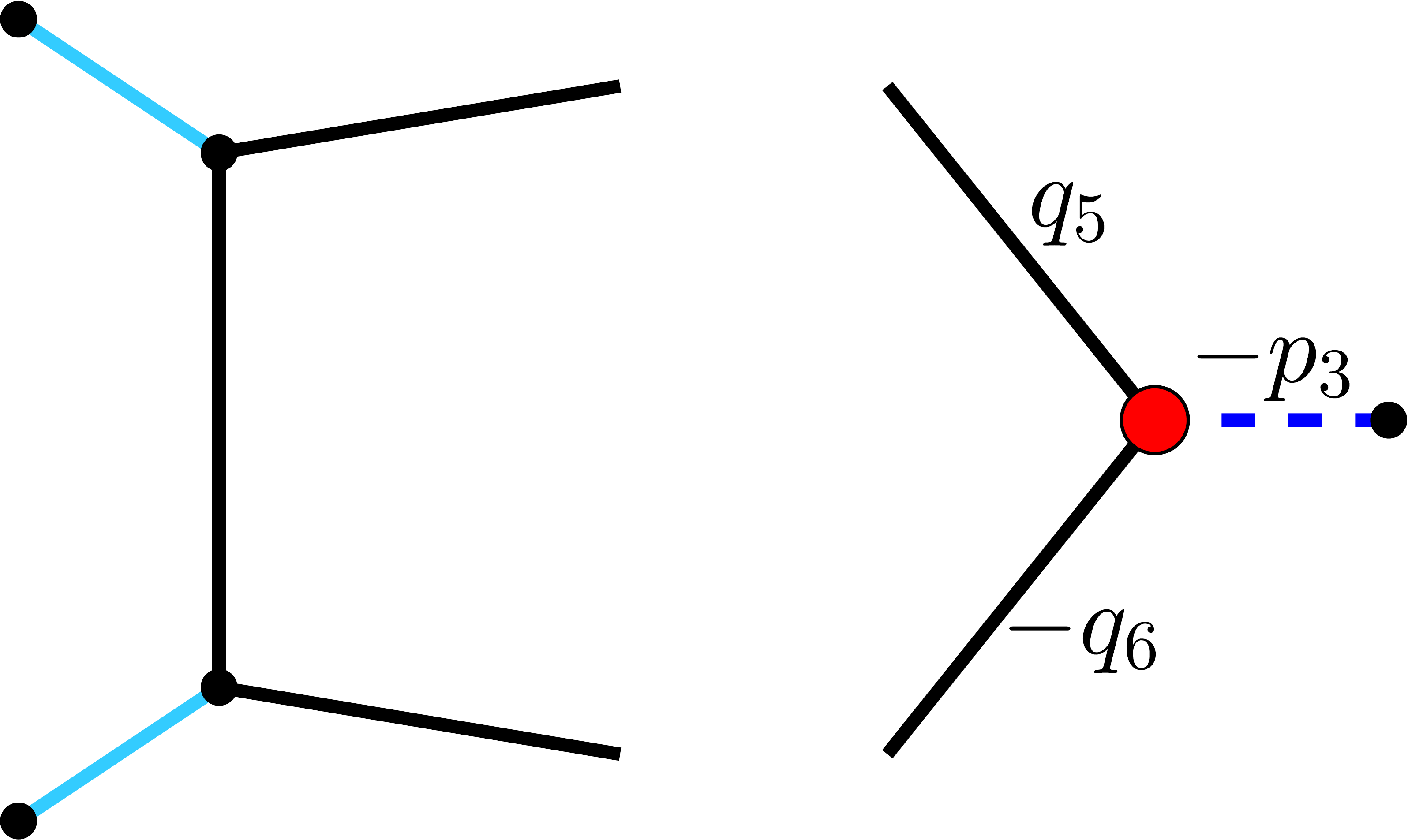}
    \caption{}
    \label{fig:inverse-kirchhoff-v3}
  \end{subfigure}\hfill
  \begin{subfigure}[t]{0.31\textwidth}
    \centering
    \includegraphics[width=\linewidth]{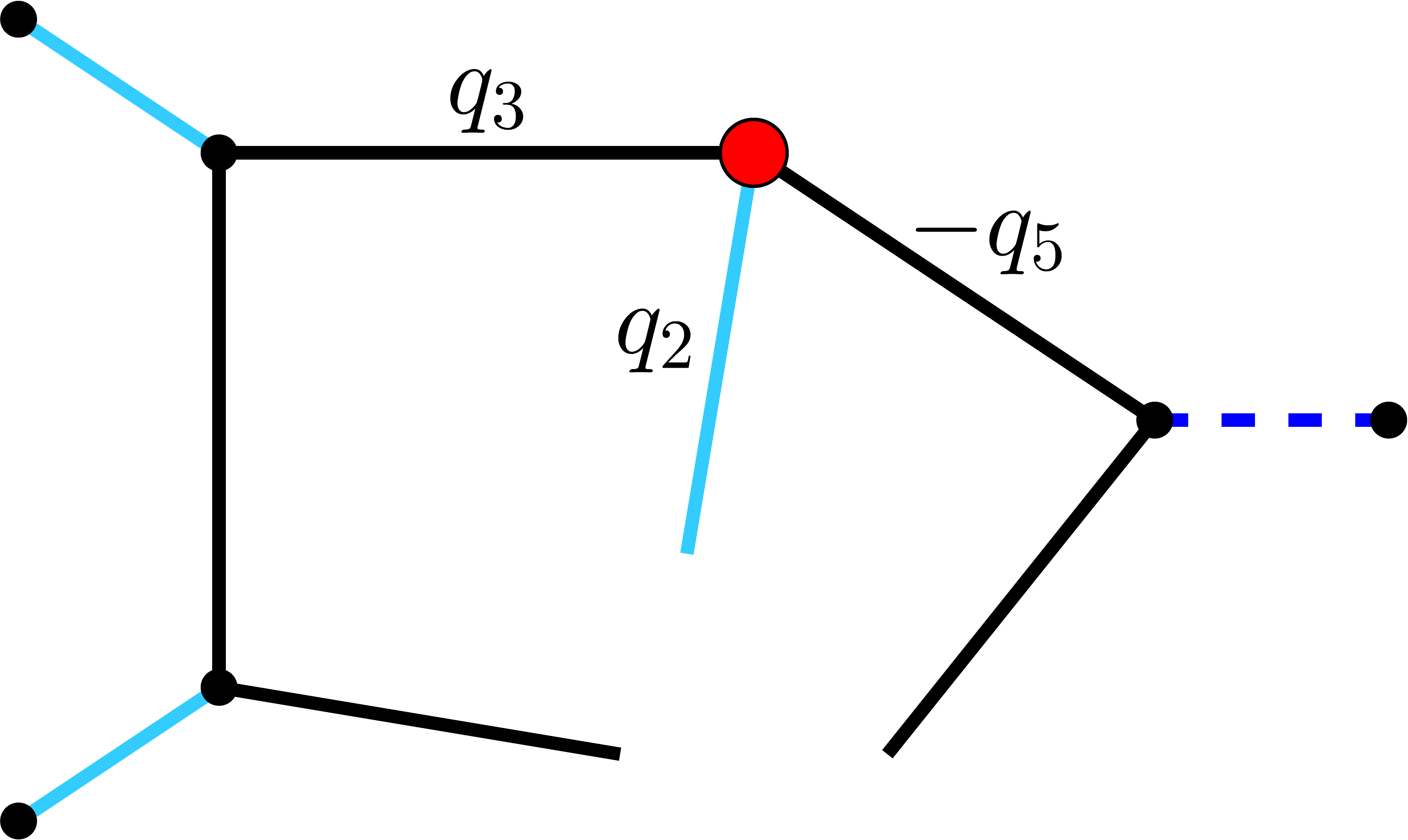}
    \caption{}
    \label{fig:inverse-kirchhoff-v4}
  \end{subfigure}\hfill
  \begin{subfigure}[t]{0.31\textwidth}
    \centering
    \includegraphics[width=\linewidth]{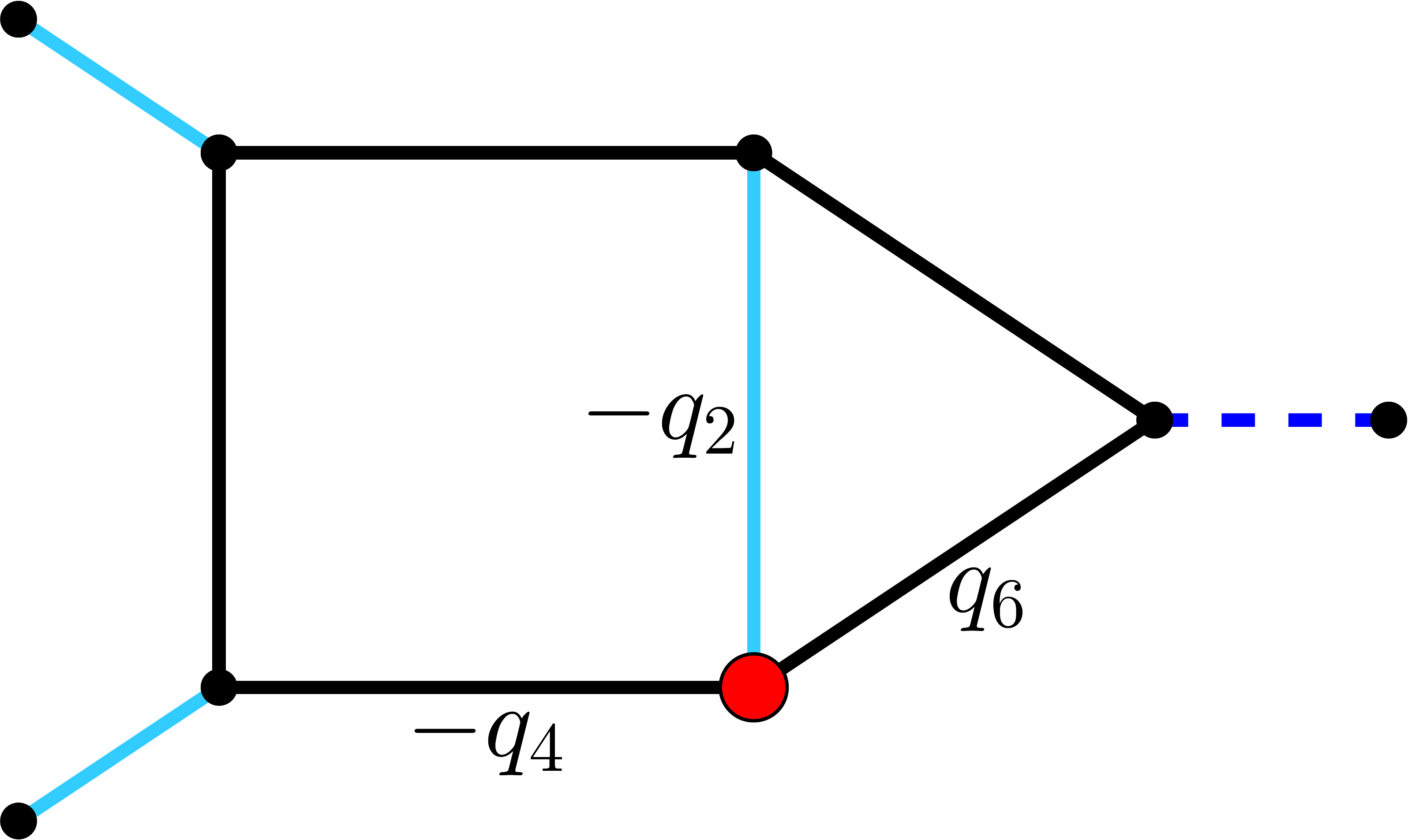}
    \caption{}
    \label{fig:inverse-kirchhoff-v5}
  \end{subfigure}
  \caption{Stepwise illustration of the inverse Kirchhoff reconstruction
  for the two-loop \mbox{$gg\to H$} example corresponding to sector
  \texttt{SIX\_LINE/63}. Panel (a) shows the three external edges before
  any internal vertices have been identified, with incoming momenta
  $p_1$, $p_2$ and outgoing momentum $p_3=p_1+p_2$. The six internal
  propagator momenta $q_1,\ldots,q_6$ are given in
  Eq.~\eqref{eq:inverse-kirchhoff:six-line}. The auxiliary denominator is
  excluded. Panels (b)--(f) display one successful branch of the
  recursive search. In each step the red dot marks the internal vertex
  identified at that stage, and the signed momentum labels on the
  incident edges sum to zero. The successive local conservation
  combinations are $(p_1,q_1,-q_3)$ in (b),
  $(p_2,-q_1,q_4)$ in (c), $(-p_3,q_5,-q_6)$ in (d),
  $(q_3,q_2,-q_5)$ in (e), and $(-q_4,-q_2,q_6)$ in (f).
  After the final step all edge incidences have been assigned and the
  connected graph is complete. Solid light-blue lines denote massless
  edges, black propagator lines carry mass label \texttt{M2}, and the
  dashed external line represents the Higgs leg. A separate
  post-processing step determines globally consistent edge orientations.}
  \label{fig:inverse-kirchhoff-steps}
\end{figure}

\paragraph{Illustration of the recursive search.}
Figure~\ref{fig:inverse-kirchhoff-steps} illustrates the reconstruction
of the two-loop \mbox{$gg\to H$} example. Initially, only the external edges
with their fixed external endpoints are drawn. The internal propagator
momenta are supplied by the integral-family definition. At each recursive
step, a trial subset must contain the first still-unused incidence.
Since the external indices occur first in the incidence list, the search
starts with vertices adjacent to external legs. In this example, the first
three accepted vertices are adjacent to $p_1$, $p_2$, and the outgoing
Higgs leg, respectively. The next two involve only internal edges.
This is not, however, a geometric growth algorithm: intermediate pieces
need not be connected, as panel (d) illustrates. If a partial assignment
cannot be completed, the search backtracks to examine alternatives.
In panels (b)--(f), the red dot marks the newly identified vertex, and only
its incident momenta are labelled, with signs chosen to give a vanishing sum.
Once all incidences have been assigned, the candidate is checked for
connectivity. The subsequent orientation procedure is described in
Section~\ref{sect:inverse-kirchhoff:routing}.

\subsection{Momentum routing and consistent edge orientations}
\label{sect:inverse-kirchhoff:routing}

Topology reconstruction and momentum routing have related but distinct
roles.  During ordinary diagram generation, \diagen{} assigns a new
momentum routing from a graph using a spanning tree, as explained in
Section~\ref{sect:preliminaries:momenta}.  In the inverse problem, by
contrast, the momentum expressions $q_e$ are already specified: what
remains after constructing the graph is to find a \emph{globally
consistent orientation} of every edge carrying those momenta.

The local test, Eq.~\eqref{eq:inverse-kirchhoff:local-test}, only proves
that some choice of signs works at each vertex separately.  It does not
yet guarantee that the two incidences of a shared propagator carry
opposite signs.  After selecting a candidate scalar diagram, the algorithm
therefore determines compatible edge orientations by a depth-first
search through the graph.  The search starts at an internal vertex
adjacent to the first external edge (or at the first vertex for a vacuum
graph), keeping one reference edge direction fixed to remove the overall
reversal freedom. At each internal vertex the search respects
edge orientations already settled at earlier vertices, explores
orientations for the remaining incident edges, and tests exact
coefficient-wise momentum conservation.  It backtracks when the local
constraint cannot be satisfied.  External degree-one vertices are not
subjected to conservation tests, and the orientation of self-loops is
irrelevant.  If no consistent set of orientations exists, an error is
reported for the selected candidate. Otherwise, the reconstructed scalar diagram is
returned with edge directions satisfying Eq.~\eqref{eq:inverse-kirchhoff:conservation}.

The six-line example makes this last distinction explicit:
the labels in Fig.~\ref{fig:inverse-kirchhoff-steps} use opposite signs
for each of $q_1,\ldots,q_6$ at its two endpoints.  In particular,
the identity at $V_5$ could equally have been found with all signs
reversed during the local search, but the global orientation step
selects the compatible convention.  The post-reconstruction procedure
therefore \emph{checks and orients the supplied momentum assignment}. It does not generate an independent set of loop momenta.

\subsection{Scope and limitations}
\label{sect:inverse-kirchhoff:limitations}

The implemented reconstruction is a graph-theoretic consistency test. It does not infer field content or validate whether an interaction is
allowed by a given model.  It assumes connected graphs without
degree-two internal vertices.  The \texttt{draw\_diagrams} interface
requires at least two external legs, although the underlying
reconstruction algorithm also supports vacuum graphs.
Selected sector entries must represent actual physical edges. Generic
irreducible numerator entries cannot automatically be treated as
propagator edges.  The sector bitmask selects edges, and its set bits
must therefore correspond to the intended physical denominators.

The YAML momentum parser supports signed sums of momentum symbols with
coefficients $+1$ and $-1$.  Expressions with explicit multiplication
or parentheses, such as \texttt{2*k1} or \texttt{k1-(p1+p2)}, are not
supported. Zero-momentum edges are rejected.  The integral-family list
is limited to 63 entries by the interface.  These restrictions arise
from the implemented reconstruction procedure, not from the \kira{}
format.  If the selected momenta cannot give rise to an admissible
connected graph, the program reports
\texttt{inconsistent set of momenta}.  Although eliminating duplicate
subsets and bounding vertex degrees reduces redundant search, the
incidence-partition problem remains combinatorial in the worst case.
The resulting scalar diagram should consequently be interpreted as one
consistent representation of the selected family sector, rather than
as a uniquely determined diagram with model-specific field assignments.

\section{Output interpretation}
\label{sect:output}
After executing the job file \texttt{jobs.yaml}, \diagen{} produces output files stored in four different directories: \texttt{amplitude}, \texttt{pictures}, \texttt{kira\_interface}, and \texttt{compute}.
\begin{itemize}
\item The \texttt{amplitude} directory contains
the generated amplitude and the associated
prototype definitions. The file
\texttt{output.diagen} contains the amplitude
in a form suitable for subsequent processing
with \form{}. The files \texttt{PR<N>definition},
where \texttt{N} is a nonnegative integer,
contain the propagator definitions of the
corresponding prototypes.
 \item The directory \texttt{pictures} contains all figures illustrating individual diagrams and prototypes.
 \item The directory \texttt{kira\_interface} contains
integral-family and kinematics definitions for the
selected top-level prototype representatives.
These are ready to be processed by \kira{}.
\item The directory \texttt{compute} contains
scripts for the user-defined \form{} treatment
and projection of the amplitude.
\end{itemize}

The main output file \texttt{output.diagen} in the directory \texttt{amplitude} contains the amplitude generated with \diagen{}.
Each term in the amplitude is associated with a diagram and one or more candidate prototypes. This file is ready for further treatment with \form{}. Each term is a product of Feynman rules as defined in the model file \texttt{model.yaml} in the directory \texttt{config}. \diagen{} multiplies each term with a fermion sign and a symmetry factor corresponding to the diagram. All terms belonging to one fermion chain are written in the correct order.

An automatic \form{} script \texttt{auto-sum.frm} is generated which takes the amplitude as input.
The \texttt{compute} directory is generated by
\diagen{}, whereas the \texttt{form} directory
containing the model-specific \form{} procedures
must be supplied by the user. The provided
examples include suitable procedures, which may
need to be adapted to the process under consideration.
In particular, a projector has to be defined in \texttt{TreatAmplitude.prc}. In \diagen{} the symbol \texttt{DS} is reserved for a scalar propagator. The user has to provide rules in \texttt{FeynmanRules.prc} to translate user-defined field functions as defined in the model file to \texttt{DS} functions.

An interface to \kira{} is established by automatically creating the necessary \kira{} \texttt{config} directory. For each top-level prototype an individual \texttt{config} directory is created. In the generated \texttt{integralfamilies.yaml} files, the physical propagators are listed first and the ISPs that complete the integral-family basis are written as \texttt{bilinear} entries, for example \Verb^{ bilinear: [ [ "k1", "p1" ], 0 ] }^. The top-level sector contains only the physical propagators.
For the combined \kira{} configuration \texttt{PRall}, the exported integral families are ordered by decreasing number of propagators in their designated top-level sector, with stable ordering for ties. This changes the order of family entries but not their denominator or mass definitions. Since family order can affect the choice between overlapping mappings, the ordering convention is part of the reproducible output configuration.

After successful execution of the \form{} processing step, the generated scripts collect the scalar integrals required by the amplitude and prepare the corresponding \kira{} reduction jobs.

\section{Model file \texorpdfstring{\texttt{model.yaml}}{model.yaml}}
\label{sect:model}

\begin{figure}[!t]
  \centering
  \begin{subfigure}[t]{0.29\textwidth}
    \vspace{0pt}\centering
    \includegraphics[width=\linewidth]{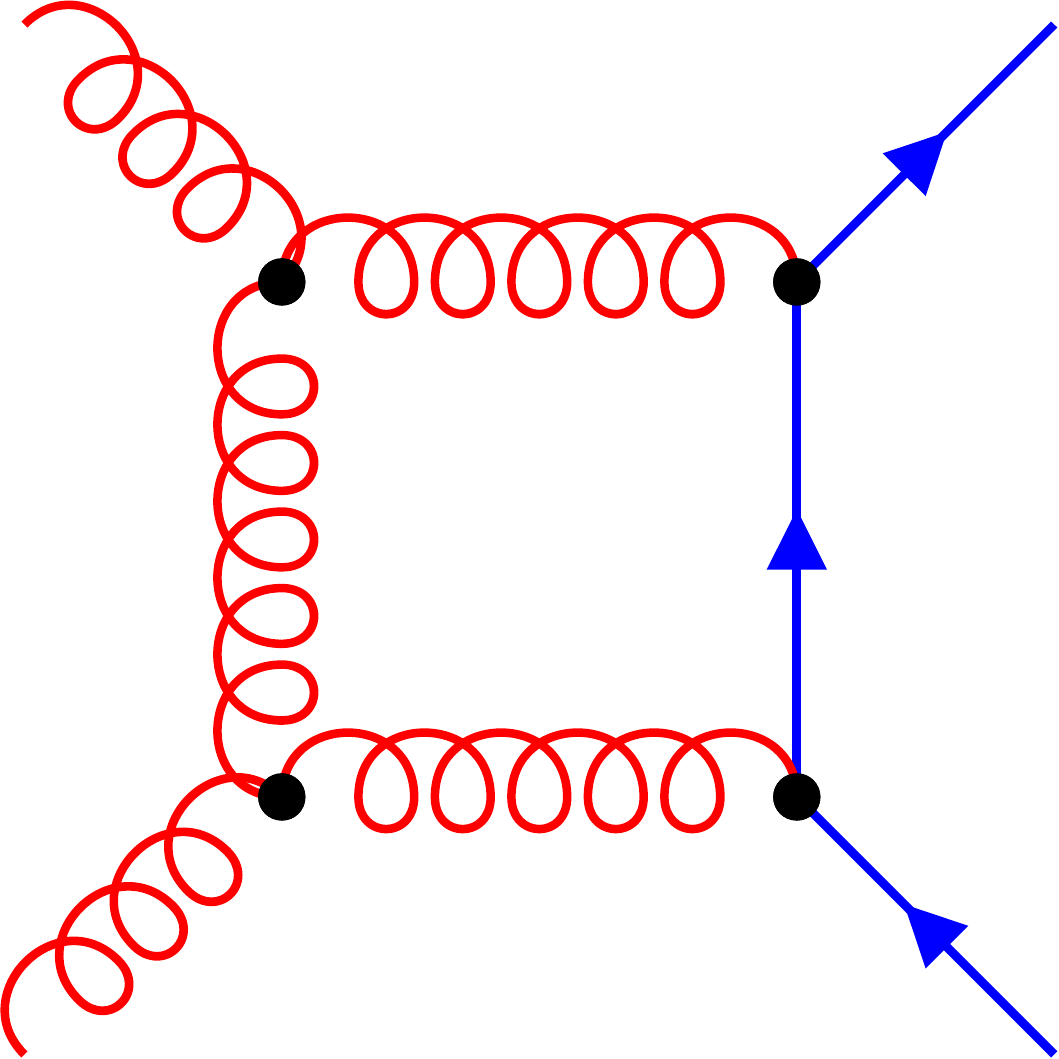}
    \caption{Feynman diagram.}
  \end{subfigure}\hfill
  \begin{subfigure}[t]{0.29\textwidth}
    \vspace{0pt}\centering
    \includegraphics[width=\linewidth]{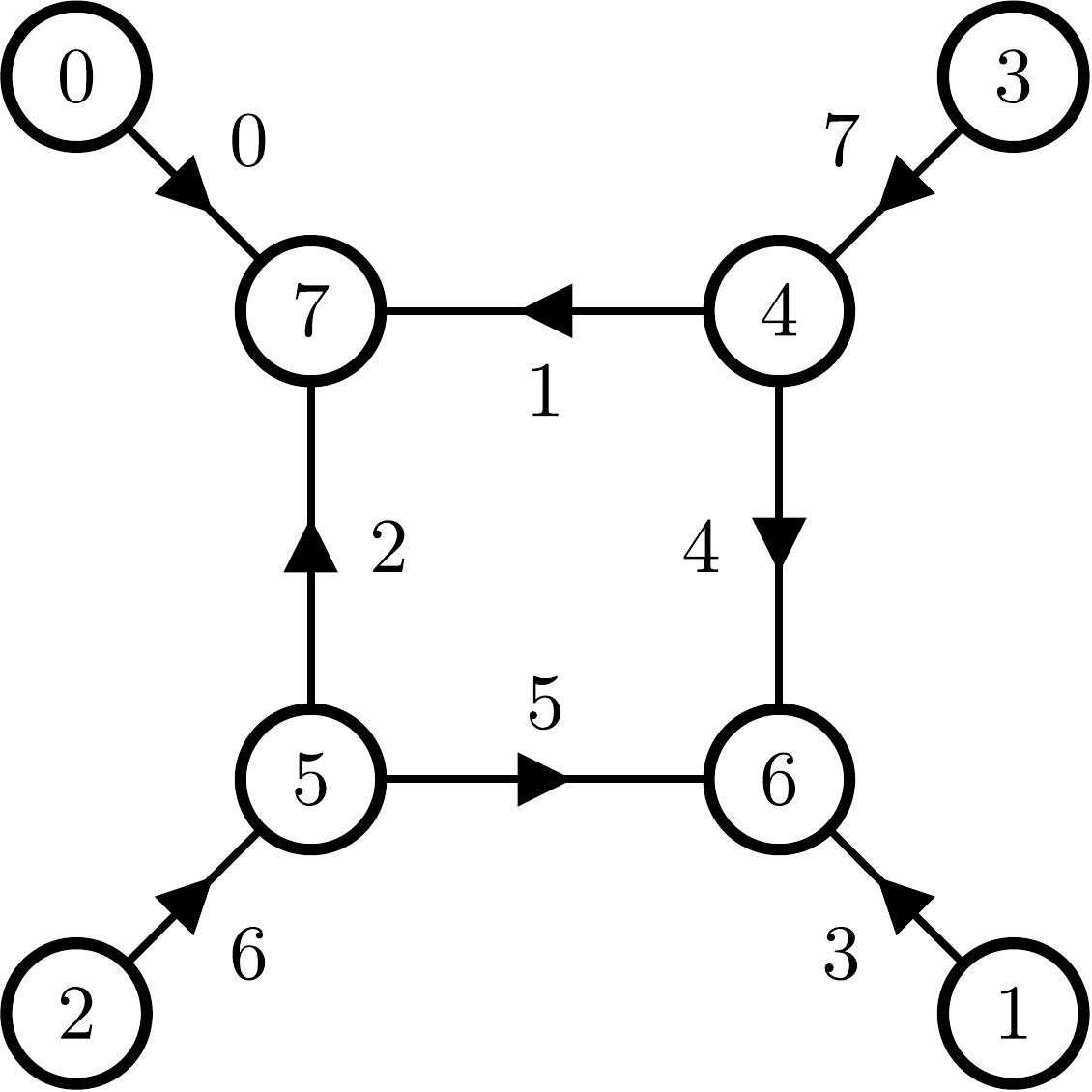}
    \caption{Internal topology: node/edge IDs and orientations.}
  \end{subfigure}\hfill
  \begin{subfigure}[t]{0.29\textwidth}
    \vspace{0pt}\centering
    \includegraphics[width=\linewidth]{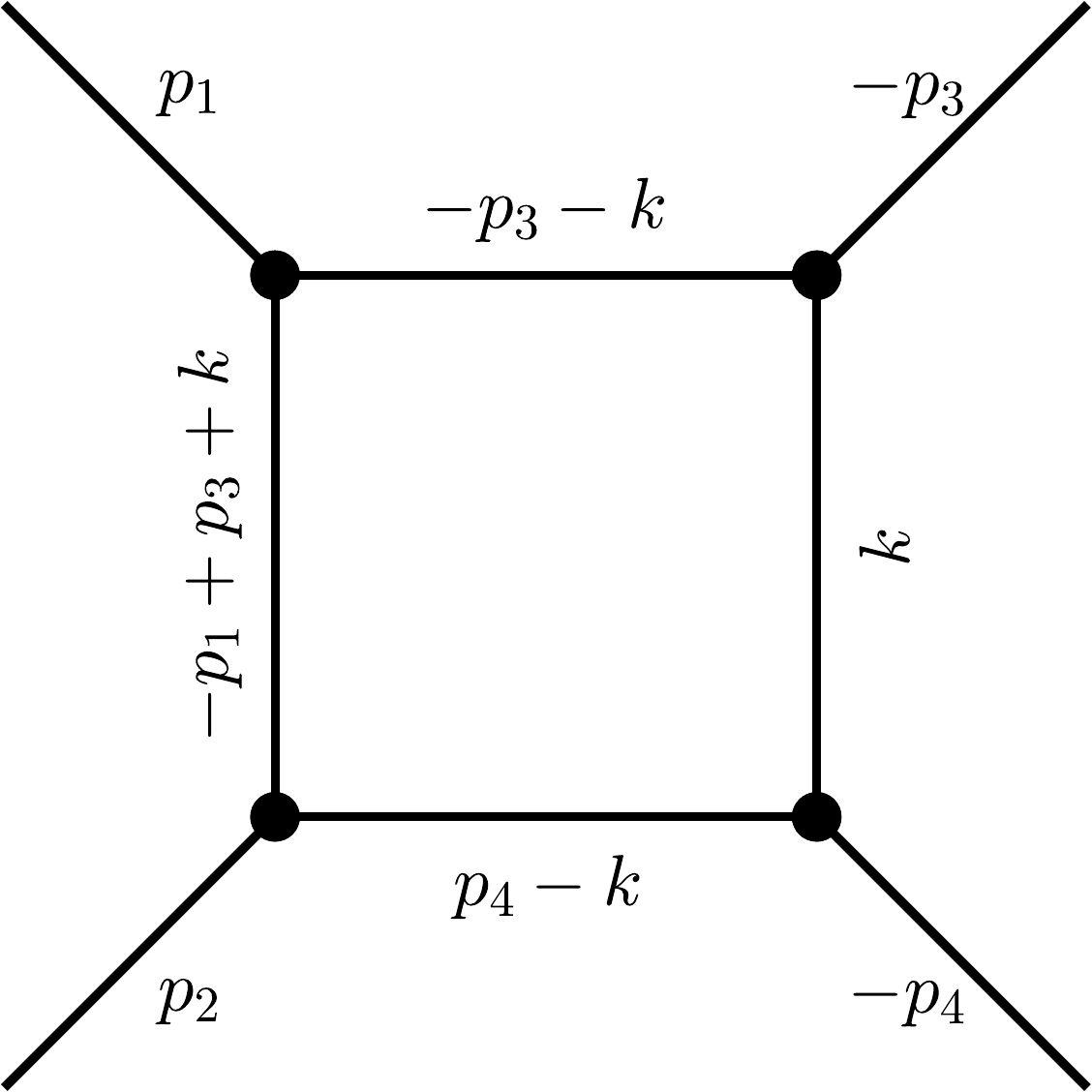}
    \caption{Graph momentum routing.}
  \end{subfigure}%

  \vspace{0.8em}

  \begin{subfigure}[t]{0.29\textwidth}
    \vspace{0pt}\centering
    \includegraphics[width=\linewidth]{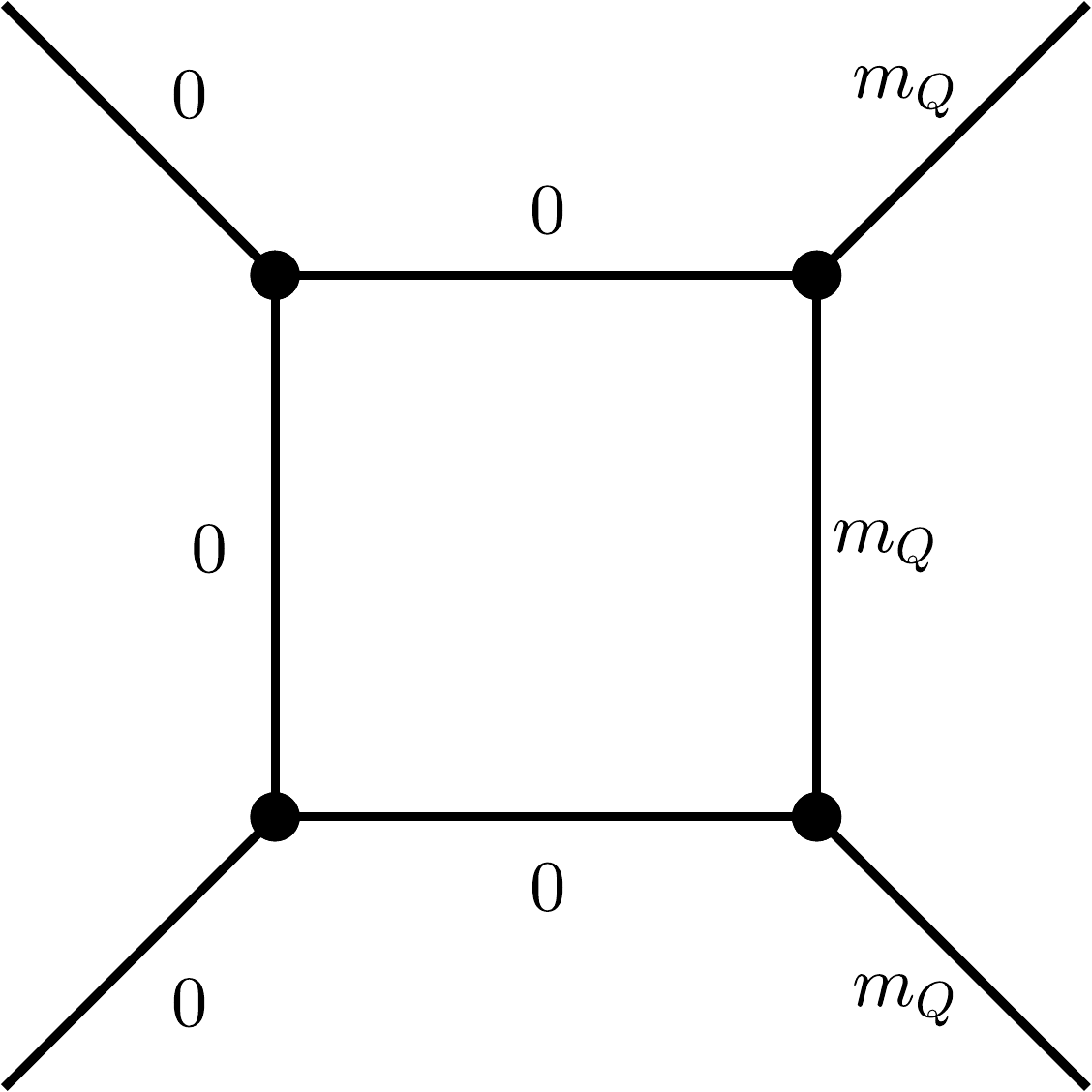}
    \caption{Mass labels (\texttt{\%m}).}
  \end{subfigure}\hfill
  \begin{subfigure}[t]{0.29\textwidth}
    \vspace{0pt}\centering
    \includegraphics[width=\linewidth]{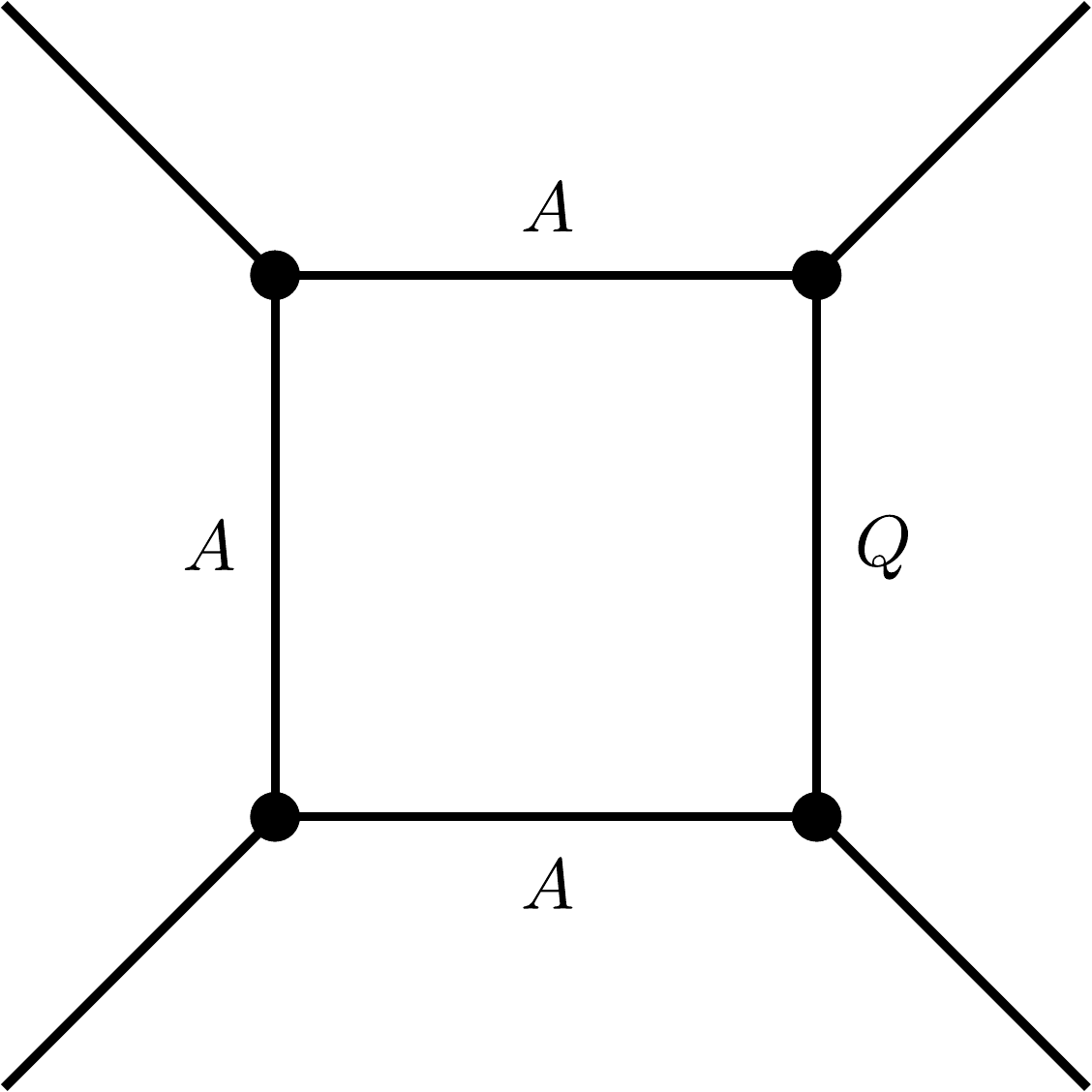}
    \caption{Field labels (\texttt{\%n}).}
  \end{subfigure}\hfill
  \begin{subfigure}[t]{0.29\textwidth}
    \vspace{0pt}\centering
    \includegraphics[width=\linewidth]{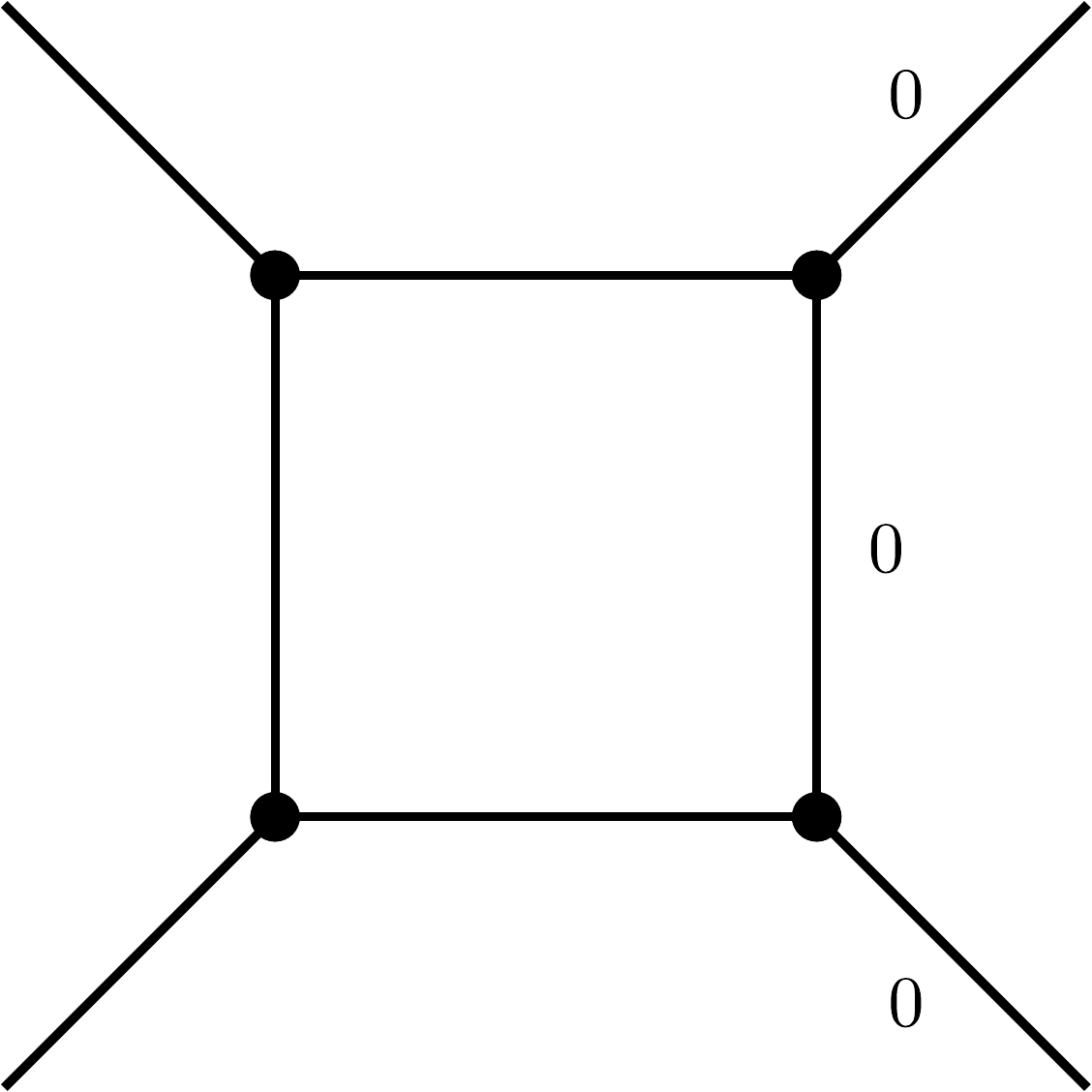}
    \caption{Fermion chains (\texttt{\%c}).}
  \end{subfigure}%

  \caption{Diagram information used to instantiate propagator, wavefunction, and vertex templates in \texttt{model.yaml}. Blue arrows indicate the chosen graph-edge orientations. Momentum arguments substituted into the templates follow the conventions of the corresponding propagators, external wavefunctions, and interaction vertices.}
  \label{fig:model-metadata}
\end{figure}

In \diagen{}, the file \texttt{model.yaml} specifies the field content,
propagators, wavefunctions, and interaction vertices. These objects are given as
template strings that may contain \texttt{\%}-macros. During diagram generation,
\diagen{} assigns unique identifiers to nodes and edges, fixes an orientation for
each edge (source $\to$ target), and determines momenta, masses, field labels,
and fermion chains. The templates are then instantiated by replacing the
\texttt{\%}-macros with the corresponding diagram-specific data.
Figure~\ref{fig:model-metadata} illustrates these ingredients for a
representative diagram from the Getting-to-Know example.

\paragraph{Internal propagator macros.}
Each internal edge connects a source node to a target node.
The edge orientation fixes the graph conventions used for
momentum routing and fermion flow. Internal propagator
templates support the following macros:
\begin{itemize}
  \item \texttt{\%s}: source node number,
  \item \texttt{\%t}: target node number,
  \item \texttt{\%e}: edge number,
  \item \texttt{\%p}: momentum argument of the propagator,
  \item \texttt{\%m}: propagator mass label,
  \item \texttt{\%n}: canonical field name assigned to the edge (the second entry of the \texttt{field} pair),
  \item \texttt{\%c}: fermion-chain index,
  \item \texttt{\%is}, \texttt{\%it}: index shortcuts
        associated with the source and target nodes,
        respectively.
\end{itemize}
The momentum supplied through \texttt{\%p} follows the
propagator conventions of the corresponding field.
In particular, for anticommuting fields, its sign may
differ from the source-to-target momentum shown in
the graph representation.

For example, the propagator template
\texttt{DV(v\%sl\%e,v\%tl\%e,\%p,\%m)}
constructs Lorentz-index labels from the source node,
target node, and edge number, while inserting the
propagator momentum and mass.

\paragraph{External wavefunction macros.}
External wavefunctions use a separate set of macros:
\begin{itemize}
  \item \texttt{\%p}: physical external momentum label,
  \item \texttt{\%m}: mass label of the external field,
  \item \texttt{\%v}: interaction node adjacent to
        the external edge,
  \item \texttt{\%l}: external edge number,
  \item \texttt{\%c}: fermion-chain index.
\end{itemize}
Here, \texttt{\%v} refers to the interaction node,
not to the degree-one external node. For example,
the wavefunction template
\texttt{Eps(v\%vl\%l,\%p,cOli\%l)}
produces \texttt{Eps(v7l0,p1,cOli0)} for the
corresponding external leg in the example below.
The macros available for external wavefunctions
are not interchangeable with those of internal
propagators.

\paragraph{Vertex macros (interaction templates).}
Each interaction vertex is assigned a unique number,
and the attached legs are ordered and indexed starting
from \(1\). Vertex templates support:
\begin{itemize}
  \item \texttt{\%v}: interaction vertex number,
  \item \texttt{\%g}: counterterm order assigned
        to the interaction vertex,
  \item \texttt{\%c}: fermion-chain index,
  \item \texttt{\%ij}: shorthand index object
        for leg \(j\),
  \item \texttt{\%lj}: edge number of leg \(j\),
  \item \texttt{\%pj}: momentum entering the
        interaction vertex through leg \(j\),
  \item \texttt{\%mj}: mass label of leg \(j\),
  \item \texttt{\%nj}: canonical field name of leg \(j\), rather than necessarily the particle or antiparticle identifier at that endpoint.
\end{itemize}
In the leg-dependent macros, \(j\in\{1,\ldots,9\}\)
denotes a single-digit leg index, not a literal
character or an arbitrary multi-digit number.
For example, \texttt{\%i3} constructs the index
associated with the third leg, while
\texttt{\%p3} supplies its momentum.
The momentum arguments \texttt{\%pj}
(with \(j\) replaced by the corresponding digit)
are defined as incoming to the interaction vertex,
independently of the source-to-target orientation
of the graph edges.

These macros provide the diagram-specific information
needed to construct interaction vertices, including
their Lorentz and color structures.

Starting from the physical diagram in
Fig.~\ref{fig:model-metadata}a, \diagen{}
constructs the internal directed graph representation
shown in Fig.~\ref{fig:model-metadata}b. Nodes and
edges carry unique identifiers, and each edge is
assigned an orientation (source $\to$ target).
This orientation determines the graph momentum-routing
conventions and is used in the treatment of fermion flow.

Given the process kinematics, \diagen{} assigns
momenta to the edges as illustrated in
Fig.~\ref{fig:model-metadata}c. The momentum arguments
substituted into the model templates depend on their
context. For internal propagators, the momentum
may differ by a sign from the oriented graph momentum,
in particular for anticommuting fields. External
wavefunctions receive the physical external momentum
label, whereas momentum arguments in vertex templates
are incoming to the corresponding interaction.

For example, the internal edge labelled
\texttt{4: 4 -> 6} carries momentum \texttt{k}
in the graph description of the one-loop
\mbox{$gg\to Q\bar{Q}$} example, while the corresponding
fermion propagator is written as
\texttt{SF(0,-k,M)} in the generated amplitude.
The difference reflects the propagator momentum
convention and does not indicate an inconsistency
in the graph routing.

Mass labels are illustrated in
Fig.~\ref{fig:model-metadata}d and field assignments
in Fig.~\ref{fig:model-metadata}e. For internal
propagators these are accessible through
\texttt{\%m} and \texttt{\%n}, respectively.
Fermionic lines are organized into chains
(Fig.~\ref{fig:model-metadata}f), whose indices
are available through \texttt{\%c}.

\section{Reproducibility and agent-driven workflows}
\label{sect:agentic}

\diagen{} is deterministic: for a fixed program version and fixed input files (\texttt{jobs.yaml}, \texttt{model.yaml}, and \texttt{kinematics.yaml}), every run produces the same diagrams in the same order, with the same field assignments, edge orientations, momentum routings, fermion signs, symmetry factors, prototype labels, and integral-family definitions. We have verified this by running the examples of Section~\ref{sect:examples} repeatedly in fresh working directories, including runs with modified memory-allocation settings, and comparing the resulting \texttt{amplitude} and \texttt{kira\_interface} directories byte by byte.

Determinism is not automatic in programs of this kind. Graph algorithms often iterate over containers whose order depends on memory addresses, which change from run to run. While preparing this release, an AI coding agent tasked with checking the reproducibility claims of this paper found exactly such a dependence: fields and vertices were ordered by their addresses in memory. As a result, the diagram numbering within a topology, the orientation of closed fermion loops, and the corresponding momentum labels could differ between two runs of the same job, although the physical content --- the number of diagrams and the set of sign and symmetry factors in each topology --- was unchanged. The ordering now follows the order in which fields and vertices are defined in the model file.

Deterministic output makes \diagen{} well suited as a building block in workflows in which an AI coding agent prepares inputs, runs the program, and processes its results:
\begin{itemize}
  \item \emph{Verification by comparison.} Two runs can be compared with standard tools such as \texttt{diff -r}. Any difference indicates a change in the input or in the program and never arises from chance.
  \item \emph{Regression tests.} Reference outputs for the supplied examples can be stored and compared automatically after every modification of the code or the model files, including modifications proposed by an agent.
  \item \emph{Separation of tasks.} The agent writes and adapts model, kinematics, and \form{} files, while the combinatorial core --- diagram generation, symmetry factors, fermion signs, and topology identification --- is carried out by a tested deterministic program rather than generated anew for each calculation.
  \item \emph{Traceability.} Diagram identifiers such as \texttt{d33} and prototype labels such as \texttt{PR19} refer to the same objects in every run, so that results, logs, and discussions can refer to them unambiguously.
\end{itemize}

Determinism concerns repeatability, not correctness: an identical result can still be wrong if the model or the \form{} procedures are wrong, and the checks recommended in Section~\ref{sect:BestPractice} remain necessary. Identifiers and orientations may change when the model, the job options, or the program version are changed. Results should therefore be quoted together with the \diagen{} version and the input files used.

\section{Benchmarks}
\label{sect:benchmarks}
Tables~\ref{tab:benchmarks:prelim} and~\ref{tab:benchmarks:final} report the wall-clock run
times for the specified example processes as a function of loop order. The timings characterize
two different configurations of the prototype search. They should not be interpreted as a
direct speed comparison with other diagram generators, since the amount of subprototype
processing differs.

The processes correspond to the example directories
\texttt{HeavyQuarkPropagator} (\mbox{$Q\to Q$}),
\texttt{ggH} (\mbox{$gg\to H$}),
\texttt{ggQQBar} (\mbox{$gg\to Q\bar{Q}$}), and
\texttt{ggQQBarH} (\mbox{$gg\to Q\bar{Q}H$}) of the source distribution,
with the loop order and the \Verb^children^ setting adjusted as described below.
All runs were performed on an x86-64 workstation equipped with an Intel Core i9-10980XE processor (18 physical cores, 36 hardware threads, base clock frequency of 3.0\,GHz) and 125\,GiB of system-reported memory, running openSUSE Leap 16.0. Diagram and prototype drawing were disabled in both configurations (\Verb^draw_graphs: false^ and \Verb^draw_protos: false^).

In the \emph{preliminary} configuration (Table~\ref{tab:benchmarks:prelim}), \Verb^children^ is
set to a high value, \Verb^children: 100^, to restrict the exploration of contracted
subprototypes. This is useful for obtaining an initial diagram and integral-family
representation at comparatively low cost.

For the \emph{final} configuration (Table~\ref{tab:benchmarks:final}), the cutoff is chosen separately for each process and loop order. Specifically, \Verb^children^ is set to the smallest number of physical propagators among the physical sectors identified in the preliminary run, as described in Section~\ref{sect:BestPractice}. This is the largest cutoff that still permits contractions down to the smallest of these sectors, so that matching among the generated physical sectors is not excluded by the cutoff. The additional subprototype construction and matching increase the computational cost. Lower cutoffs may be necessary if mappings to further subsectors are required.

\begin{table}[th]
    \centering
    \begin{tabular}{@{}lccccc@{}}
        \toprule
        Process & \multicolumn{5}{c@{}}{Loop order} \\
        \cmidrule(lr){2-6}
        & 1 & 2 & 3 & 4 & 5 \\
        \midrule
        \mbox{$Q \to Q$}            & {$<1$\,s} & {$<1$\,s} & {$<1$\,s} & {$5$\,s}  & {$1027$\,s} \\
        \mbox{$gg \to H$}           & {$<1$\,s} & {$<1$\,s} & {$<1$\,s} & {$36$\,s} & -- \\
        \mbox{$gg \to Q\bar{Q}$}    & {$<1$\,s} & {$<1$\,s} & {$66$\,s} & --        & -- \\
        \mbox{$gg \to Q\bar{Q}H$}   & {$<1$\,s} & {$5$\,s}  & --        & --        & -- \\
        \bottomrule
    \end{tabular}
    \caption{Wall-clock run times for the preliminary benchmark runs (\texttt{children:~100}) as a function of loop order.}
    \label{tab:benchmarks:prelim}
\end{table}

\begin{table}[th]
    \centering
    \begin{tabular}{@{}lccccc@{}}
        \toprule
        Process & \multicolumn{5}{c@{}}{Loop order} \\
        \cmidrule(lr){2-6}
        & 1 & 2 & 3 & 4 & 5 \\
        \midrule
        \mbox{$Q \to Q$}            & {$<1$\,s} & {$<1$\,s} & {$<1$\,s} & {$35$\,s}   & {$19533$\,s} \\
        \mbox{$gg \to H$}           & {$<1$\,s} & {$<1$\,s} & {$1$\,s}  & {$1624$\,s} & -- \\
        \mbox{$gg \to Q\bar{Q}$}    & {$<1$\,s} & {$2$\,s}  & {$1295$\,s} & --        & -- \\
        \mbox{$gg \to Q\bar{Q}H$}   & {$<1$\,s} & {$109$\,s} & --        & --         & -- \\
        \bottomrule
    \end{tabular}
    \caption{Wall-clock run times for the final benchmark runs as a function of loop order, with the cutoff set to the smallest physical sector found in the corresponding preliminary run.}
    \label{tab:benchmarks:final}
\end{table}

\section{Best practice}
\label{sect:BestPractice}

\begin{itemize}
 \item Enable diagram and/or topology drawing during debugging to visually inspect the generated output. Note that rendering can significantly slow down the generation.
  \item To choose the \Verb^children^ cutoff for a production run, first perform a preliminary run with a value larger than the number of physical propagators in any generated prototype (e.g.\ \Verb^children: 100^ for the examples considered here). This suppresses the generation of contracted subprototypes. Determine the smallest number $t_{\min}$ of physical propagators among the sectors obtained in this run. For example, inspect the \Verb^top_level_sectors^ entries in the generated \path{kira_interface/PRall/config/integralfamilies.yaml}: the number of set bits in each sector identifier gives the number of physical propagators, excluding the auxiliary bilinears. Set \Verb^children^ to $t_{\min}$ for the production run. This is the optimal cutoff with respect to the range of contractions needed to match the generated physical sectors: a larger cutoff can miss such relations, whereas a smaller one additionally explores lower subsectors. If mappings involving sectors below $t_{\min}$ are needed, lower the cutoff accordingly. Repeat the procedure when changing the process, loop order or diagram-selection criteria.
 \item To validate that the symmetry factors implied by your Feynman rules agree with those used by \diagen{}, generate a tree-level sample by setting \Verb^loop: 0^. Comparing \texttt{amplitude/output.diagen} with your expectations provides a direct consistency check.
  \item When \diagen{} is run by scripts or AI coding agents, keep the input files under version control, record the program version, and confirm reproducibility by running a job twice and comparing the output directories with \texttt{diff -r}.
  \item Report the chosen prototype-search cutoff alongside the resulting numbers of generated diagrams, retained prototypes, and exported top-level integral families. A smaller family collection is useful only when the mapping remains valid for all scalar integrals required by the calculation.
\end{itemize}

\section{Installation}
\label{sect:installation}

\subsection{Obtaining \diagen{}}

The source code of \diagen{} is available from the project's public \gitlab{} repository at \url{https://gitlab.com/diagen-cpp/diagen}. For reproducibility, we recommend using the release version~1.0 associated with this publication, available under the corresponding release tag in the repository. Development versions may contain additional features and changes not included in the release.

\subsection{Prerequisites}
\label{sect:prerequisites}

\noindent
\textit{Platform requirements}\smallskip\\
\code{Linux x86\_64} or \code{macOS}.

\medskip\pagebreak[2]
\noindent
\textit{Compiler requirements}\smallskip\\
A \code{C++} compiler supporting the \code{C++14} standard.

\medskip\pagebreak[2]
\noindent
\textit{Build system requirements}\smallskip\\
\diagen{} is built with the \code{Autotools} build system.

\medskip\pagebreak[2]
\noindent
\textit{Dependencies}\smallskip\\

\diagen{} requires the following libraries:
\begin{itemize}
  \item \code{GiNaC} and its dependency \code{CLN},
  \item \code{yaml-cpp},
  \item \code{GMP},
  \item \code{zlib}.
\end{itemize}

The build system uses \code{Autotools} and \code{pkg-config}. When regenerating the build configuration, the corresponding \code{pkg.m4} macros must be available.

Diagram and prototype visualization requires \code{Graphviz}, including \code{neato}. Additionally, \code{Ghostscript} (\code{gs}) is required to combine individual diagram and topology drawings into the files \texttt{d-all.pdf} and \texttt{t-all.pdf}, respectively. Without Ghostscript, individual drawings can still be generated, but the combined PDF files are not produced.

The supplied amplitude-processing examples require \form{} and standard shell and text-processing utilities. Subsequent integration-by-parts reductions require \kira{}.

Dependencies can generally be installed using the system package manager. When using libraries from nonstandard installation directories, the compiler and linker search paths can be specified through \code{CPPFLAGS} and \code{LDFLAGS}, respectively, and \code{PKG\_CONFIG\_PATH} can be used to locate package metadata.

The libraries must be compatible with the target architecture and \texttt{C++} application binary interface (ABI). Using exactly the same compiler for every dependency is not required, provided binary compatibility is maintained.

\subsection{Compiling \diagen{} with the Autotools build system}
\label{installwithautotools}

First run
\begin{verbatim}
  autoreconf -i
\end{verbatim}
and then compile and install with
\begin{verbatim}
  ./configure --prefix=/install/path
  make
  make install
\end{verbatim}
where the optional \code{-{}-prefix} argument sets the installation prefix.

The source distribution also contains the
\texttt{examples} directory, which provides the
configuration files and \form{} procedures needed
to reproduce the calculations discussed in this
paper. These examples are not installed by
\texttt{make install}. Users should copy the
desired example directory from the unpacked
source distribution to a writable working directory
before running \diagen{}.

\section{Scope and limitations}
\label{sect:limitations}

The present diagram generator assumes a set of external fields and propagators with unambiguous field-to-conjugate-field pairing. Vacuum diagrams without external particles, mixed-field propagators, and Majorana-fermion contractions are not handled by the general diagram-assignment algorithm described here. These limitations concern the model classes supported directly by the program and should not be confused with restrictions of the subsequent \form{} or IBP reduction steps.

Furthermore, the fermion-chain and sign-handling
algorithms assume that each interaction traversed
by a fermion chain contains a pair of anticommuting
fields. More general multi-fermion contact
interactions are not covered by the validated
implementation and may require extensions of
these algorithms.

Model-specific Lorentz structures, color algebra, and projector definitions are supplied through user-defined templates and \form{} procedures. The program organizes the diagrammatic and scalar-integral structures, but it does not automatically validate the physical normalization of arbitrary model inputs. Similarly, graph-based matching identifies equivalences within the implemented representation and is not a proof of global minimality under all possible integral identities.

The graph-theoretical library also includes routines for enumerating heavy subgraphs for large-mass expansions and constructing reduced graphs. A user-facing expansion-by-subgraphs workflow, including the subsequent Taylor expansion and assembly of the amplitude, is outside the scope of this release. Similarly, although the generation options described in Section~\ref{sect:examples:ggh} allow diagrams to be selected according to the field content of admissible cuts, the construction of individual cut diagrams and their adapted momentum routings is not yet exposed as a dedicated user workflow. Support for the generation and export of cut integral families is planned for a future release.

These restrictions define the scope of the present release. In particular, applications involving more general field mixing or nonstandard fermion contractions require additional validation and, where appropriate, extensions of the underlying generation rules.

\section{Conclusions}
\label{sect:conclusions}

We have presented \diagen{}, a diagram-generation framework that connects model-level amplitudes to symmetry-aware representations of their scalar-integral content. Its distinctive component is the construction of a prototype map during generation, including graph-based matching, integral-level canonical comparison for vacuum and two-point prototypes, admissible subtopologies, and the momentum and index transformations needed to express contributions in common integral representations. The output combines \form{}-readable amplitudes with integral-family and kinematic data suitable for subsequent IBP reduction. Conversely, the inverse Kirchhoff mode reconstructs connected scalar diagrams from integral-family sectors and determines orientations consistent with the supplied momenta.

The examples demonstrate the workflow and the measured cost of different prototype-search settings. The benchmarks also show why the scope of subprototype matching must be specified when quoting generation times. Further quantitative comparisons should separate diagram-generation cost from matching cost and assess the impact of the generated mappings on the downstream integral representation. Such tests provide a reproducible basis for evaluating the method as multiloop calculations become more demanding.

Because its output is deterministic and plain-text, \diagen{} can serve as a reliable component of agent-driven workflows, in which AI coding agents orchestrate a calculation while exact steps such as diagram generation and topology identification are delegated to reproducible programs. We expect this division of labour to become increasingly important as such agents are adopted in precision calculations.

\section*{Acknowledgments}

J.U. thanks Arnd Behring for insightful discussions on diagram generation with \texttt{QGRAF}, and Andrzej Czarnecki, who encouraged us to finalize this work.
The work of M.C. was supported by the Deutsche Forschungsgemeinschaft (DFG) through the grant 396021762 - TRR 257: Particle Physics Phenomenology after the Higgs Discovery.
The work of M.N. was supported by the European Research Council (ERC) under the European Union's Horizon 2020 research and innovation program grant agreement 101019620 (ERC Advanced Grant TOPUP). 
The work of J.U. has been supported in part by the Deutsche Forschungsgemeinschaft (DFG, German Research Foundation) Projektnummer 417533893/GRK2575 ``Rethinking Quantum Field Theory'' and by the European Union through the European Research Council under grant ERC Advanced Grant 101097219 (GraWFTy).
The Feynman diagrams in this paper were drawn with \texttt{FeynGame}~\cite{Harlander:2020cyh,Harlander:2024qbn,Bundgen:2025utt}.
AI tools based on large language models (Claude by Anthropic and ChatGPT by OpenAI) assisted in writing and editing this paper. The authors reviewed and verified all content and take full responsibility for it.


\bibliographystyle{JHEP}
\bibliography{bib}

\end{document}